\documentclass[11pt, a4paper]{article}
\usepackage{rotating}
\usepackage{authblk}
\usepackage[margin=1in]{geometry}
\usepackage{graphicx}
\usepackage{subcaption}
\usepackage{amsthm}
\DeclareCaptionFont{my_9pt}{\fontsize{9}{10}\small}
\usepackage[justification=centering]{caption}
\usepackage{tabulary}
\usepackage{setspace}
\usepackage{amsmath}
\usepackage{multirow}
\usepackage[nocomma]{optidef}
\usepackage[titletoc,title]{appendix}
\usepackage[ruled,vlined,linesnumbered]{algorithm2e}
\usepackage[colorlinks,linkcolor=black,anchorcolor=black,citecolor=black,urlcolor=black]{hyperref}
\usepackage{lineno}
\usepackage[shortlabels]{enumitem}
\usepackage{palatino}
\usepackage{newpxtext}
\usepackage{newpxmath}
\usepackage{soul}
\usepackage{longtable}
\usepackage{stackengine}
\usepackage{titlesec}

\usepackage{natbib}
 \bibpunct[, ]{(}{)}{,}{a}{}{,}%

\usepackage{xcolor}
\usepackage{mleftright}
\usepackage{dsfont}
\usepackage{mathtools}
\usepackage{threeparttable}
\titleformat{\paragraph}
{\normalfont\normalsize\bfseries}{\theparagraph}{1em}{}
\titlespacing*{\paragraph}
{0pt}{3.25ex plus 1ex minus .2ex}{1.5ex plus .2ex}

\newcommand\numberthis{\addtocounter{equation}{1}\tag{\theequation}}
\newtheorem{definition}{Definition}
\newtheorem{lemma}{Lemma}
\newtheorem{proposition}{Proposition}
\newtheorem{remark}{Remark}
\newtheorem{fact}{Fact}
\newtheorem{assumption}{Assumption}
\newtheorem{theorem}{Theorem}

\newtheorem{example}{Example}

\DeclareMathOperator*{\argmin}{arg\,min}

\newcommand{\x}{\boldsymbol{x}}
\newcommand{\y}{\boldsymbol{y}}
\newcommand{\xt}[1]{\x^{#1,(t)}}
\newcommand{\xtt}[1]{\x^{#1,(t+1)}}
\newcommand{\bp}{\boldsymbol{\pi}}
\newcommand{\pxtt}[1]{\bp^{#1,(t+1)}}
\newcommand{\xa}[1]{\x^{#1,\ast}} 
\newcommand{\xl}[1]{\x^{#1,\lozenge}} 
\newcommand{\xd}[1]{\x^{#1,\dagger}} 

\newcommand{\xs}[1]{\x^{#1,\star}} 
\newcommand{\xc}[1]{\x^{#1,\circ}} 

\newcommand{\pxa}[1]{\bp^{#1,\ast}} 
\newcommand{\pxl}[1]{\bp^{#1,\lozenge}} 
\newcommand{\pxd}[1]{\bp^{#1,\dagger}} 

\newcommand{\pxs}[1]{\bp^{#1,\star}} 
\newcommand{\pxc}[1]{\bp^{#1,\circ}} 
\newcommand{\tx}{\tilde{\x}}
\newcommand{\txt}{\tx^{(t)}}
\newcommand{\txa}{\tx^{\ast}}
\newcommand{\txl}{\tx^{\lozenge}}
\newcommand{\txd}{\tx^{\dagger}}

\newcommand{\txs}{\tx^{\star}}
\newcommand{\txc}{\tx^{\circ}}
\newcommand{\bz}{\boldsymbol{z}}
\newcommand{\bmu}{\boldsymbol{\mu}}
\newcommand{\btau}{\boldsymbol{\tau}}
\newcommand{\bd}{\boldsymbol{d}}
\newcommand{\indone}{\mathds{1}}
\newcommand{\bone}{\boldsymbol{1}}
\newcommand{\bzero}{\boldsymbol{0}}
\newcommand{\bigzero}{\mbox{\normalfont\large\bfseries 0}}
\newcommand{\cc}{\boldsymbol{c}}
\newcommand{\ccop}[1]{\cc\mleft( #1 \mright)}
\newcommand{\cpxtt}[1]{\cc\mleft( \bp^{#1,(t+1)} \mright)}
\newcommand{\zt}[1]{\bz^{#1,(t)}}
\newcommand{\halpha}{\hat{\alpha}}
\newcommand{\htheta}{\hat{\theta}}
\newcommand{\hgamma}{\hat{\gamma}}
\newcommand{\hzeta}{\hat{\zeta}}
\newcommand{\bQ}{\bar{Q}}
\newcommand{\bv}{\boldsymbol{v}}
\newcommand{\pder}[2]{\frac{\partial #1}{\partial#2}}
\newcommand{\distance}[2]{\mleft\lVert{#1}\mright\rVert^{#2}}
\newcommand{\bigbracket}[1]{\mleft( #1 \mright)}
\newcommand{\bigsbracket}[1]{\mleft[ #1 \mright]}
\newcommand{\bigcbracket}[1]{\mleft\{ {#1} \mright\}}
\newcommand{\bigvbracket}[1]{\mleft\vert {#1} \mright\vert}
\newcommand{\sQop}[1]{Q\bigsbracket{#1}}
\newcommand{\dQop}[2]{Q_{#1}\bigsbracket{#2}}
\newcommand{\dPhiop}[2]{\Phi^{#1}\bigsbracket{#2}}
\newcommand{\tPhiop}[3]{\Phi^{#1}_{#2}\bigsbracket{#3}}
\newcommand{\dUpop}[2]{\Upsilon^{#1}\bigsbracket{#2}}
\newcommand{\sDop}[1]{D\bigsbracket{#1}}
\newcommand{\sJop}[1]{J\bigsbracket{#1}}
\newcommand{\dHop}[2]{H_{#1}{\bigsbracket{#2}}}
\newcommand{\sPop}[1]{P\bigsbracket{#1}}
\newcommand{\dPop}[2]{P_{#1}{\bigsbracket{#2}}}

\newcommand{\bolda}{\boldsymbol{a}}
\newcommand{\mycomment}[1]{}

\title{\vspace{-1.5cm}Cognitive Hierarchy in Day-to-day Network Flow Dynamics}
\author[a]{Minyu Shen}
\author[b]{Feng Xiao\footnote{Corresponding author. Email: evan.fxiao@gmail.com}}
\author[c]{Weihua Gu}
\author[c]{Hongbo Ye}

\affil[a]{School of Management Science and Engineering, Southwestern University of Finance and Economics, China}
\affil[b]{Business School, Sichuan University, China}
\affil[c]{Department of Electrical and Electronic Engineering, The Hong Kong Polytechnic University}

\date{\vspace{-4.5em}}

\begin{document}
\maketitle
\noindent\hrulefill

\doublespacing

\section*{Abstract}

When making route decisions, travelers may engage in a certain degree of reasoning about what the others will do in the upcoming day, rendering yesterday's shortest routes less attractive. This phenomenon was manifested in a recent virtual experiment that mimicked travelers' repeated daily trip-making process. Unfortunately, prevailing day-to-day traffic dynamic models failed to faithfully reproduce the collected flow evolution data therein. To this end, we propose a day-to-day traffic behavior modeling framework based on the Cognitive Hierarchy theory, in which travelers with different levels of strategic-reasoning capabilities form their own beliefs about lower-step travelers' capabilities when choosing their routes. Two widely-studied day-to-day models, the Network Tatonnement Process dynamic and the Logit dynamic, are extended into the framework and studied as examples. Calibration of the virtual experiment is performed using the extended Network Tatonnement Process dynamic, which fits the experimental data reasonably well. Our analysis reveals that both extended dynamics exhibit multiple equilibria, one of which corresponds to the classical user equilibrium. We further analyze and characterize these non-user-equilibrium states. While analyzing global stability is intractable due to the presence of multiple equilibria, local stability criteria near equilibria are developed analytically. General insights on how key parameters affect the stability of user equilibria are unveiled.
\\
\noindent \textbf{Keywords:} day-to-day traffic dynamics; cognitive hierarchy; strategic thinking; route choice behavior; multiple equilibria; experiment calibration

\section{Introduction}
Wardrop's first principle characterizes a steady-state User Equilibrium (UE) wherein no driver can reduce their travel time by unilaterally changing the route. However, due to the ever-changing nature of a transportation network (e.g., induced by traffic incidents, capacity modifications, and network structure changes), traffic states may always be in a disequilibrium state \citep{kumar2015day}. To explain how travelers adjust their route-choice behaviors and predict those out-of-equilibrium states, various day-to-day traffic-network flow models have been proposed in the literature over the years.\par

Most day-to-day models, whether continuous or discrete, concentrate on modeling how the \textit{past} evolution of traffic flow influences travelers' route choice behavior \textit{today}. The differences among the experienced travel times are the internal driving force that induces the flow change. For example, a rational behavior adjustment process (RBAP) introduced by \citet{yang2009day} generalizes a flow evolution process wherein the total travel cost today will decrease based on yesterday's experienced cost. This framework encompasses numerous day-to-day models, including the proportional-switch adjustment \citep{smith1984stability}, the projected dynamical system \citep{nagurney1997projected}, the network tatonnement process \citep{friesz1994day}, the evolutionary traffic dynamic \citep{sandholm2010population}, and the simplex gravity flow dynamic \citep{smith1983existence}. Some studies model travelers' memory processes by updating the perceived cost through an exponential moving average of past perceived and actual costs \citep{horowitz1984stability,cantarella1995dynamic,bie2010stability,xiao2016physics, ye2021day}.


Largely ignored in the existing modeling efforts is the fact that travelers may predict what other travelers will do in the \textit{future} (upcoming day). Substantial evidence in psychological studies demonstrates that when making decisions, people engage in a certain degree of reasoning about what others will do. This reasoning process is referred to as ``theory of mind,'' the ability to understand another person's mental state \citep{lo2017adaptive}. In a route-choice context, suppose now a transportation network is in a disequilibrium state, and travelers in the costlier route have the incentive to switch to the shortest route with the minimum travel time. A traveler may reason like: ``if many travelers choose the shortest route, then I will try to avoid being on that route''. This phenomenon was somehow manifested in a recent day-to-day virtual experiment \citep{ye2018exploration}, but the models calibrated in that study did not take it into account. Intuitively, if all travelers acted with perfect rationality, using only yesterday's complete information without predicting what others might do today, they would all choose yesterday's shortest route. Such behavior would result in permanent flow oscillations between routes. Yet real-world traffic patterns rarely exhibit such extreme fluctuations, suggesting other behavioral factors at play. Previous research has addressed this discrepancy through modeling travelers' inertia --- their reluctance to change routes completely from day to day. However, the strategic prediction of others' route choices remains largely unexplored.

To model the travelers' prediction behaviors, this paper employs the idea of Cognitive Hierarchy (CH) theory from \citet{camerer2004cognitive}, in which travelers form their beliefs of their opponents using an iterated reasoning process. In the CH theory, the heterogeneous players differ in their strategic-reasoning levels (i.e., cognitive capacities). Lower-step players do not carefully think through the whole game. Meanwhile, higher-step players would try to reap benefits by predicting how these lower-level ``careless'' players respond to the current situation. The CH model provided superior explanations in many game-theoretic settings, including guessing games \citep{costa2006cognition} and extensive-form games \citep{ho2013dynamic}. It can also explain many phenomena in economics, marketing, and operation management, including market entry competition \citep{goldfarb2011thinks}, platform competition \citep{hossain2013markets}, information disclosure in movie markets \citep{brown2013estimating} and capacity allocation games \citep{cui2017cognitive}. It is worth mentioning that although the idea stemmed from \citet{camerer2004cognitive}, our proposed model's novelty lies in its dynamic nature. In contrast, most applications of the CH model in the literature are static one-shot games such as the $p$-beauty game.\par

In the literature, two pieces of work have modeled such prediction behavior; see \citet{he2012modeling} and \citet{he2016marginal}. However, both have their limitations. The former work predicted future congestion only once, with prediction memory gradually vanishing over time. The latter work focused on the case where all the travelers are homogenous 1-step travelers who believe others are 0-step travelers making myopic switches based on previous experience. Moreover, this model uses ordinary differential equations which, despite their tractability for mathematical analysis, bring challenges in calibrating with real-world discrete daily trip data. Stability in continuous time may represent weaker results compared to the discrete-time counterpart \citep{li2018managing}. Under certain scenarios, conclusions from continuous- and discrete-time systems may even contradict each other \citep{ye2021day}.\par

In this paper, we establish a general framework of day-to-day network flow dynamics based on the CH theory to analyze travelers' repeated routing behaviors. In this framework, travelers are heterogeneous in their strategic-reasoning capabilities and form their own beliefs about other travelers' strategic-reasoning capabilities and actions. We extend two widely-studied dynamics in the literature into our CH framework: the Network Tatonnement Process (NTP) dynamic for deterministic UE (DUE) and the Logit dynamic for stochastic UE (SUE). The two extended models are referred to as CH-NTP and CH-Logit dynamics, respectively. A previous online virtual experiment in \citet{ye2018exploration}, which conventional day-to-day models could not fit, can now be calibrated by our proposed CH-NTP dynamic reasonably well. Moreover, the two models predict that strategic-reasoning behaviors would lead the system to multiple equilibria, one of which being the classical UE. We formally characterize these equilibria and show that those which differ from UE emerge when travelers excessively anticipate others' behaviors. Jacobian matrices at the equilibrium points are derived and used to form analytical criteria on local stability around UE. Theoretical results and numerical experiments unveiled several insights into how key parameters affect the local stability.\par

The rest of the paper is organized as follows. The following section describes the general modeling framework. The CH-NTP dynamic is presented in Section \ref{sec:NTP} as the DUE example of the framework. Calibration of the real-world experiment with CH-NTP dynamic is furnished in Section \ref{sec:virtual_experiment}. As the SUE example, the CH-Logit dynamic is presented in Section \ref{sec:Logit}. Conclusions are drawn in Section \ref{sec:conclusion} with discussions on future research directions.

\section{Modeling Framework}
\subsection{Setups}\label{sec:setups}
We consider a strongly-connected transportation network $G(N,L,W)$ consisting of a set $N$ of nodes, a set $L$ of links, and a set $W$ of origin–destination (OD) pairs. Each OD pair, $w\in W$, has a travel demand of $d_w$ and is connected by a set $R_w$ of routes. Travelers in the network are categorized into a set $K$ (with $|K|$ denoting the cardinality) of classes (steps) by their cognitive-capacity levels. (Throughout this paper, the terms ``class'' and ``step'' are used interchangeably.) We assume for simplicity that the proportion of $k$-step travelers, $p^k$ ($k\in K$ and $\sum_{k=0}^{|K|-1}{p^k}=1$), is the same across all the OD pairs. This assumption allows us to write the scaled feasible route flow set of $k$-step travelers as $\Omega_{p^k} \equiv \{ \x^k| \Gamma\x^k=p^k\boldsymbol{d},\x^k\geq0\}$, where $\x^k\equiv(x_{rw}^k, r\in R_w, w\in W)^T$ is the route flow vector (pattern) of class $k\in K$, $\boldsymbol{d}\equiv(d_w, w\in W)^T$ is the OD demand vector and $\Gamma$ is an  OD-route incidence matrix indicating if a route belongs to an OD pair. The superscript ``$T$'' represents the transpose operation. With a little abuse of notation, we denote $\Omega_{\eta}$ as the feasible route flow set with \textit{scaled} OD demand $\eta\boldsymbol{d},\eta\in(0,1]$. Denote $c_{rw}(\x)$ as the actual travel time of route $r\in R_w, w\in W$, under any route flow pattern $\x\in \Omega_{1}$, and $\cc(\x) \equiv (c_{rw}(\x),r\in R_w, w\in W)^T$. Assume that $\cc(\x)$ is continuous in $\x$.\par
\subsection{Flow update rules}\label{sec:flow_update_rules}
On day $t$, the route flow pattern of $k$-step is denoted as $\xt{k}$, and the aggregate observed route flow pattern $\txt=\sum_k{\xt{k}}$. Throughout the paper, we use the notation ``$\sim$'' above a variable to denote the aggregation of route flows across all classes, resulting in a total flow pattern. By observing $\txt$ (or receiving it from the advanced traveler information system), a $k$-step traveler before trip will predict the route flow pattern on ($t+1$)-th day as per her cognitive capacity level (see subsequent Section \ref{sec:cognitive_level}). The predicted route flow pattern and its corresponding travel cost pattern are denoted as $\pxtt{k}$ and $\cpxtt{k}$, respectively. With $\xt{k}$ and $\cpxtt{k}$, the flow pattern of $k$-step travelers on the next day $t+1$ can be updated as:
\begin{equation}\label{eq:general_dynamic}
	\xtt{k} = \dHop{\Omega_{p^k}}{\xt{k},\cpxtt{k};\alpha,\zeta} \equiv \alpha \y\bigsbracket{\xt{k}, \cpxtt{k};\zeta} + (1-\alpha)\xt{k},
\end{equation}
where $\y$ represents the ``target'' flow pattern in $\Omega_{p^k}$ determined by a specific day-to-day dynamical model parameterized by $\zeta$, and $\alpha\in(0,1]$ is a parameter that reflects travelers' inertia or reluctance to change. Note that the predicted flow pattern for any $k$, $\pxtt{k}$, is always in $\Omega_1$. 

Eq.~\eqref{eq:general_dynamic} describes a general day-to-day adjustment process in which travelers partially adjust their route choices toward a target flow pattern $\y$ based on current predicted costs, with $\alpha$ controlling the extent of this adjustment (larger $\alpha$ indicates smaller inertia). This convex combination structure provides a memory and learning process by weighing current decisions against previous ones, a form widely adopted in day-to-day models \citep{friesz1994day,he2010link,han2012link,guo2015link} to represent the learning and adapting process before reaching a steady state \citep{wu2024multiday}. Through instantiating different target flow patterns, we can derive many traditional day-to-day models in the literature from this general formulation.\par

While Eq.~\eqref{eq:general_dynamic} provides a standard form of memory and learning process, we deliberately omit the more complex memory effects presented in previous day-to-day models \citep{horowitz1984stability,cantarella1995dynamic,bie2010stability,xiao2016physics, ye2021day}. These studies represent another line of research that explicitly models the updates of travelers' perceptions of travel costs and the resulting route choices \citep{wu2024multiday}. Our decision to implement the memory and learning mechanism in Eq.~\eqref{eq:general_dynamic} rather than the more complex memory process is based on the following considerations: (i) it aligns with our empirical findings from experimental data \citep{ye2018exploration} analyzed in Section~\ref{sec:virtual_experiment} where incorporating the complex memory process showed minimal improvements in explanatory power; (ii) it helps distinguish the effects of strategic prediction from those of more complicated memory effect; and (iii) it maintains analytical tractability while still capturing important behavioral insights, including the emergence of multiple equilibria (Section~\ref{sec:NTP}) and substantially improved goodness-of-fit to experimental data (Section~\ref{sec:calibration_ch_ntp}).\par

By describing the dynamic in such a general way, most day-to-day models in the literature can be applied to the framework. For example, the day-to-day models satisfying the RBAP property \citep{yang2009day} can be applied, i.e., any $\y[\cdot]$ satisfying:
\begin{equation}\label{eq:rbap_set}
	\y\bigsbracket{\xt{k},\cpxtt{k}; \zeta}
	\left\{
	\begin{aligned}
	& \in \Psi^{k,(t)}, \Psi^{k,(t)} \neq \emptyset \\
	& =\x^{k,(t)}, \Psi^{k,(t)} = \emptyset
	\end{aligned}
	\right.,
\end{equation}
where $\Psi^{k,(t)}\equiv \bigcbracket{\y|\y\in \Omega_{p^k}, \y^T\cpxtt{k}< (\xt{k})^T\cpxtt{k}} $ is the feasible route flow set of $k$-step for which the $k$-step travelers' total travel costs based on the \textit{predicted} costs decreases. This dynamic process describes how, at each time step $t$, the current path flow pattern of $k$-step travelers, $\xt{k}$, evolves toward a target flow pattern $\y$ that reduces the total travel cost of the $k$-step travelers based on their predicted cost $\cpxtt{k}$. If such a cost-reducing flow $\y$ does not exist (i.e., $\Psi^{k,(t)}$ is empty), the flow remains unchanged. This formulation defines the evolution direction from $\xt{k}$ as a set rather than a single point, allowing it to encompass many day-to-day models as special cases.\par

Moreover, applications of the framework are not confined to deterministic day-to-day models; i.e., the $\y[\cdot]$ can also be in a stochastic sense. For example, Logit-based stochastic day-to-day models contained in the ``SUE version'' of the RBAP framework \citep{xiao2019day} can be adopted. In this paper, we study two typical examples for the DUE and SUE, respectively.
\subsection{Cognitive hierarchy levels}\label{sec:cognitive_level}
We now use the CH theory to model how travelers with different cognitive capacities form their beliefs on the next day's flow pattern. To start, 0-step travelers do not think strategically at all. They deem that the flow pattern of day $t$ will remain unchanged on day $t+1$:
\begin{equation}
	\pxtt{0}=\txt.
\end{equation}
Here, $\txt$ represents the aggregate observed route flow pattern on day $t$, which is the sum of route flows across all the classes. In a CH model for the one-shot games, 0-step players were assumed to either randomize equally across all strategies or choose a salient strategy using \textit{ex-ante} information \citep{camerer2004cognitive}. Here, we adopt the latter idea by modeling 0-step travelers using conventional day-to-day models without prediction. This approach offers two advantages: it incorporates conventional day-to-day models as special cases (when $p^0=1$) and provides a tractable, interpretable baseline for higher-level travelers to build their predictions upon.\par

As per the CH theory \citep{camerer2004cognitive}, the $k (k\geq 1)$-step travelers try to take advantage by predicting how lower-step players respond to the current flow pattern. However, they are \textit{overconfident} and do not realize others are using exactly as many thinking steps as they are. Denote $q_k^h = \frac{p^h}{\sum_{i=0}^{k-1}{p^i}}$ as $k$-step travelers' belief about the normalized proportion of $h$-step travelers and $q_k^{h}=0$ for $\forall h\geq k$. As consistent with the CH theory, such a setting means that travelers do not know the exact distribution of lower-step travelers --- they only confidently assume the normalized distribution. For practical implementation, modelers can treat these strategic level proportions ($p^k$) as parameters to be calibrated alongside other model parameters. The calibration can be performed by minimizing the root mean squared error (RMSE) between predicted and observed flow patterns across multiple days. We demonstrate this approach in Section~\ref{sec:calibration_ch_ntp}.

Based on the above idea, a 1-step traveler thinks that all the others are 0-step travelers, i.e.,  $q_1^0=\frac{p^0}{p^0}=1$. Her predicted flow on day $t+1$ is thus formed by looking one step ahead:
\begin{equation}\label{eq:general_pred_flow_step_1}
	\pxtt{1} = \dHop{\Omega_{q_1^0}}{q_1^0\txt, \cpxtt{0};\halpha,\hzeta} = \dHop{\Omega_{1}}{\txt, \ccop{\txt};\halpha,\hzeta},
\end{equation}
where $\halpha\in(0,1)$ and $\hzeta$ are the predicted coefficients in travelers' minds.\par
With this predicted flow pattern, the 1-step travelers will switch the route choice, and the resultant flow pattern becomes:
\begin{equation}
	\xtt{1} = \dHop{\Omega_{p^1}}{\xt{1}, \cpxtt{1};\alpha,\zeta} = \dHop{\Omega_{p^1}}{\xt{1}, \ccop{\dHop{\Omega_{1}}{\txt, \ccop{\txt};\halpha,\hzeta}};\alpha,\zeta }.
\end{equation}\par
2-step travelers will jointly predict both 0-step and 1-step travelers' responses. The normalized proportion of these two types are $q_2^0=\frac{p^0}{p^0+p^1}$ and $q_2^1=\frac{p^1}{p^0+p^1}$, respectively. They anticipate that the flow pattern of the next day will be:
\begin{equation}\label{eq:general_pred_flow_step_2}
	\pxtt{2} = \dHop{\Omega_{q_2^0}}{q_2^0\txt, \cpxtt{0};\halpha,\hzeta} + \dHop{\Omega_{q_2^1}}{q_2^1\txt,\cpxtt{1}; \halpha,\hzeta}.
\end{equation}\par
Here we assume that the travelers of $k$-step ($k\geq 1$) share the same predicted parameters, $\halpha$, and $\hzeta$. This assumption allows us to \textit{qualitatively} model whether the higher-step travelers over- or under-predict lower-step travelers' behaviors. It also keeps our model parsimonious. Despite this simple treatment, as we will see in the subsequent Section \ref{sec:virtual_experiment}, the model still captures the virtual-experiment data quite well, even when we further set $\halpha=\alpha$ and $\hzeta=\zeta$!\par

Based on $\pxtt{2}$, the 2-step travelers update their route choices as follows:
\begin{equation} 
	\begin{aligned}
	\xtt{2} &= \dHop{\Omega_{p^2}}{\xt{2}, \cpxtt{2}; \alpha,\zeta} \\
	&= \dHop{\Omega_{p^2}}{\xt{2}, \ccop{\dHop{\Omega_{q_2^0}}{q_2^0\txt,\cpxtt{0};\halpha,\hzeta}+ \dHop{\Omega_{q_2^1}}{q_2^1\txt, \cpxtt{1};\halpha,\hzeta}}; \alpha,\zeta}.
	\end{aligned}
\end{equation}\par
The rest types of travelers ($k>2$) can be done in a recursive manner as follows:
\begin{equation}
	\left\{
	\begin{aligned}
	& \pxtt{k} = \sum_{0\leq h<k}{\dHop{\Omega_{q_k^h}}{q_k^h \txt, \cpxtt{h}}}; \halpha,\hzeta]; \\
	& \xtt{k} = \dHop{\Omega_{p^k}}{\xt{k}, \cpxtt{k}; \alpha,\zeta}.
	\end{aligned}
	\right.
\end{equation}\par
Substantial experiments in the behavioral game theory found that the average thinking step across all the players lies between 1 and 2 \citep{camerer2004cognitive,chong2016generalized}. Hence, the rest of this paper focuses on the case where $|K|\leq 3$. Note that these experiments were designed for one-shot games with few subjects (atomic players), unlike our dynamic setting with many non-atomic participants. It is also worth mentioning that the methodology in this paper can also be applied to cases where $|K|>3$.

\subsection{Justifications and interpretations of applying the strategic thinking into the day-to-day traffic dynamics}

Strategic thinking behavior is the core component of our proposed modeling framework. Stemming from the CH theory \citep{camerer2004cognitive}, we model heterogeneous travelers with different strategic-reasoning capabilities. Higher-level travelers predict the actions of lower-level travelers but assume no one else possesses their level of sophistication. This fundamental premise has been successfully applied in various domains such as platform competition markets \citep{hossain2013markets}, information disclosure in movie markets \citep{brown2013estimating}, and capacity allocation games in operations management \citep{cui2017cognitive}. We preserve this essential CH mechanism while adapting it to the \textit{non-atomic, dynamic} day-to-day traffic setting.

\subsubsection{Justifications for strategic thinking}

We offer four reasons for incorporating strategic thinking into day-to-day traffic dynamics:

1. Our virtual experiment (which we will discuss in detail in Section~\ref{sec:virtual_experiment}) revealed an interesting phenomenon: during the early stages of the experiment, participants sometimes chose yesterday's second-shortest route over the shortest route. This phenomenon contradicts conventional day-to-day models, which would predict travelers gravitating toward the shortest path. Instead, this phenomenon suggests strategic thinking: participants anticipated that others would choose yesterday's shortest route and deliberately avoided it to escape potential congestion. Moreover, interviews with a subset of the 268 participants revealed that more than one-third described using such strategic route evaluations in their decision-making process.

2. Regarding model calibration for our experiment, we conducted ablation studies that removed higher-level strategic thinking ($k\geq 1$) and treated all travelers as 0-step players (as in conventional day-to-day models). The calibrated results consistently failed to reproduce the experimental data. This failure is shared by all the five conventional day-to-day models examined in \citet{ye2018exploration}. Such consistent failure across multiple models demonstrates the necessity of incorporating strategic thinking.

3. In addition to our experiment, \citet{selten2007commuters}'s comprehensive 200-period route choice experiments also demonstrated that participants explicitly engage in strategic prediction of others' choices. The authors identified a ``contrary'' response mode where participants switch routes after experiencing good outcomes because they anticipated that others would be attracted to those routes in subsequent periods. Similar prediction behavior was also clearly documented in a more recent experiment conducted by \citet{liu2020experimental}.

4. Beyond laboratory experiments, \citet{he2012modeling} analyzed real-world citywide detector data following the I-35W Bridge collapse in Minneapolis, Minnesota. The authors found that drivers made route choices based on predicted traffic flow patterns rather than merely reacting to experienced congestion. Notably, even immediately after the disruption, when drivers had not yet experienced the new traffic conditions, they already adjusted their routes by anticipating potential congestion on what would typically be considered shortest paths. This real-world evidence demonstrates that strategic thinking in route choice extends beyond laboratory settings to actual travel behavior.

\subsubsection{Model interpretations}


Following the CH theory, travelers form beliefs about the distribution of lower-level types using normalized proportions $q_k^h$ rather than exact population distributions. We interpret the normalized proportions as a simplified heuristic representing the ``strength'' of underlying cognitive processes travelers use to anticipate others' behavior. These normalized proportions captures whether prediction tendencies are relatively stronger or weaker within the population. They might be viewed as aggregate \textit{beliefs} held by a class of travelers, rather than exact knowledge possessed by individuals. Furthermore, our day-to-day model operates at a macroscopic level with non-atomic users. At this aggregate level, the proportions can be viewed as emergent properties arising from collective behavior. Finally, our modeling framework aligns with the ``theory of mind'' concept from cognitive psychology --- the ability to attribute mental states and intentions to others \citep{lo2017adaptive}. In our transportation context, the strategic levels and associated beliefs represent a manifestation of travelers' theory of mind about other road users.

\subsubsection{Practical relevance}
Our modeling framework may also be used to characterize the strategic thinking behaviors of major decision-making entities that each control a portion of traffic flow. In this context, the flow distribution vector of one class can be viewed as a decision of one large entity. This approach has particular relevance in modern transportation networks where route choices are increasingly determined by sophisticated navigation systems and centralized management centers operating at different levels of strategic sophistication:

\textbf{Basic reactive systems ($k=0$)}: Simple navigation apps and individual drivers react to current traffic conditions without strategic anticipation. 

\textbf{Predictive navigation platforms ($k=1$)}: Advanced navigation services like Google Maps provide route recommendations based not only on current traffic conditions but also on predictive models of future traffic states \citep{google_maps_ai_2020}. Since recommending the current faster routes to too many users may create new congestion on these routes \citep{sun2020dynamic}, navigation services might adopt a strategic approach. They anticipate the actions of other road users and potential congestion points. Based on these predictions, they recommend shorter routes for their users. Anecdotal evidence from online forums suggests the navigation app Waze may already implement such predictive traffic distribution algorithms to distribute traffic and reduce overall network congestion \citep{reddit2024waze,reddit2025waze}. While such observations are not conclusive, they suggest that principles similar to strategic prediction may already operate in real-world navigation systems. Similarly, Gaode Maps (from China) has reported implementing what they refer to as ``traffic load-balancing'' when planning routes, which distributes traffic across multiple paths to prevent congestion on any single route \citep{pengpai2020gaode}.

\textbf{Strategic traffic management ($k=2$ or above)}: Advanced traffic management centers and mobility service providers may attempt to optimize overall traffic flow by considering multiple levels of interaction between different traveler groups. Their decisions may require anticipating the responses of these groups and adjusting traffic control measures accordingly.

\section{CH-NTP Dynamic}\label{sec:NTP}
The previous section describes the idea using a general day-to-day operator $H[\cdot]$. This section specifies $H[\cdot]$ as the particular NTP dynamic that admits DUE as a fixed point. The NTP dynamic's behavioral explanation is identical to a link-based day-to-day model proposed in \citet{he2010link}: travelers seek to minimize their travel costs while exerting some efforts (or incurring costs) to deviate from incumbent routes. In addition, the NTP dynamic has a unified closed-form formula with the projected dynamical system when route flows only evolve in the interior of the feasible route flow set \citep{nagurney1997projected,guo2015link,xiao2016physics}.\par
In a discrete version, for every $k$-step traveler, the CH-NTP dynamic reads
\begin{equation}\label{eq:tp_model}
	\xtt{k} - \xt{k} = \alpha\bigbracket{\dPop{\Omega_{p^k}}{\xt{k}-\gamma\cpxtt{k}} - \xt{k}}, k\in K,
\end{equation}
where $\dPop{\Omega_{\eta}}{\bz}$ is the projection operator that solves the optimization $\argmin_{\boldsymbol{h}\in \Omega_{\eta}} \distance{\boldsymbol{h}-\bz}{2}$. The $\zeta$ in Eq.~\eqref{eq:general_dynamic} is now replaced by $\gamma (>0)$, which captures the costs incurred by deviations from the incumbent routes. With a larger $\gamma$, travelers have smaller inertia and are prone to switching.\par

As per Section \ref{sec:cognitive_level}, the predicted flow pattern of each class $k\in \{0,1,2\}$ is given by:
\begin{align}
	\label{eq:tp_k0_pred_cost}
	& \pxtt{0} = \txt, \\
	\label{eq:tp_k1_pred_cost}
	& \pxtt{1} = \halpha \dPop{\Omega_1}{\txt-\hgamma\cpxtt{0}} + (1-\halpha)\txt, \\
	\label{eq:tp_k2_pred_cost}
	& \pxtt{2} = \halpha \dPop{\Omega_{q_2^0}}{q_2^0\txt-\hgamma\cpxtt{0}} + \halpha \dPop{\Omega_{q_2^1}}{q_2^1\txt-\hgamma\cpxtt{1}} + (1-\halpha)\txt,
\end{align}
where the predicted parameter $\hzeta$ in Eqs.~\eqref{eq:general_pred_flow_step_1} and \eqref{eq:general_pred_flow_step_2} is replaced by $\hgamma(>0)$. Eqs.~\eqref{eq:tp_k0_pred_cost}-\eqref{eq:tp_k2_pred_cost} indicate that for each $k$, $\pxtt{k}$ is a function of $\txt$.\par

The following lemma introduces a fundamental property of the projection operator, which will be useful in the following analyses.
\begin{lemma}\label{lemma:projection_property}
	\citep{facchinei2007finite}. Let $C$ be a nonempty and closed convex subset of $\mathcal{R}^n$. Then for any $\bz\in \mathcal{R}^n$ and $\boldsymbol{w}\in C$,
	\begin{equation}
		(\dPop{C}{\bz}-\boldsymbol{z})^T(\boldsymbol{w}-\dPop{C}{\bz})\geq 0.
	\end{equation}
	Moreover, $\dPop{C}{\bz}$ is the only point in $C$ satisfying the above relation.
\end{lemma}

\subsection{Mixed prediction-based equilibria}\label{sec:tp_fixed_point}
Fixed points of the dynamical system in Eqs.~\eqref{eq:tp_model}-\eqref{eq:tp_k2_pred_cost} are termed as mixed prediction-based equilibria (MPE) and they are characterized in the following proposition.\par

\begin{proposition}\label{prop:MPE}
	When the route cost function $\ccop{\x}$ is continuous, the dynamical system in Eqs.~\eqref{eq:tp_model}-\eqref{eq:tp_k2_pred_cost} admits at least one MPE (i.e., one fixed point). Moreover, a vector $\x^{\lozenge}\equiv(\xl{k}, \xl{k}\in\Omega_{p^k}, k=0,1,2)^T$ is an MPE if and only if the following variational inequality (VI) holds:
\begin{equation}\label{eq:MPE}	
	\sum_{k=0,1,2}{ \ccop{\pxl{k}\bigbracket{\txl}} ^T \bigbracket{\x^{k} - \xl{k}} } \geq 0, \forall \x\equiv (\x^0,\x^1,\x^2)^T\in \prod_{k=0,1,2}{\Omega_{p^k}},
\end{equation}
where $\bp^{k,\lozenge}$ is a function of  the aggregate flow pattern $\tilde{\x}^{\lozenge} = \sum_k{\xl{k}}$, given by \eqref{eq:tp_k0_pred_cost}-\eqref{eq:tp_k2_pred_cost}.
\end{proposition}
\proof{
	Continuity of the RHS of the dynamical system in Eqs.~\eqref{eq:tp_model}-\eqref{eq:tp_k2_pred_cost} is guaranteed by the continuity of both the projection operator \citep{penot2005continuity} and the route cost function. According to Brouwer's Fixed Point Theorem, at least one fixed point exists for the dynamical system.\par

		Any fixed point of dynamical system in Eqs.~\eqref{eq:tp_model}-\eqref{eq:tp_k2_pred_cost}, denoted as $\x^{\lozenge}\equiv(\xl{k}, \xl{k}\in\Omega_{p^k}, k=0,1,2)^T$, must satisfy:
		\begin{equation}\label{eq:fixed_point_naive_ch_ntp}
		\xl{k} = \dPop{\Omega_{p^k}}{\xl{k}-\gamma\ccop{\pxl{k}\bigbracket{\txl}}}, \forall k,
		\end{equation}
		where $\pxl{k}$ is the predicted flow pattern under $\txl= \sum_k{\xl{k}}$, as defined by Eqs.~\eqref{eq:tp_k0_pred_cost}-\eqref{eq:tp_k2_pred_cost}. We can interpret $\xl{k}-\gamma\ccop{\pxl{k}}$ as $\boldsymbol{z}$ and $\Omega_{p^k}$ as $C$ in Lemma~\ref{lemma:projection_property}, which allows us to rewrite Eq.~\eqref{eq:fixed_point_naive_ch_ntp} as:
		\begin{equation}
			\gamma\ccop{\pxl{k}\bigbracket{\txl}}^T \bigbracket{\x^{k}-\xl{k}} \geq 0,\forall \x^{k}\in \Omega_{p^k}, \forall k.
		\end{equation}
		 These inequalities take the form of VIs. We can combine them because sets $\Omega_{p^k} (k\in K)$ are mutually disjoint \citep{kinderlehrer2000introduction}.
}\endproof

We next define a $|K|$-class DUE in Definition \ref{def:class_ue} (see also in \citealp{nagurney2000multiclass}, and \citealp{zhou2020discrete}) and show that it is one of the MPE in Proposition~\ref{prop:tp_multi_eqs}.
\begin{definition}\label{def:class_ue}
	\citep{nagurney2000multiclass,zhou2020discrete} A route flow pattern $\x^{\ast}\equiv(\xa{k}, \xa{k}\in \Omega_{p^k}, k\in K)^T$ is said to be a $|K|$-class DUE if the following VI holds:
	\begin{equation}\label{eq:class_ue}
		\sum_{k\in K}{\cc^{k}\mleft(\txa\mright)^T \bigbracket{\x^{k}-\xa{k}} } \geq 0, \forall \x\equiv (\x^k,k\in K)^T\in \prod_{k\in K}{\Omega_{p^k}},
	\end{equation}
	where $\txa = \sum_k{\xa{k}}$ and $\cc^k(\txa)$ is the experienced route cost vector of class $k$.
\end{definition}
Note that the aggregate flow $\txa$ equals the classical DUE of a single class in the normal sense. The superscript $k$ can also be dropped from $\cc^k(\txa)$ because all the travelers have the same experience on each route. In the remainder of the paper, the prefix ``$|K|$-class'' will be omitted if the context allows.\par

\begin{proposition}\label{prop:tp_multi_eqs}
	A $|K|(=3)$-class DUE in Definition \ref{def:class_ue} is an MPE of the dynamical system in Eqs.~\eqref{eq:tp_model}-\eqref{eq:tp_k2_pred_cost}, but not vice versa.
\end{proposition}
\proof{

We prove the sufficiency, and the necessity can be easily negated by a counter-example, as will be demonstrated in Section \ref{sec:when_non_due_mpe_emerge}).\par

	At the DUE, Eq.~\eqref{eq:tp_k0_pred_cost} indicates that $\pxa{0}$ equals the aggregate flow pattern, $\txa$. It is well-known that the classical single-class DUE $\txa\in \Omega_1$ satisfies the following VI:
	\begin{equation}\label{eq:mixed_UE_VI}
	\ccop{\txa}^T \bigbracket{\x-\txa} \geq 0, \forall\x\in \Omega_{1}.
	\end{equation}
	We multiply Eq.~\eqref{eq:mixed_UE_VI} by $-\hgamma$. We then add $(\txa)^T (\x-\txa)$ to both sides and rearrange the terms. This yields:
		\begin{equation}
			\bigbracket{\txa-\bigbracket{\txa-\hgamma\ccop{\txa}}}^T \bigbracket{\x-\txa} \geq 0, \forall\x\in \Omega_{1},
		\end{equation}
	which, by Lemma~\ref{lemma:projection_property}, implies that:
		\begin{equation}\label{eq:ue_full_proj}
			\txa = \dPop{\Omega_{1}}{\txa-\hgamma\ccop{\txa}}.
		\end{equation}
	Hence, according to Eq.~\eqref{eq:tp_k1_pred_cost}, at the DUE, the predicted flow pattern of class-1 player, $\pxa{1} = \txa$.\par
	In fact, for any scaling factor $\eta > 0$ and $\hgamma > 0$,
	\begin{align}\label{eq:scale_proj}
		\dPop{\Omega_{\eta}}{\eta\txa-\hgamma\ccop{\txa}} &= \argmin_{y\in \Omega_{\eta}} \distance{\y-\eta\txa+\hgamma \ccop{\txa}}{2} = \argmin_{\y\in \Omega_{\eta}} \distance{\frac{\y}{\eta}-\txa+\frac{\hgamma}{\eta} \ccop{\txa}}{2}\nonumber \\
		& = \eta \argmin_{\y\in \Omega_{1}} \distance{\y-\txa+\frac{\hgamma}{\eta} \ccop{\txa}}{2} = \eta\txa.
	\end{align}
	In Eq.~\eqref{eq:scale_proj}, the first equality applies the definition of projection. The second equality occurs because dividing the objective by a constant $\eta$ does not affect the optimal solution. The third equality holds because the optimization problem commutes with scaling of the demand ($\eta$) due to the linear structure of $\Omega_\eta$. The fourth equality follows from Eq.~\eqref{eq:ue_full_proj}.
	
	Applying Eq.~\eqref{eq:scale_proj} to the definition of the 2-step travelers' predicted cost in Eq.~\eqref{eq:tp_k2_pred_cost}, we obtain that the predicted flow pattern of class-2 player, $\pxa{2}=\txa$.\par
	Summarizing the above, the predicted flow patterns of all the three classes at DUE are exactly $\txa$, and thus Eq.~\eqref{eq:MPE} becomes Eq.~\eqref{eq:class_ue}.
}\endproof

\subsection{Local stability}
The MPE's non-uniqueness makes global stability analysis difficult, if not impossible, because we cannot construct a global convex Lyapunov function that decreases as time passes. We therefore focus on analyzing its local stability. The definitions and theorems regarding local stability used in this paper are relegated to the Appendix \ref{apdx:stabilities}. In this subsection, we first give the Jacobian of the CH-NTP dynamic at the MPE and then analyze the stability near the DUE.

\subsubsection{Jacobian matrix}

The following lemma describes the property of the Jacobian matrix of a projection operator.

\begin{lemma}\label{lemma:Q}
	For a projection of a vector $\bz$ onto $\Omega_\eta \equiv \{ \x| \Gamma\x=\eta\boldsymbol{d},\x\geq0\}$, its Jacobian is a diagonal block matrix:
	\begin{equation}
		\sQop{\bz} =
		\begin{bmatrix}
   			\dQop{w=1}{\bz_1} &  & \bigzero \\ 
   			& \ddots & \\
   			\bigzero &  & \dQop{w=|W|}{\bz_{|W|}}
 		\end{bmatrix},
	\end{equation}
with each block element $\dQop{w}{\bz_w}$ being $\text{Diag}(\indone_w) - \frac{\indone_w\indone_w^T}{|E(\bz_w)|}$, where $\indone_w$ is an indicator vector of size $|R_w|$ whose $r$-th entry is 1 if $r\in E(\bz_w)$ and 0 otherwise. $E(\bz_w)$ is the set of routes on $w$ that exhibit positive flow after applying the operator onto $\Omega^w_\eta \equiv \{ \x| \Gamma_w\x=\eta d_w,\x\geq0\}$, where $\Gamma_w$ is the OD-incidence matrix of OD pair $w$. Each $\dQop{w}{\bz_w}$ is positive semidefinite (PSD) whose eigenvalues are either 0 or 1. It is also idempotent; i.e., $(\dQop{w}{\bz_w})^n=\dQop{w}{\bz_w}, \forall n\in \mathbb{N}=\{1,2,\ldots\},\forall w$. $\text{Rank}(\dQop{w}{\bz_w}) = tr(\dQop{w}{\bz_w}) = |E(\bz_w)|-1$. It has a repeated eigenvalue 1 with an algebraic and geometric multiplicity of both $|E(\bz_w)|-1$. $\text{Rank}(\dQop{w}{\bz_w}-I)=|R_w|-|E(\bz_w)|+1$.\par

If the projected vector lies in the interior of $\Omega^w_\eta,\forall w$ (i.e., all the route flows after projection are positive), $\sQop{\bz}$ degrades to $\bQ$ with each block of OD pair $w$ being $\bar{Q}_w = \text{Diag}(\bone_w) - \frac{\bone_w\bone_w^T}{|R_w|}$ where $\bone_w$ is an all-one vector of size $|R_w|$.
\end{lemma}\par
\proof{Please see Appendix \ref{apdx:jp_jacob}.
}\endproof\par

\begin{remark}
	The algebraic and geometric multiplicities of $\sQop{\bz}$ in Lemma~\ref{lemma:Q} are used to identify the local stability (instead of local \textit{asymptotic} stability) when the Jacobian matrix has eigenvalue(s) of 1; see the requirement of the Jordan block of order 1 in condition (ii) of Theorem~\ref{thm:eigen_stability}.
\end{remark}\par

With Lemma~\ref{lemma:Q} established, we can now derive the Jacobian matrix for the CH-NTP dynamic and utilize it to analyze local stability at any MPE by examining its eigenvalues.

\begin{fact}\label{fact:tp_jacob}
	The Jacobian matrix for the CH-NTP dynamic with $|K|=3$ is a 3-by-3 block matrix:
\begin{equation}\label{eq:tp_jacob}
	JP = \begin{bmatrix}
   	JP_{0,0} & JP_{0,1} & JP_{0,2} \\
   	JP_{1,0} & JP_{1,1} & JP_{1,2} \\
   	JP_{2,0} & JP_{2,1} & JP_{2,2} \\
    \end{bmatrix},
\end{equation}
where each block $JP_{i,j}$ is defined as follows.
{\footnotesize
\begin{align*}
	& JP_{0,0} = \alpha Q_0\cdot\bigbracket{I -\gamma D_0}+(1-\alpha)I, \\
	 & JP_{0,j\neq0}=\alpha Q_0\cdot\bigbracket{-\gamma D_0},\\
	 & JP_{1,1} = \alpha Q_1\cdot\bigbracket{ I-\gamma \bigbracket{ D_1\cdot ( \halpha \hat{Q}_1^0 \cdot (I-\hgamma D_0) + (1-\halpha)I ) } } + (1-\alpha)I,\\
	 & JP_{1,j\neq1}=\alpha Q_1\cdot\bigbracket{-\gamma \bigbracket{ D_1\cdot ( \halpha \hat{Q}_1^0 \cdot (I-\hgamma D_0) + (1-\halpha)I ) } }, \\
	 & JP_{2,2} = \alpha Q_2 \bigbracket{I-\gamma D_2 \bigsbracket{ \halpha \hat{Q}_2^0 (q_2^0I-\hgamma D^0) + (1-\halpha)I + \halpha\hat{Q}_2^1\bigbracket{ q_2^1I-\halpha\hgamma D_1\bigbracket{ \halpha \hat{Q}_1^0 (I-\hgamma D_0)+(1-\halpha)I}}} } + (1-\alpha)I,\\
	 & JP_{2,j\neq2} = \alpha Q_2 \bigbracket{-\gamma D_2 \bigsbracket{ \halpha \hat{Q}_2^0 (q_2^0I-\hgamma D^0) + (1-\halpha)I + \halpha\hat{Q}_2^1\bigbracket{ q_2^1I-\halpha\hgamma D_1\bigbracket{\halpha \hat{Q}_1^0 (I-\hgamma D_0)+(1-\halpha)I} }} }, 
\end{align*}
}
where $D_k\equiv \sDop{\pxtt{k}}$ representing the Jacobian of the route cost functions evaluated at $\pxtt{k}$; $Q^k\equiv \sQop{\xt{k}-\gamma\cpxtt{k}}$; $I$ the identity matrix of size $\sum_{w}{|R_w|}$; $\hat{Q}_1^0 \equiv \sQop{q_1^0\txt-\hgamma\cpxtt{0}}$; $\hat{Q}_2^0 \equiv \sQop{q_2^0\txt-\hgamma\cpxtt{0}}$; and $\hat{Q}_2^1 \equiv \sQop{q_2^1\txt-\hgamma\cpxtt{1}}$.
\end{fact}\par
\proof{
The derivation is relegated to Appendix \ref{apdx:tp_jacob}.
}\endproof\par

As stated in Section \ref{sec:tp_fixed_point}, the strategic-reasoning behavior causes the original NTP dynamic to have multiple MPE, including those non-DUE ones. Local stability regarding all the MPE can be analyzed by the Jacobian in Fact~\ref{fact:tp_jacob}.

\subsubsection{Stability around the DUE}
We are particularly interested in how the travelers' prediction behaviors would affect the stability near the DUE. Numerically, it is always tractable to calculate $JP$'s eigenvalues at the DUE and check its local stability. However, general analytical insights on the parametric space may rely on certain assumptions. Recognizing this, we assume that as the day-to-day process evolves, the projection always yields a positive flow on each route. Under this assumption, the NTP dynamic reduces to a closed-form expression \citep{sandholm2010population,xiao2016physics}. The $\alpha$ and $\gamma$ (and $\halpha$ and $\hgamma$) can be simply combined (specifically, multiplied) to represent the sensitivity to the cost difference between two routes; see Appendix A in \citet{xiao2016physics}. We can therefore fix $\alpha=\halpha=1$ and vary $\gamma$ and $\hgamma$ while the behavioral explanation is not compromised.\par
We divide the discussion into two parts: (i) when higher-step travelers can exactly predict lower-step travelers' switching tendency, i.e., $\hgamma=\gamma$; and (ii) when the prediction is inaccurate, i.e., $\hgamma\neq\gamma$. The following assumption is made on the route cost function.
\begin{assumption}\label{assumption:cost_psd}
	The Jacobian of the route cost function $\cc(\x)$ w.r.t. route flow pattern $\x\in \Omega_1$, denoted by $\sDop{\x}$, is symmetric and positive semidefinite.
\end{assumption}
A widely-adopted case of Assumption \ref{assumption:cost_psd} is when link travel time functions are separable, differentiable, increasing, and additive (i.e., the route travel time is equal to the sum of travel time on all links that constitute the route).

\paragraph{Perfect prediction $\hgamma=\gamma$}
\begin{proposition}\label{prop:tp_ue_stability_eq_gamma}
For a dynamical system with Eqs.~\eqref{eq:tp_model}-\eqref{eq:tp_k2_pred_cost} featuring any $|K|\in \{1,2,3\}$, if the projection operator can always generate positive flow on each route (hence $\halpha$ and $\alpha$ can be set to 1) and $\hgamma=\gamma$, its DUE under Assumption \ref{assumption:cost_psd} is locally stable if all the moduli of the eigenvalues of matrix $\bQ(I-\gamma D_{\ast})$ are less than 1, where $D_{\ast}$ is the Jacobian of the route cost function evaluated at the aggregate DUE, $\txa$, and $\bQ$ is defined in Lemma~\ref{lemma:Q}.
\end{proposition}
\proof{
Please see Appendix \ref{apdx:tp_ue_stability_eq_gamma}.
}\endproof

\begin{remark}
	Proposition~\ref{prop:tp_ue_stability_eq_gamma} shows that conditions for ensuring local stability near the DUE are equivalent for any $|K|\in \{1,2,3\}$ when $\hgamma=\gamma$. In other words, thinking multiple steps ahead does not affect the local stability at the DUE when higher-step travelers can exactly predict lower-step ones' switching behavior. Moreover, the stability is independent of the distribution of heterogeneous travelers.\par
	
	The key distinction between models with different $|K|$ values appears in their response to instability. Our numerical experiments show that when instability occurs with $|K|=1$, the system tends to permanently oscillate near the DUE, repeatedly crossing this equilibrium point. Conversely, systems with $|K|\geq 2$ are more likely to evolve toward and remain at an MPE that differs from the DUE.
\end{remark}\par
The following lemma describes the property of the stability criterion used in Proposition~\ref{prop:tp_ue_stability_eq_gamma}, $\sQop{\bz}(I-\gamma \sDop{\x})$, by relating it to the PSD matrix $\sQop{\bz}\sDop{\x}$, where $\sQop{\bz}$ and $\sDop{\x}$ are not necessarily equal to $\bQ$ and $D_{\ast}$, respectively.
\begin{lemma}\label{lemma:A_property}
	Under Assumption \ref{assumption:cost_psd}, for any given $\bz$ and $\x$, denote $\mu_i$ as $\sQop{\bz}\sDop{\x}$'s $i$-th eigenvalue and $\mu_i \geq 0, \forall i$. The corresponding eigenvalue of $\sQop{\bz}\bigbracket{I-\gamma\sDop{\x}}$ is $1-\gamma\mu_i$ if $\mu_i\neq 0$. Therefore, as $\gamma$ increases from 0 to $\infty$, the modulus of the $i$-th eigenvalue of $\sQop{\bz}(I-\gamma \sDop{\x})$, $|1-\gamma\mu_i|$, will first decrease from $1$ to $0$ and then increase to $\infty$. As a result, there exists a $\bar{\gamma}=\frac{2}{\max_i{\mu_i}}$ such that when $\gamma < \bar{\gamma}$, the maximum modulus $\max_i{|1-\gamma\mu_i|} < 1$ and that when $\gamma > \bar{\gamma}$, $\max_i{|1-\gamma\mu_i|} > 1$.
\end{lemma}
\proof{
Please see Appendix \ref{apdx:A_property}.
}\endproof

\begin{remark}
	To provide realistic context for our stability conditions, note that the critical threshold $\bar{\gamma}\equiv \frac{2}{\max_i{\mu_i}}$ reflects the network's inherent properties, specifically how steeply costs increase with traffic flow. Networks with steeper cost functions have larger eigenvalues $\mu_i$ of $\bQ D_{\ast}$, resulting in a smaller $\bar{\gamma}$ and making them naturally more prone to instability. The parameter $\gamma$ represents travelers' sensitivity to cost differences between routes. When $\gamma < \bar{\gamma}$, the system maintains stability at equilibrium, while larger values of $\gamma$ may lead to oscillatory traffic patterns as travelers rapidly shift between routes.
\end{remark}

\paragraph{Imperfect prediction when $|K|=2$}\label{sec:neq_gamma}
The consistency of stability conditions under different $|K|$ breaks down when $\hgamma\neq\gamma$. We first turn to analyze a simple toy network to get some intuitions. Consider an OD pair served by two routes with $|K|=2$. According to Proposition~\ref{prop:tp_multi_eqs}, at the DUE, $\pxtt{0}=\pxtt{1}$ and thus the Jacobians of route cost function $D_1=D_0=D_{\ast}$. Without loss of generality, we assume $D_{\ast}=\begin{bsmallmatrix} a&b\\b&c\end{bsmallmatrix}$ with $a,b,c>0$ and $ac-b^2\geq 0$ (positive semidefiniteness in Assumption \ref{assumption:cost_psd}). Using the similar derivation in Proposition~\ref{prop:tp_ue_stability_eq_gamma}, we can obtain that $JP$ with $|K|=2$ has four eigenvalues $0$, $0$, $1$ and $\frac{1}{4}\gamma  \hgamma(a-2b+c)^2-\gamma(a-2b+c)+1$, with the last one being denoted as $f(a-2b+c;\gamma,\hgamma)$. Note that the stability only depends on $f(a-2b+c;\gamma,\hgamma)$ and that $a-2b+c \geq 0$ because $ac-b^2 \geq 0$. We refer to stable/unstable region as region of $\gamma$ and $\hgamma$ that makes the DUE stable/unstable (i.e., whether $f(a-2b+c;\gamma,\hgamma)\in (-1,1)$ holds or not).\par
The following observations can be made:
\begin{itemize}
	\item When $\hgamma=\gamma$, $-1< f(a-2b+c;\gamma,\hgamma) < 1$ simplifies to $a-2b+c < \frac{4}{\gamma}$.
	\item When $\hgamma > \gamma$, the stable region is still $a-2b+c < \frac{4}{\hgamma}$. Compared to the case of $\hgamma=\gamma$, the size of the stable region is shrunk.
	\item When $\frac{\gamma}{2} < \hgamma < \gamma$, the minimum point of $f(a-2b+c;\gamma,\hgamma)$, $1-\frac{\gamma}{\hgamma}$, is always $>-1$. Since the zero points of $f(a-2b+c;\gamma,\hgamma)=1$ are 0 and $\frac{4}{\hgamma}$, respectively, the stable region becomes $a-2b+c < \frac{4}{\hgamma}$. Compared to the case of $\hgamma=\gamma$, the size of the stable region expands because $\hgamma < \gamma$.
	\item When $\hgamma < \frac{\gamma}{2}$, $f(a-2b+c;\gamma,\hgamma)=-1$ will have two zero points. The stable region is separated. One region ranges from $0$ to the first zero point, $2\hgamma-\frac{2\sqrt{\gamma(\gamma-2\hgamma)}}{\gamma\hgamma}$, and the other ranges from the second zero point, $2\hgamma+\frac{2\sqrt{\gamma(\gamma-2\hgamma)}}{\gamma\hgamma}$, to $\frac{4}{\hgamma}$. The combined size of these two regions is $\frac{4}{\hgamma} - \frac{4\sqrt{\gamma(\gamma-2\hgamma)}}{\gamma\hgamma}$. By the inequality of arithmetic and geometric means and after algebraic simplification, we can see that $\frac{4}{\hgamma} - \frac{4\sqrt{\gamma(\gamma-2\hgamma)}}{\gamma\hgamma} \geq \frac{4}{\gamma}$. Hence, the total size of the two separated regions is increased.
\end{itemize}\par
The above discussion shows that when higher-step travelers over-predict lower-step ones' switching tendency, the stability condition is more restricted. In contrast, a ``mild'' under-prediction helps relax the stability condition.\par

Although these findings emerge from a simple two-route network analysis, they generalize to arbitrary networks. This generalization is formally established in the following proposition.

\begin{proposition}\label{prop:tp_ue_stability_neq_gamma}
Consider a dynamical system with Eqs.~\eqref{eq:tp_model}-\eqref{eq:tp_k1_pred_cost} with $|K|=2$ where the projection can always generate positive flow on each route (hence $\halpha$ and $\alpha$ can be set to 1). Denote $\beta_i$ as $\bQ D_{\ast}$'s $i$-th eigenvalue and $\beta_i \geq 0, \forall i$. The DUE is locally stable if and only if $\bigvbracket{\gamma\hgamma\beta_i^2-2\gamma\beta_i+1} < 1, \forall i$.
\end{proposition}
\proof{
Please see Appendix \ref{apdx:tp_ue_stability_neq_gamma}.
}\endproof
\begin{remark}\label{remark:tp_neq_gamma_lemma}
	Solving $\bigvbracket{\gamma\hgamma\beta_i^2 - 2\gamma\beta_i+1} < 1$ for the quadratic function of $\beta_i$ with parameters $\gamma$ and $\hgamma$ generates the same conclusion on how over- and under-predictions affect stability as the simple two-route-network example.\par
	Moreover, when $\hgamma$ is greater than $\bar{\gamma}=\frac{2}{\max_i{\beta_i}}$ defined in Lemma~\ref{lemma:A_property}, $\max_i{\beta_i} > \frac{2}{\hgamma}$. Since the right root of $\gamma\hgamma\beta_i^2 - 2\gamma\beta_i+1=1$ is $\frac{2}{\hgamma}$, the quadratic equation is always greater than 1 when $\hgamma>\bar{\gamma}$ and thus the system with $|K|=2$ is always unstable, regardless of $\gamma$.
\end{remark}

\subsection{Analyses of non-DUE MPE (MPE that are not corresponding to DUE) }\label{sec:analyses_non_DUE_MPE}

This section analyzes the properties and implications of MPE that do not correspond to DUE. Section~\ref{sec:characterize_no_due_mpe} characterizes a fundamental property of non-DUE MPE that each traveler class must abandon at least one route due to strategic anticipation of congestion. This property suggests that non-DUE MPE are more likely to emerge when travelers extensively anticipate others' behaviors. We confirm this hypothesis in Section~\ref{sec:when_non_due_mpe_emerge} through numerical experiments. Section~\ref{sec:total_cost_no_due_mpe} examines the total network cost implications of these equilibria.

\subsubsection{Characterization of non-DUE MPE}\label{sec:characterize_no_due_mpe}


\begin{proposition}\label{prop:non_ue_mpe_zero_flow}\label{prop:non_due_mpe}
	Consider the CH-NTP dynamic of Eqs.~\eqref{eq:tp_model}-\eqref{eq:tp_k1_pred_cost} with $|K|=2$. At any non-DUE MPE, denoted as $\x^{\dagger}\equiv(\xd{k}, \xd{k}\in\Omega_{p^k}, k=0,1)^T$, each traveler class has at least one unused route $j(k)$ where $x_{j(k)}^{k,\dagger} = 0$, despite the original DUE admitting strictly positive flow on each route.
\proof{
	Please see Appendix~\ref{apdx:non_due_mpe}.\endproof
}
\end{proposition}

\begin{remark}\label{remark:abandonment_analysis}
Intuitively, this abandonment stems from travelers' \textit{extensive} prediction capabilities: higher-step travelers avoid seemingly costless routes due to their reasoning that these routes would become congested through widespread adoption by lower-step travelers; this creates a strategic redistribution where higher-step travelers select longer routes to avoid predicted congestion. These patterns become self-reinforcing once established: persistent over-predictions of lower-step travelers' responses confines higher-step travelers to some routes, even as actual costs remain elevated on these chosen routes.
\end{remark}

\subsubsection{When non-DUE MPE emerge?}\label{sec:when_non_due_mpe_emerge}

From the analysis in Remark~\ref{remark:abandonment_analysis}, we expect non-DUE MPE to emerge only when $\hat{\gamma}$ exceeds a certain threshold. This is confirmed through numerical experiments on a parallel three-route network from \citet{wang2021day}, using CH-NTP dynamics with $|K|=2$, $p^0=p^1=0.5$ and $\halpha=\alpha=1$. (We have conducted additional tests across numerous parameter configurations, and the general patterns and conclusions remain consistent.)

\begin{figure}[b!]
	\begin{subfigure}[tp]{0.48\columnwidth}
	\includegraphics[width=\columnwidth]{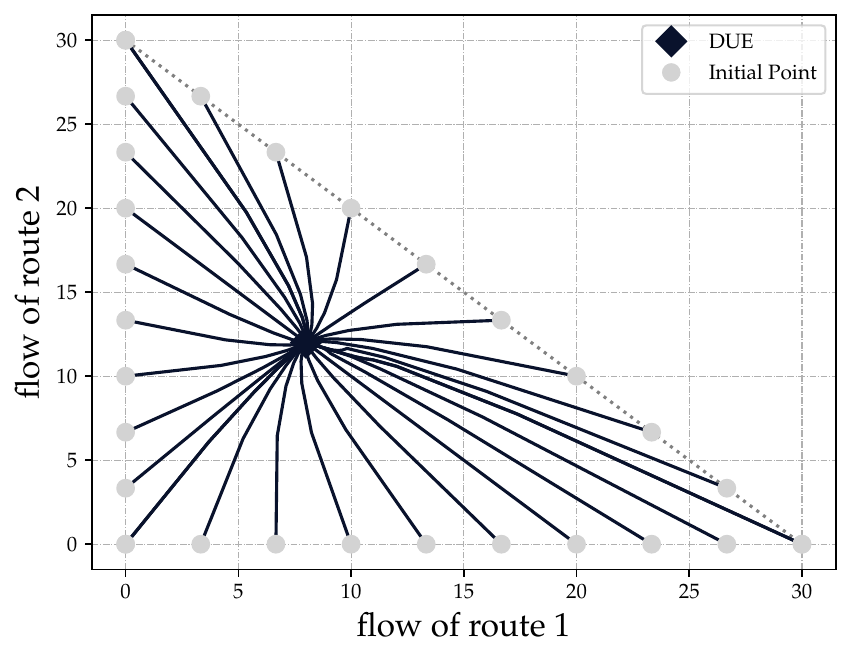}
        \caption{$\hgamma=0.8$}
        \label{fig:vis_evolution_k2_hgamma_80}
    \end{subfigure} 
    \begin{subfigure}[tp]{0.48\columnwidth}
        \includegraphics[width=\columnwidth]{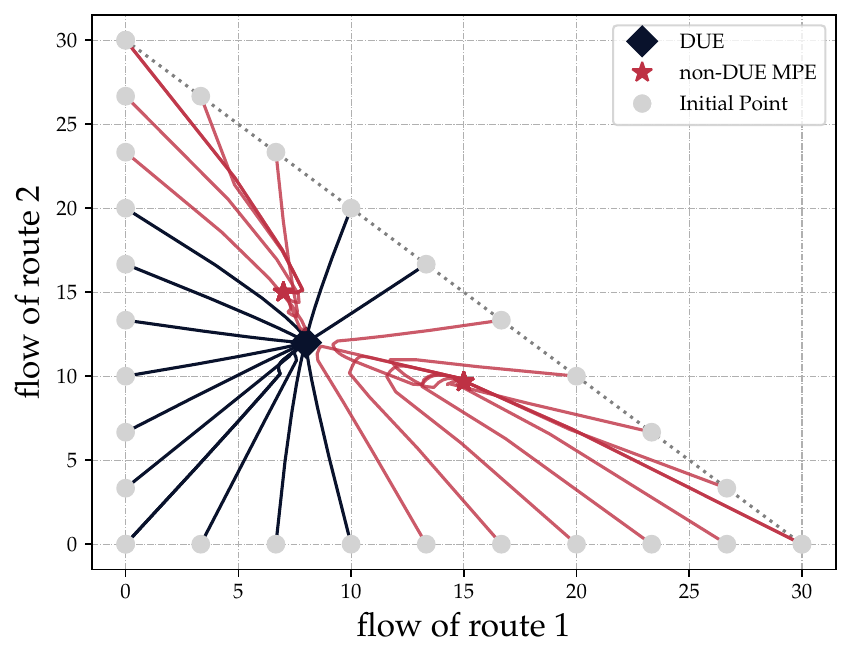}
        \caption{$\hgamma=1.3$}
        \label{fig:vis_evolution_k2_hgamma_130}
    \end{subfigure}\\
    \begin{subfigure}[tp]{0.48\columnwidth}
        \includegraphics[width=\columnwidth]{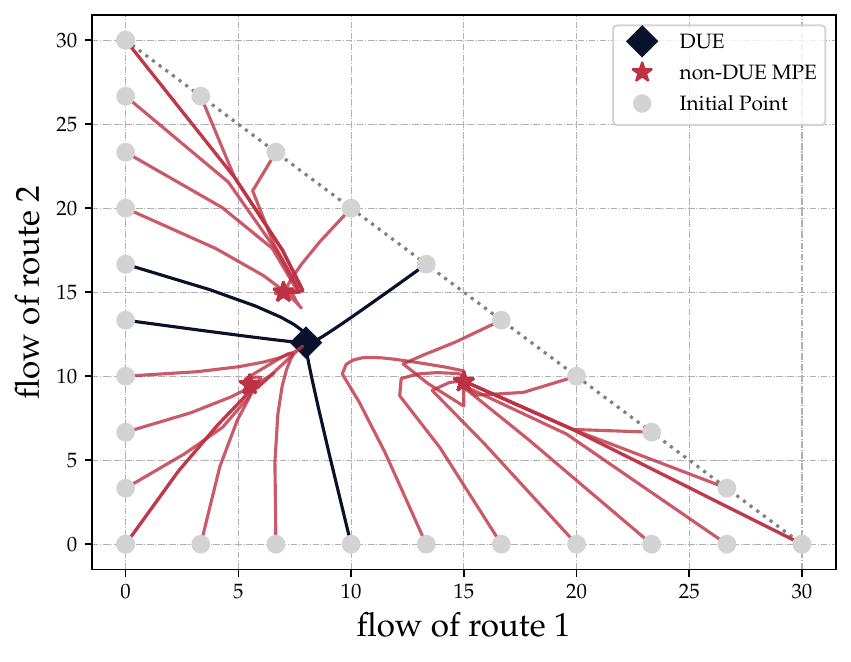}
        \caption{$\hgamma=1.8$}
        \label{fig:vis_evolution_k2_hgamma_180}
    \end{subfigure}
    \begin{subfigure}[tp]{0.48\columnwidth}
        \includegraphics[width=\columnwidth]{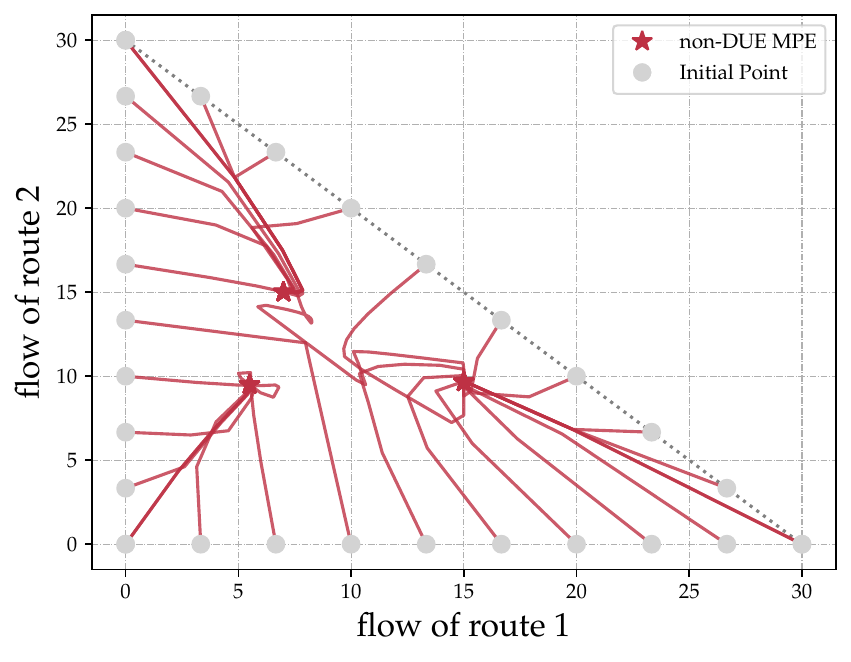}
        \caption{$\hgamma=2.5$}
        \label{fig:vis_evolution_k2_hgamma_250}
    \end{subfigure}
	\caption{\centering Evolution trajectories of aggregate route flows under varying $\hgamma$ values for CHT-NTP dynamics with $|K|=2$, $p^0=p^1=0.5$, and $\halpha=\alpha=1$ in the parallel three-route network from \citet{wang2021day}. Each trajectory originates from distinct boundary points where at least one route having zero flow.}
\end{figure}

Figures~\ref{fig:vis_evolution_k2_hgamma_80}-\subref{fig:vis_evolution_k2_hgamma_250} depict the evolution trajectories of aggregate route flows from different initial points under increasing $\hat{\gamma}$ values. Figure~\ref{fig:vis_evolution_k2_hgamma_80} shows that with a small $\hat{\gamma}(0.8)$, all initial points converge to the DUE. However, as $\hat{\gamma}$ increases to 1.3, 1.8, and 2.5 (Figures~\ref{fig:vis_evolution_k2_hgamma_130}-\subref{fig:vis_evolution_k2_hgamma_250}), multiple non-DUE MPE emerge, and the probability of evolution toward these non-DUE MPE increases. At $\hat{\gamma}=2.5$, all trajectories evolve exclusively toward non-DUE MPE.\par

\subsubsection{Total network cost at non-DUE MPE}\label{sec:total_cost_no_due_mpe}


When the system settles at a non-DUE MPE, does it result in higher or lower total network costs compared to DUE? The answer proves to be nuanced and depends critically on the network parameters. This complexity becomes evident even in the perhaps simplest case of a two-route parallel network with linear cost functions.
\begin{example}\label{example:two_route_cost_comparison}
	Consider the CH-NTP dynamic with Eqs.~\eqref{eq:tp_model}-\eqref{eq:tp_k1_pred_cost} with $|K|=2$ in a two-parallel-route network with linear cost functions $c_1(\tilde{x}_1) = a_1 \tilde{x}_1 + b_1$ and $c_2(\tilde{x}_2) = a_2 \tilde{x}_2 + b_2$. At the DUE, both routes exhibit positive aggregate flows. There exist two non-DUE MPE: one is $\x^{\dagger}\equiv (x_1^{0,\dagger},x_2^{0,\dagger},x_1^{1,\dagger},x_2^{1,\dagger}) = (p^0d,0,0,p^1d)^T$, and the other is $\x^{\dagger}=(0,p^0d,p^1d,0)^T$. The non-DUE MPE produces lower total costs than DUE when either: 
	\begin{itemize}
		\item[(i)] The dynamic finally settles at $\x^{\dagger}=(p^0d,0,0,p^1d)^T$, $b_1<b_2$ and $p^0\in \bigbracket{\frac{a_2}{a_1+a_2}, \frac{b_2-b_1+a_1d}{(a_1+a_2)d}}$; or
		\item[(ii)] The dynamic finally settles at $\x^{\dagger}=(0,p^0d,p^1d,0)^T$, $b_1>b_2$ and $p^1\in \bigbracket{\frac{b_2-b_1+a_2d}{(a_1+a_2)d}, \frac{a_2}{a_1+a_2}}$.
	\end{itemize}
	
	\proof{Please see Appendix~\ref{apdx:two_route_cost_comparison}.
	}\endproof
\end{example}
	
Example~\ref{example:two_route_cost_comparison} reveals that when the proportion of strategic thinkers falls within specific critical ranges that depend on the cost function parameters, the system may settle at a non-DUE MPE with flow pattern that favor routes with lower marginal costs rather than average costs. This phenomenon resembles aspects of system optimal behavior yet emerges endogenously from cognitive hierarchy dynamics. In all other cases not covered by the conditions in Example~\ref{example:two_route_cost_comparison}, strategic thinking becomes detrimental to overall system efficiency.

\section{Validation Using a Virtual Experiment}\label{sec:virtual_experiment}

\begin{figure}[h!]
\centering
\begin{minipage}{.4\textwidth}
  \centering
  \includegraphics[width=0.9\textwidth]{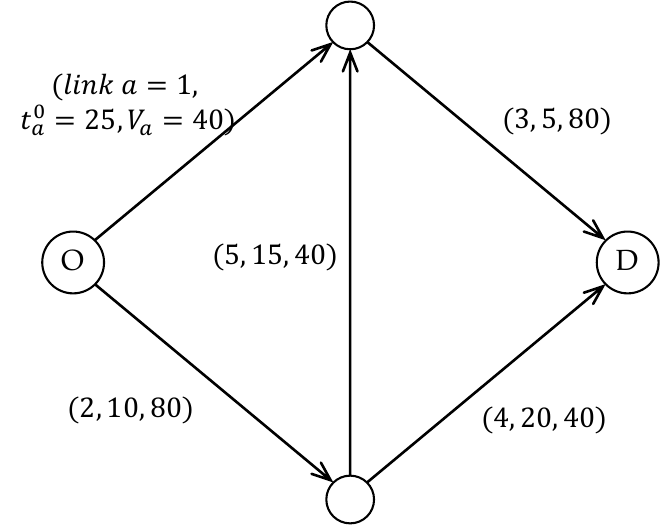}
  \caption{\centering Braess network used in the virtual experiment \citep{ye2018exploration}.}\label{fig:braess_network}
\end{minipage}%
\begin{minipage}{.6\textwidth}
  \centering
  \includegraphics[width=0.85\textwidth]{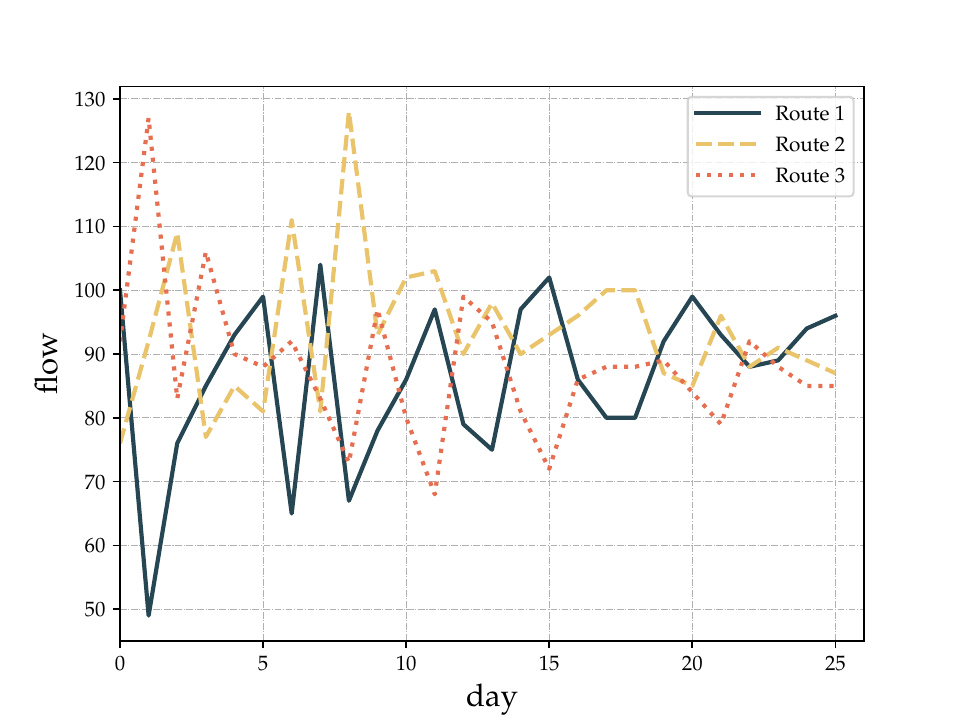}
  \caption{\centering Collected route flow evolutions in the virtual experiment \citep{ye2018exploration}.}\label {fig:experiment_flow}
\end{minipage}
\end{figure}

We conducted an online route choice experiment to mimic travelers’ decision-making processes from day to day \citep{ye2018exploration}. The experiment collected 268 participants' route choices on 26 rounds, where each round corresponded to a true calendar day. It was implemented on the well-known Braess paradox network (Figure~\ref{fig:braess_network}) with one OD pair served by three routes: Route 1 as $1\to3\to5$; Route 2 as $2\to5\to3$; and Route 3 as $2\to4$. The travel time on each link $a\in L$ followed the standard BPR function: $t_a(v_a)=t_a^0\bigbracket{1+0.15\bigbracket{\frac{v_a}{V_a}}^4}$. The parameters are marked in the figure in order as ($a$, $t_a^0$, $V_a$), where $t_a^0$ represents the free-flow travel time and $V_a$ the capacity. The observed route flow (i.e., how many participants selected that route) on each day is visualized in Figure~\ref{fig:experiment_flow} with the days indexed from 0 to 25. Note the large oscillations in the trajectories in Figure~\ref{fig:experiment_flow}. Readers can refer to \citet{ye2018exploration} for more details of the experiment.

Before proceeding with the analysis, it is worth noting why we use experimental data rather than field observations. Obtaining real-world route choice data for day-to-day analysis presents significant challenges due to limitations in sensor coverage, numerous confounding factors in traffic conditions, and difficulties in extracting route-specific decisions from aggregate traffic counts. These challenges have led transportation researchers to utilize controlled experiments when studying route choice behavior \citep{qi2023investigating}. While experiments cannot perfectly replicate real-world conditions, our experiment described above notably narrows the gap between laboratory conditions and real-world environments in two ways: (i) it includes a large sample of 268 participants, whereas previous studies typically involve fewer than 100 participants per scenario; and (ii) it spans 26 actual calendar days, rather than condensing all rounds within a few hours as commonly practiced in laboratory settings. This allows for more natural learning and adaptation processes similar to travelers' day-to-day experiences.

\subsection{Difficulty of prevailing models without prediction in reproducing the observed pattern}\label{sec:failure_previous}

A natural way of calibrating a day-to-day dynamical model is to find the optimal parameters that minimize the RMSE between simulated and observed flow evolution trajectories. In doing so, \citet{ye2018exploration} found that the prevailing day-to-day models could not produce fluctuated trajectories fitting the experimental data; see Figure~3 of that paper. (Due to the intractability of the above ``simulation-based'' method, they had to turn to a relaxed problem using regression analysis.) Below we give an explanation of the prevailing models' difficulties.\par

\begin{figure}[h!]
    \centering
    \includegraphics[scale=0.7]{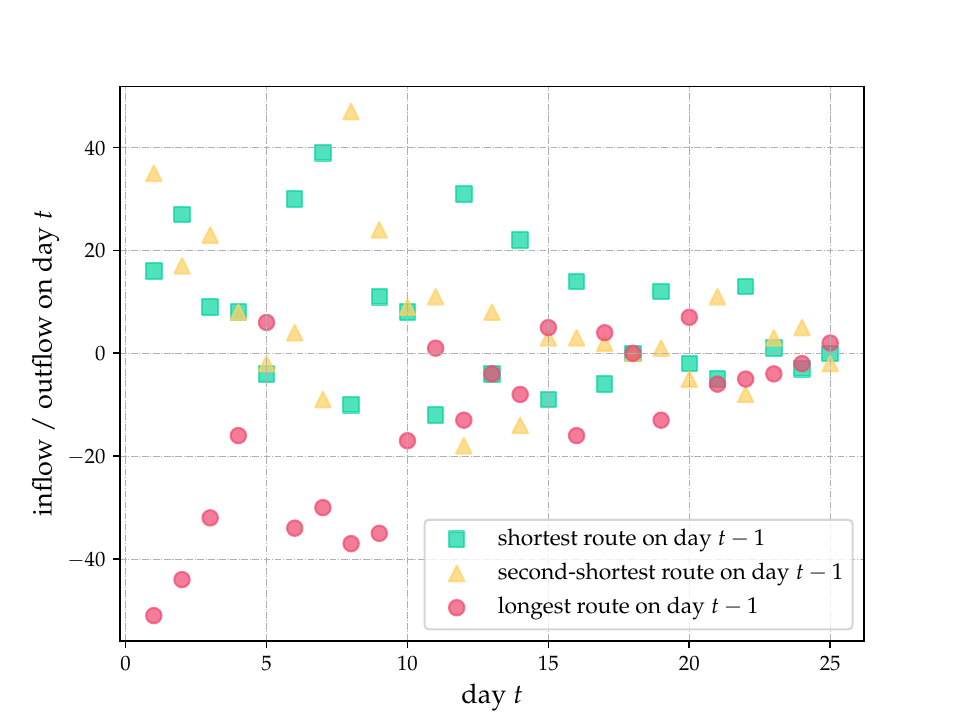}
    \caption{\centering Net flow on day $t$ of the routes sorted by travel cost on day $t-1$ in the virtual experiment.}\label{fig:in_out}
\end{figure}\par

To explain, we plot the net flow of three routes on each day $t$ in Figure~\ref{fig:in_out}. We use different colors and shapes to mark different routes ranked by the travel cost on the day $t-1$: green square points for the shortest route, yellow triangular points for the second-shortest route, and red circular points for the longest route.\par

First note that for day $t\in[1, 16]$, the longest route on the day $t-1$ would by-and-large out-flow to the other two routes on the day $t$. This observation is in line with most day-to-day models' assumption that travelers would select the routes with lower costs. However, there was no distinction between the shortest and the second-shortest routes. On quite a few days (i.e., 1, 3, 8, 9, 10, and 13), the previous day's second-shortest route received more inflow than the shortest route. This indiscrimination between the shortest and the second-shortest routes contradicted the best-response-based day-to-day models' underlying hypothesis that the shortest route would be the most popular \citep[e.g.,][]{friesz1994day,he2010link,xiao2016physics}. We argue that this indiscrimination was a consequence of strategic thinking. Specifically, a 1-step traveler might think that other travelers would go for the shortest route of the day $t-1$ and thus avoided switching to that route on the day $t$. As a result, the second-best route becomes her preference.\par
For day $t\in [17,25]$, a keen-eyed reader may find that travelers' preferences seem to become independent of yesterday's route flow pattern. For example, the net in- and out-flows of the three routes on day 18 are 0, regardless that the DUE had not been achieved. We conjecture that the participants learned after 18 days’ experience that the occurrence of the shortest route was highly non-predictable and hence used a uniformly-random mixed strategy to respond. (We learned this phenomenon from conversations with some participants in the final days.)\par

Calibration results of our proposed hierarchical model confirm the above discussions, which are presented in the sub-sections to follow.

\subsection{Calibrations of the CH-NTP dynamic}\label{sec:calibration_ch_ntp}
To minimize the number of parameters, we consider special cases with perfect prediction (i.e., $\hgamma=\gamma$) and $\halpha=\alpha=1$. By doing so, we intend to demonstrate our model's general explanation capability rather than ``overfit'' data using more parameters. (Relaxing these constraints will further reduce the fitting errors.) The 0-step model is exactly the so-called XYY dynamics used in \citet{xiao2016physics} and \citet{ye2018exploration}, which has only one parameter, $\gamma$, to be estimated. (The other day-to-day models (without prediction) calibrated in \citet{ye2018exploration} exhibited very similar (bad) performance as the 0-step model and hence are omitted here for simplicity. We aim to show how introducing cognitive hierarchy can significantly improve model interpretation.) The 1-step model has one more parameter $p^0$ because $p^1$ equals $1-p^0$. And the 2-step model has three parameters, namely, $\gamma$, $p^0$, and $p^1$. Note that we limit our analysis to cognitive hierarchies with $|K|\leq 3$, as substantial experiments in behavioral game theory have found that the average thinking step across players typically lies between 1 and 2 \citep{camerer2004cognitive,chong2016generalized}.

\subsubsection{Preparations}
We calibrated by minimizing the RMSE between the predicted flows and the ground truth:
\begin{equation}\label{eq:rmse_def}
	RMSE = \sqrt{\frac{\sum_{t=1}^{M}\sum_{r=1}^{3}{\bigbracket{\hat{x}_r^{(t)}-\bar{x}_r^{(t)}}^2}}{3M}},
\end{equation}
where $M$ is the number of days (i.e., days 1 to $M$) used for calibration, $\hat{x}_r^{(t)}$ is the model-predicted aggregate flow on route $r$ of day $t$, and $\bar{x}_r^{(t)}$ is the observed flow. The RMSE function may be highly non-linear. To avoid being trapped in local optima, we performed a grid search over the parametric space with $\gamma$ varying from 0.01 to 1.0 with step 0.002, $p^0$ and $p^1$ ranging from 0.01 to 1.0 with step 0.01. For models with $|K|\geq 2$, given all the parameters, the initial route flow pattern (on day 0) was chosen such that the predicted aggregate route flow pattern on day 1 was closest to the observed one in terms of RMSE.\par

While our day-to-day model operates at a ``macroscopic'' level using continuous route flow variables, it has a ``microscopic'' probabilistic interpretation. For each cognitive level $k$, we can divide each route flow of that level $k$ by corresponding OD demands and interpret it as the probability of choosing that route. (This can be viewed as playing mixed strategies with these probabilities in their route choice process in \citealp{krichene2015online}.)

With probabilistic interpretation, we can perform likelihood ratio tests. Specifically, for each CH-NTP dynamic with different $|K|$, we first obtain the optimal parameters by minimizing the RMSE in Eq.~\eqref{eq:rmse_def}. With these optimized parameters, we calculate the model's predicted probability of a traveler $l$ from class $k$ choosing route $j$ on day $t$, denoted as $g^{k}\bigbracket{r_l^{j,(t)} \mid \gamma, p^0,\dots,p^{|K|-1}}$. Then the final total predicted probability, $G^{(t)}(r_l^j)$, is an aggregation of all thinking steps weighted by the proportions; i.e., $G^{(t)}(r_l^j) = \sum_{k}p^{k}g^k\bigbracket{r_l^{j,(t)}\mid \gamma,p^0,\dots,p^{|K|-1}}$. Finally, we can form a log-likelihood function by combining all the travelers on all the routes and days:

\begin{equation}\label{eq:ll_func}
	LL\bigbracket{\gamma, p^0,\dots,p^{|K|-1}} = \sum_{t=1}^{M} \sum_{l=1}^{268} \sum_{j=1}^{3} {I\bigbracket{r_i^{j,(t)}}} \cdot \text{ln}G^{(t)}\bigbracket{r_i^j},
\end{equation}
where $268$ denotes the number of participants in the experiment and $I\bigbracket{r_l^{j,(t)}}$ is the indicator function being 1 if traveler $l$ chooses route $j$ on day $t$ and 0 otherwise.

\subsubsection{Results}\label{sec:calibration_results}
\begin{figure}[t]
    \centering
\begin{subfigure}[tp]{0.48\columnwidth}
	\includegraphics[width=\columnwidth]{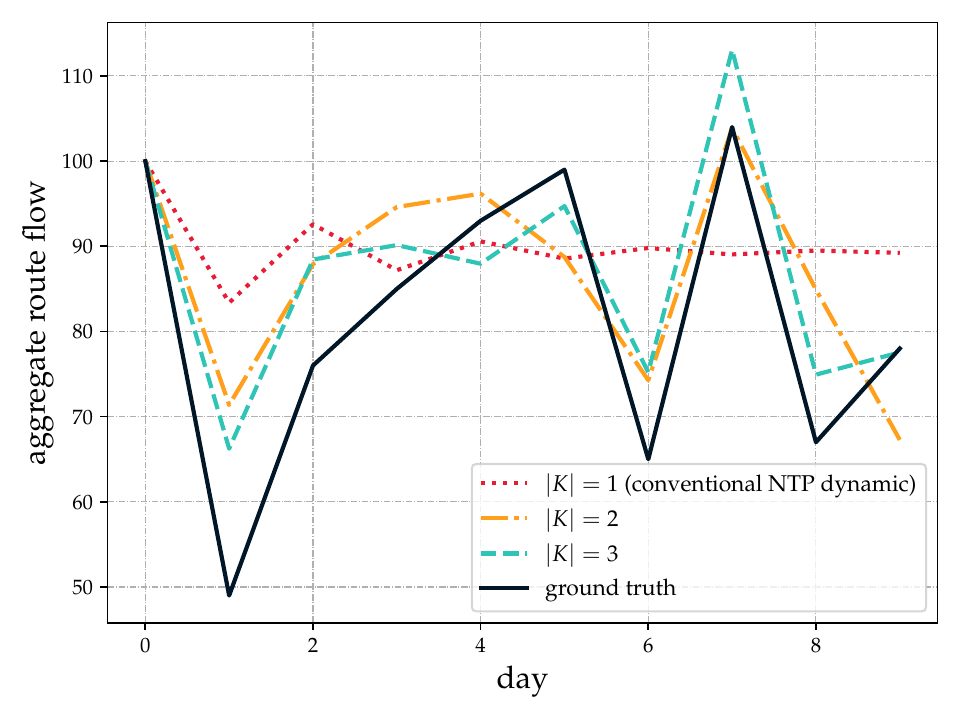}
        \caption{the first 9 days}
        \label{fig:day_9_route0}
    \end{subfigure} 
    \begin{subfigure}[tp]{0.48\columnwidth}
        \includegraphics[width=\columnwidth]{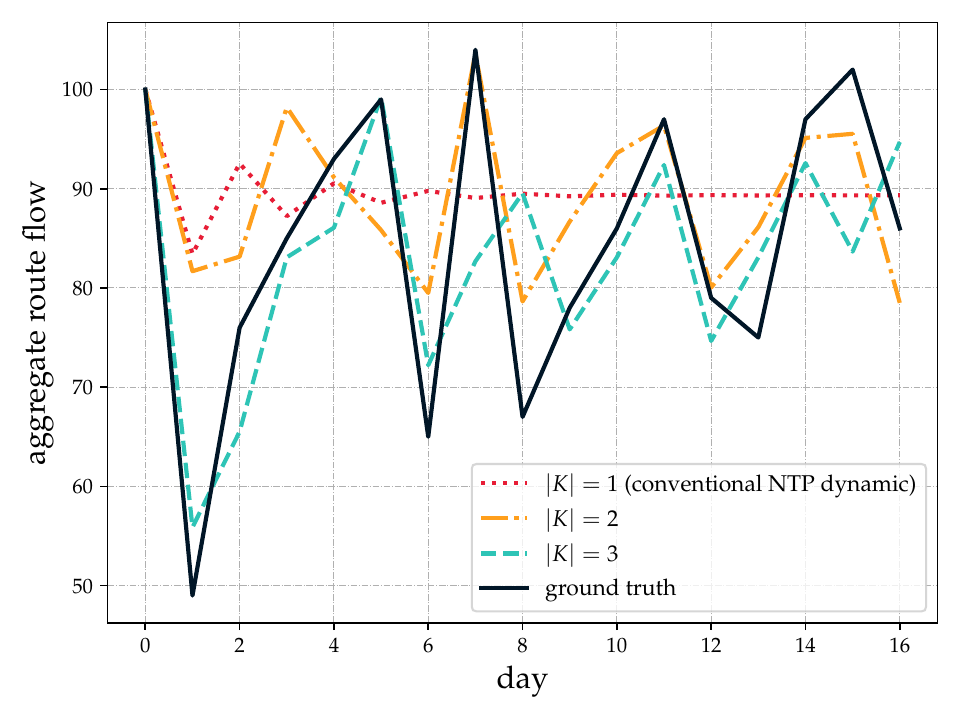}
        \caption{the first 16 days}
        \label{fig:day_16_route0}
    \end{subfigure}\\
    \begin{subfigure}[tp]{0.48\columnwidth}
        \includegraphics[width=\columnwidth]{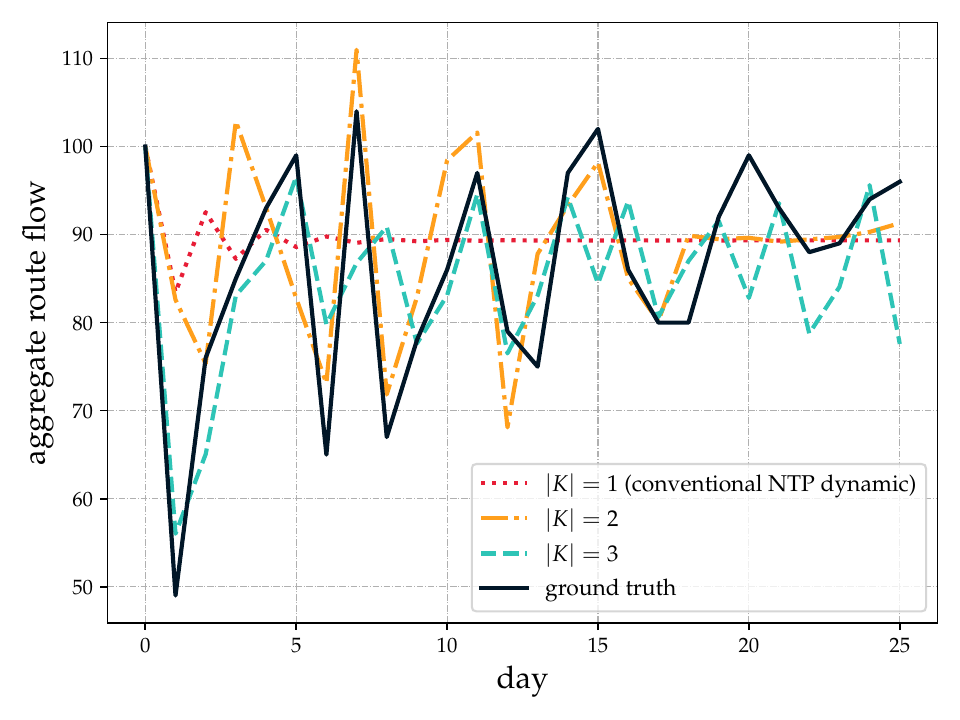}
        \caption{all the 25 days}
        \label{fig:day_25_route0}
    \end{subfigure}
	\caption{\centering Comparisons between the calibrated and ground-truth flow evolutions of Route 1.}
\end{figure}

The simulated flow evolutions of Route 1 with the parameters that minimized the RMSE are presented in Figures~\ref{fig:day_9_route0}-\subref{fig:day_25_route0} for $M=9$, $16$, and $25$, respectively. Each figure uses different line styles to represent the CH-NTP dynamic with varying values of $|K|\in\{1,2,3\}$, while solid lines depict the ground-truth observed data. (The other two routes exhibit similar results and hence are omitted to conserve space.) The likelihood ratio test using the optimal parameters is reported in Table \ref{table:ll_test}.

\addtolength{\tabcolsep}{-2.5pt}
\begin{table}[h!]
\footnotesize
\begin{threeparttable}
	\centering
	\caption{Calibration results of the CH-NTP dynamic}\label{table:ll_test}
	\vspace{1.0em}
	\begin{tabulary}{\textwidth}{cc|ccc|ccc|ccc}
		\hline
		\multicolumn{2}{c|}{Number of days} & \multicolumn{3}{c|}{9} & \multicolumn{3}{c|}{16} & \multicolumn{3}{c}{25}\\[0.15em]
		\multicolumn{2}{c|}{CH-NTP dynamic with $|K|=$} & 1 & 2 & 3 & 1 & 2 & 3 & 1 & 2 & 3 \\
		\hline
		\multicolumn{2}{c|}{Estimated log-likelihood\tnote{$\lozenge$} } & -2646.2 & -2631.3 & -2618.8 & -4706.1 & -4693.8 & -4680.2 & -7356.0 & -7344.3 & -7334.4\\
		\multicolumn{2}{c|}{Maximum log-likelihood\tnote{$\star$}} & \multicolumn{3}{c|}{-2604.0} & \multicolumn{3}{c|}{-4652.6} & \multicolumn{3}{c}{-7297.7}\\[0.15em]
		\multicolumn{2}{c|}{Likelihood Ratio Test} &-  & 29.8 & 52.8 & - & 24.5 & 51.7 & - & 23.3 & 43.0\\
		\multicolumn{2}{c|}{$p$-value} &- & 4.77-e8 & 1e-12 & - & 7.58-e7 & 6e-12 & - & 1.35e-6 & 4.62e-10\\
		\multirow{3}{*}{Parameters}&$\gamma$ &0.358 &0.708 &0.522 &0.358 &0.570 &0.508 &0.358 &0.566 &0.492 \\
		&$p^0$ &- &0.88 &0.37 &- &0.94 &0.31 &- &0.90 &0.31 \\
		&$p^1$ &- &- &0.23 &- &- &0.05 &- &- &0.05 \\
		\hline
	\end{tabulary}
	\begin{tablenotes}
      \small
      \item[$\lozenge$] Estimated log-likelihood using the optimal parameters that minimized RMSE in Eq.~\eqref{eq:rmse_def}. 
      \item[$\star$] The maximum that the Log-likelihood function can achieve, i.e., the entropy of the ground-truth route choice distribution.
    \end{tablenotes}
\end{threeparttable}
\end{table}

\normalfont

First note that in all the figures, the nonhierarchical model with $|K|=1$, i.e., the conventional NTP dynamic, fails to produce a significant fluctuation pattern fitting the ground-truth trajectory; see the dotted curves. This result is in agreement with \citet{ye2018exploration}. The corresponding flow trajectories for $M=9$, $16$, and $25$ are identical and the optimal $\gamma=0.358$.\par

Further note that we have implemented a multi-class day-to-day model with distinct sensitivity parameter for each class \citep{zhou2020discrete}. Despite increasing the number of classes to three, this multi-class model still fails to reproduce the observed fluctuation pattern. In fact, the calibrated flow evolution from this model closely resembles the pattern seen in the $|K|=1$ case (illustrated by the red dotted curve in Figures \ref{fig:day_9_route0}-\subref{fig:day_25_route0}). This outcome strongly suggests that our improved fitting performance is not merely a result of additional parameters, but rather stems from the enhanced explanatory capability of our proposed model.\par

Happily, significant fluctuation patterns emerge when some travelers think one step further, as revealed by the yellow dash-dotted curves. The fitting error decreases as a consequence. Note in Figures~\ref{fig:day_16_route0} and \subref{fig:day_25_route0} how the predicted curves faithfully capture the ground-truth trend from day 5 to 16. This is also confirmed by observing that the hierarchical model of $|K|=2$ improves the log-likelihood by 24.5 and 23.3 for $M=16$ and $25$, respectively. Moreover, the hierarchical model with $|K|=3$ almost doubles the log-likelihood improvement and exhibits better fitting performance from the very first day until day 16. All the improvements are statistically significant under the likelihood ratio test with the $p$-value less than $0.0002\%$.\par
Fitting performance is less satisfying for days 17-25, as shown by the trailing edges of the curves in Figure~\ref{fig:day_25_route0}. This decline may be explained by participant behaviors in the final eight days. As the experiment continued, participants found little advantage in trying to predict how others would behave. This led them to take somehow uniformly-random behaviors. Our time-invariant parameters cannot capture such a pattern.


\section{Numerical Experiments on Local Stability near the DUE}\label{sec:numerical_experiments}

\begin{figure}[h!]
\centering
\includegraphics[scale=0.6]{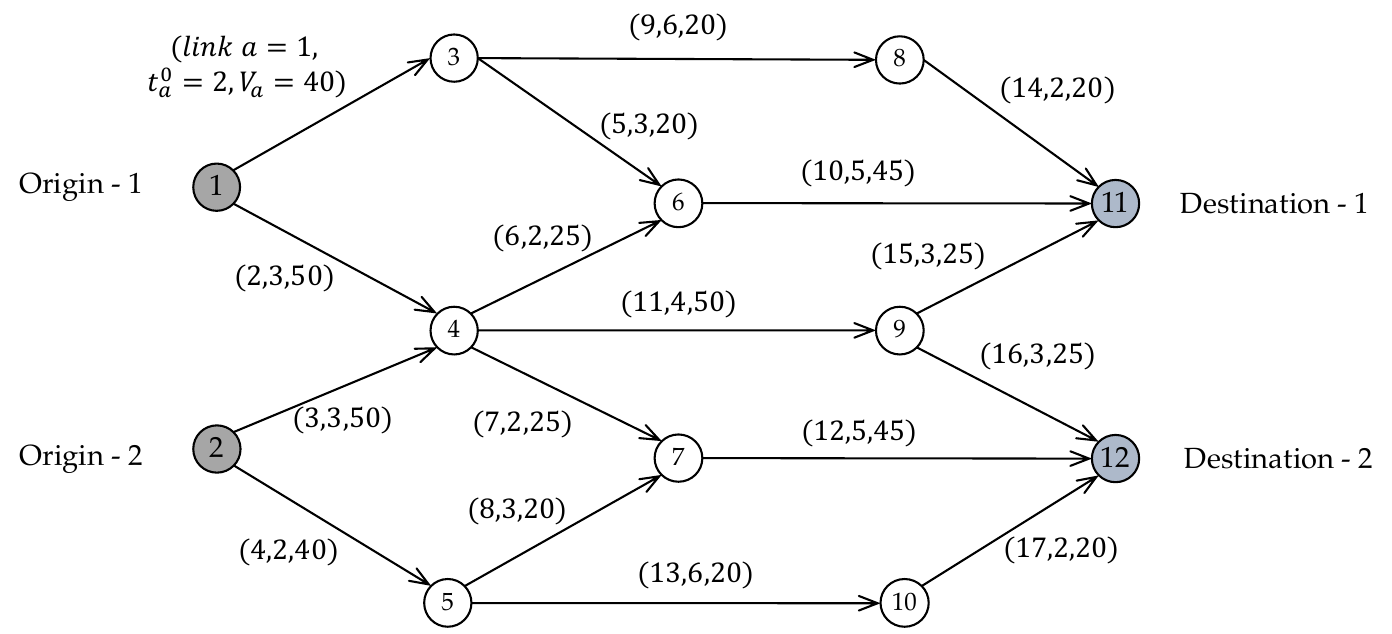}
\caption{A larger testing network (from \citealp{zhang2015nonlinear}).}\label{fig:zhang_network}
\end{figure}

In this section, we conduct numerical experiments to verify the stability conditions using a larger network in Figure~\ref{fig:zhang_network} \citep{zhang2015nonlinear}. The network contains 8 routes that connect two OD-pairs with demand $(90,90)^T$: Route 1 as $1\to9\to14$; Route 2 as $1\to5\to10$; Route 3 as $2\to6\to10$; Route 4 as $2\to11\to15$; Route 5 as $3\to11\to16$; Route 6 as $3\to7\to12$; Route 7 as $4\to8\to12$; and Route 8 as $4\to13\to17$. Link travel times are again given by the well-known BPR function with link numbers, free-flow travel times, and capacities marked in the figure in order as ($a$, $t_a^0$, $V_a$). The aggregate DUE is $(20,20,25,\allowbreak 25,25,25,20,20)^T$.

\subsection{In the interior of the feasible route flow set}
\begin{figure}[h!]
    \centering
\begin{subfigure}[t]{0.52\linewidth}
	\includegraphics[width=\columnwidth]{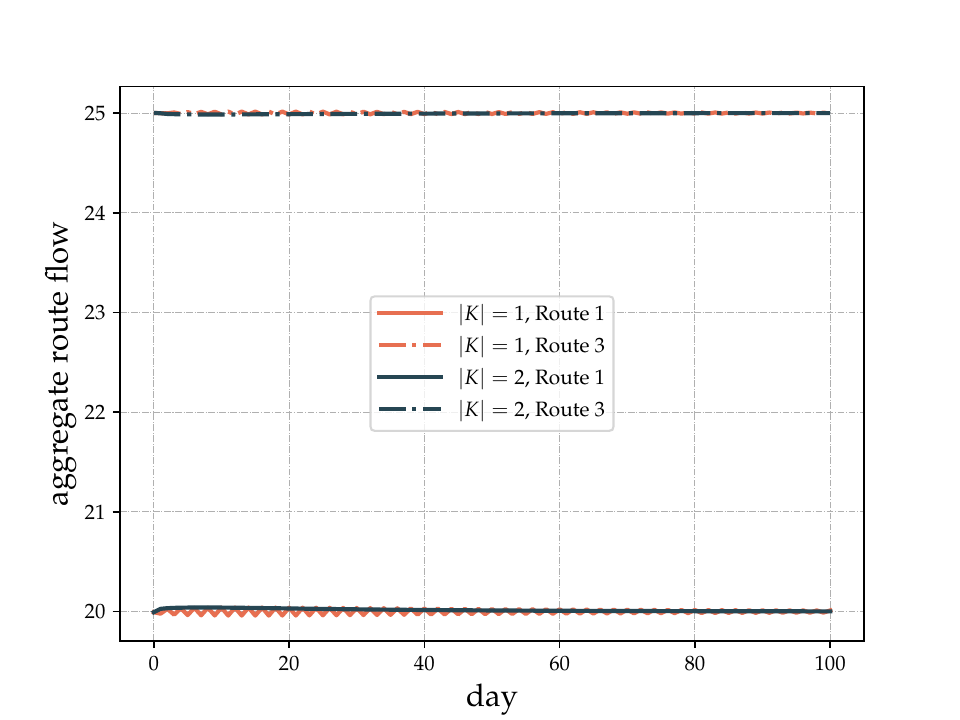}
        \caption{$\gamma=0.78$}
        \label{fig:zhang_stability:small_gamma}
    \end{subfigure}
    \begin{subfigure}[t]{0.47\linewidth}
        \includegraphics[width=\columnwidth]{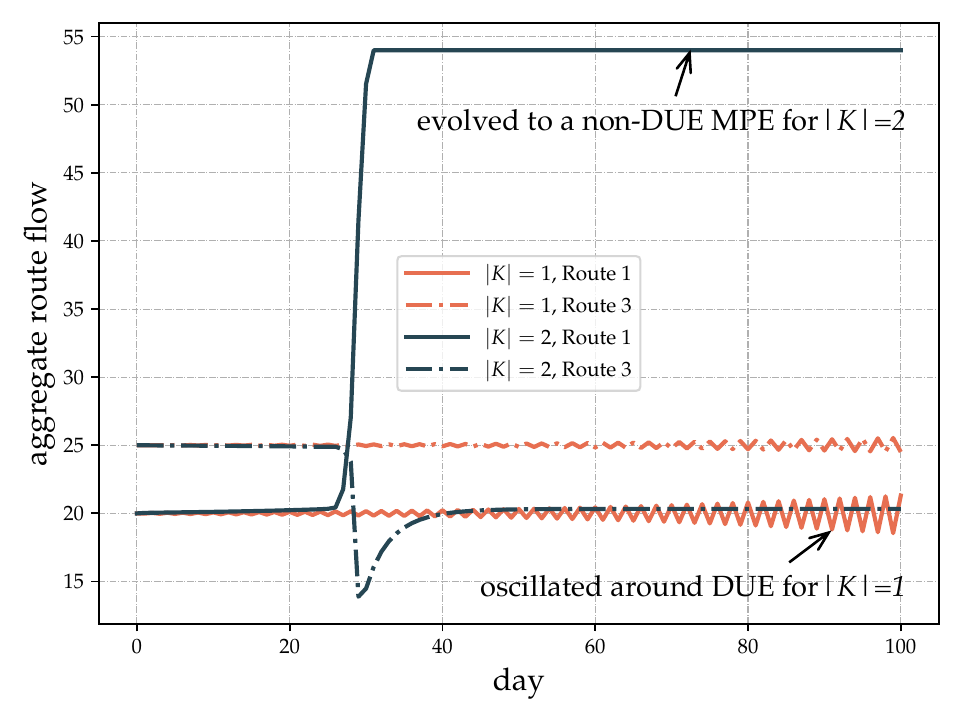}
        \caption{$\gamma=0.8$}
        \label{fig:zhang_stability:large_gamma}
    \end{subfigure}\\
	\caption{\centering Flow evolutions of the CH-NTP dynamic with $|K|\in \{1,2\}$ under different $\gamma$.}
\end{figure}

We first assess Proposition~\ref{prop:tp_ue_stability_eq_gamma} where predictions can be perfectly made, and the projection operators yield positive flows on all the routes. Since we focus on the local stability, the initial aggregate route flow pattern in the following experiments is slightly perturbed from the DUE. We set $p^0=0.4, p^1=0.6$ for all the routes at the initial point, $\halpha=\alpha=1.0$ and $\hgamma=\gamma$. We calculate that $\bar{\gamma}\approx 0.79$ in Lemma~\ref{lemma:A_property}. According to Proposition~\ref{prop:tp_ue_stability_eq_gamma} and Lemma~\ref{lemma:A_property}, the equilibrium is stable when $\gamma < \bar{\gamma}$ and unstable when $\gamma > \bar{\gamma}$. Figures~\ref{fig:zhang_stability:small_gamma} and  \subref{fig:zhang_stability:large_gamma} plot two selected routes' flow trajectories for $\gamma=0.78$ and $0.8$, respectively. It is expected that for both $|K|=1$ and $2$, a smaller $\gamma(=0.78)$ eliminates the small initial perturbations while a larger $\gamma(=0.8)$ renders the system deviating from the DUE. Note that for $|K|=1$, such a deviation results in a permanent oscillation near the DUE, while for $|K|=2$, the system evolves to a new MPE that is not DUE.
\par
\begin{figure}[h!]
\centering
\includegraphics[scale=0.65]{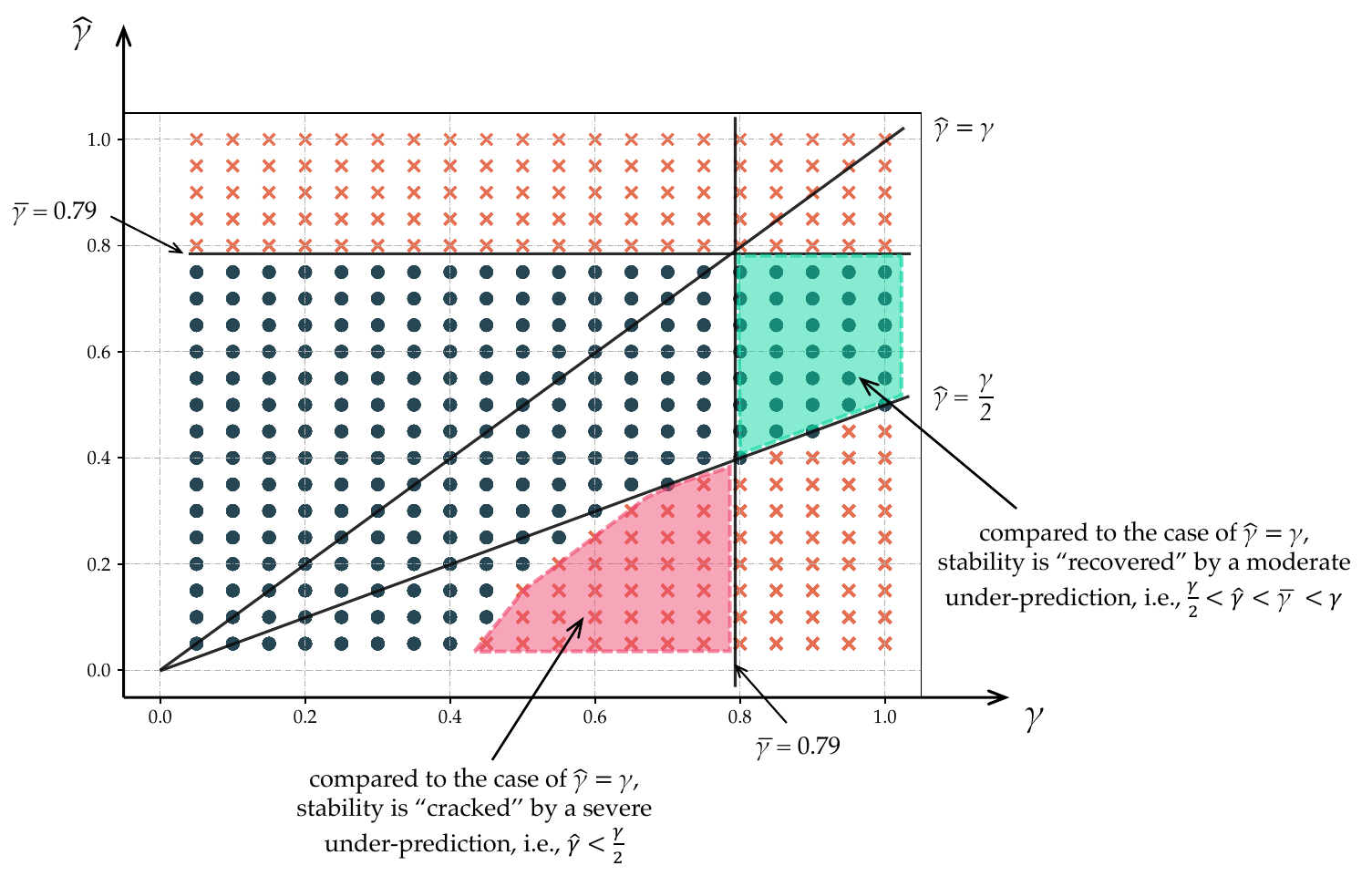}
\caption{\centering Stable and unstable regions w.r.t. combinations of $\gamma$ and $\hgamma$ for the CH-NTP dynamic with $|K|=2$, $p^0=0.4, p^1=0.6$.}\label{fig:zhang_neq_gamma}
\end{figure}

We next verify Proposition~\ref{prop:tp_ue_stability_neq_gamma} where predictions are inaccurate. All the parameters and initial flow patterns are the same as the above case except for the values of $\gamma$ and $\hgamma$. Figure~\ref{fig:zhang_neq_gamma} shows whether the system deviates from the DUE after a small perturbation across various combinations of $\gamma$ and $\hgamma$ at a resolution of 0.05. Solid circles represent stable systems, while crosses indicate unstable systems. The points on the $45^{\circ}$ solid curve represent the case of $\hgamma=\gamma$ (perfect prediction), and the solid curve lying beneath represents the case of $\hgamma = \frac{\gamma}{2}$. The approximate $\bar{\gamma}=0.79$ by Lemma~\ref{lemma:A_property} is also marked.
\par
First note the green shaded area where a moderately small $\hgamma$, $\frac{\gamma}{2} < \hgamma < \bar{\gamma} < \gamma$, helps stabilize the dynamic even when $\gamma$ exceeds $\bar{\gamma}$. This justifies the finding in Proposition~\ref{prop:tp_ue_stability_neq_gamma} that, a mild under-prediction enlarges the stable region. In contrast, when $\hgamma$ is very small (i.e., $<\frac{\gamma}{2}$), the stable region is separated into two pieces. The eigenvalues of $\bQ D_{\ast}$ that previously satisfy the stability condition when $\hgamma=\gamma<\bar{\gamma}$ fall into the unstable region due to the separation when $\hgamma<\frac{\gamma}{2}$; see the red shaded area. In addition, as expected in Remark~\ref{remark:tp_neq_gamma_lemma}, a $\hgamma > \bar{\gamma}$ will make the system permanently unstable, regardless of $\gamma$; see all the crosses above the horizontal line of $\bar{\gamma}=0.79$. Finally, for this particular numerical example, the stable region diminishes when over-predictions occur ($\gamma < \hgamma < \bar{\gamma}$). Nevertheless, the eigenvalues of $\bQ D_{\ast}$ still satisfy the stability condition in Proposition~\ref{prop:tp_ue_stability_neq_gamma}, which explains why the points above the curve $\hgamma=\gamma$ remain stable.

\subsection{On the boundary of the feasible route flow set}
\begin{figure}[h!]
\centering
\includegraphics[scale=0.6]{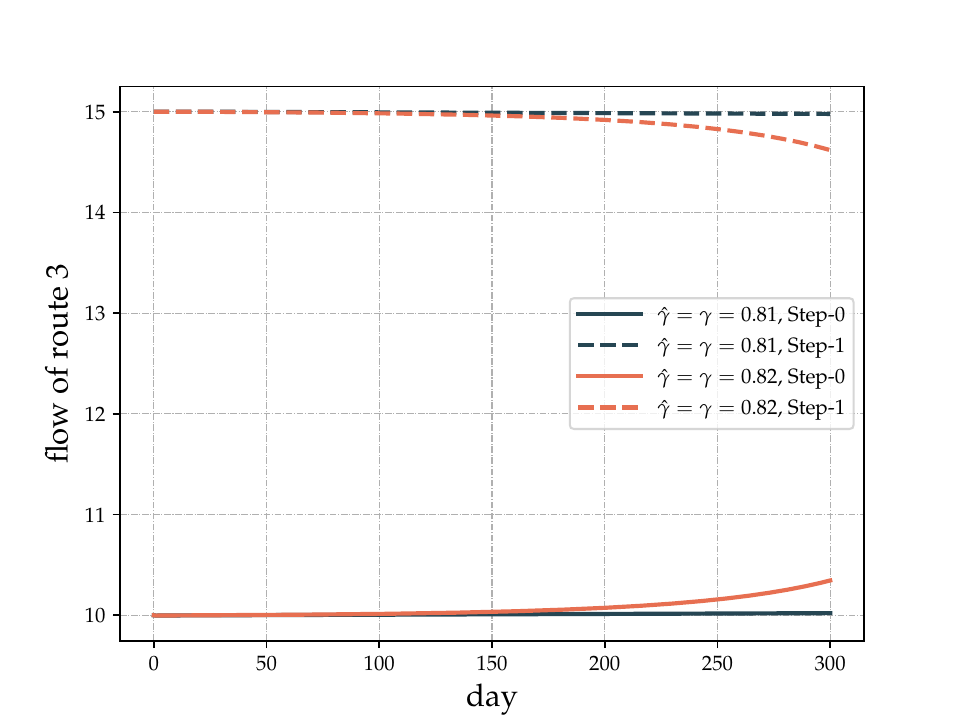}
\caption{Validation of Fact~\ref{fact:tp_jacob} for $|K|=2$ when any projected flow is on the boundary of the feasible set.}\label{fig:zhang_boundary}
\end{figure}

Proposition~\ref{prop:tp_ue_stability_eq_gamma} becomes invalid when the projected flow pattern is on the boundary of the feasible set. Under this circumstance, the general analytical results are invalid, and we have to resort to Fact~\ref{fact:tp_jacob} for numerically checking the stability. To test this fact, we set $|K|=2$, $\halpha=\alpha=0.5$, $\hgamma=\gamma\in\{0.81,0.82\}$, $p^0=0.4, p^1=0.6$ and the initial route flow pattern as $\x^{k=0,(t)=0}=(0,\allowbreak 7.999,10,18, 9.98,10.01,8.01,8.01)^T$ and $\x^{k=1,(t)=0} = (20, 11.99, 15, 7, 14.98, 15.01, 12.01, 12.01)^T$, such that $JP_{0,0}$ in Fact~\ref{fact:tp_jacob} is evaluated on the boundary. From Fact~\ref{fact:tp_jacob}, we have that the maximum moduli of $JP$ of $|K|=2$ for $\gamma=0.81$ and $0.82$ are 1 and 1.013, respectively. This indicates that the former should exhibit stability while the latter should demonstrate instability. Figure~\ref{fig:zhang_boundary} confirms these expectations, with different colors distinguishing the two $\gamma$ values and distinct line styles marking the route flows of different classes. The case of $|K|=3$ is also consistent with the numerical result of Fact~\ref{fact:tp_jacob}, which is omitted here for simplicity.

\section{CH-Logit Dynamic}\label{sec:Logit}

While the deterministic CH-NTP dynamic captures strategic reasoning under perfect information, real-world route choices involve perceptual errors and decision uncertainty. We now replace the day-to-day operator $H[\cdot]$ with a stochastic Logit dynamic where randomness arises from quantal responses --- a concept in which travelers are more likely to choose routes with lower costs yet may still select higher-cost alternatives due to inherent perception errors \citep{fudenberg1998learning}. This extension enables us to examine whether key findings about strategic prediction behavior (such as multiple equilibria and stability conditions) remain consistent across different dynamic formulations, including both deterministic and stochastic ones.

If the day-to-day operator $H[\cdot]$ is replaced by a ``stochastic'' Logit dynamic \citep{fudenberg1998learning,sandholm2010population,xiao2019day}, we have the following CH-Logit dynamic:
\begin{equation}\label{eq:logit_model}
	\xtt{k}- \xt{k} = \alpha\bigbracket{p^{k} \dPhiop{\theta}{\cpxtt{k}} - \xt{k}},
\end{equation}
where $\dPhiop{\theta}{\bz} \equiv \bigbracket{d_w\tPhiop{\theta}{w}{\bz_w},w\in W}^T$, $\tPhiop{\theta}{w}{\bz_w} \equiv (d_w\varphi_{rw}(\bz_w),r\in R_w)^T$, $\varphi_{rw}(\bz_w) = \frac{\exp(-\theta z_{rw})}{\sum_{s\in R_w}{\exp(-\theta z_{sw})} }$, and $\theta$ is a dispersal parameter capturing travelers' perception errors. For each $w\in W$, as $\theta \to 0$, route choices become equally probable across all routes in $R_w$. Conversely, as $\theta \to \infty$, choices become extremely concentrated on the minimum cost route of $R_w$. Note that for the Logit dynamic, the flow variable (i.e., the first variable) in $\y[\cdot]$ of Eq.~\eqref{eq:general_dynamic} is not explicitly required.\par
As per Section~\ref{sec:cognitive_level}, the predicted flow pattern of each class $k\in \{0,1,2\}$ reads:
\begin{align}
	\label{eq:logit_0_step_pred_cost}
	& \pxtt{0} = \txt, \\
	\label{eq:logit_1_step_pred_cost}
	& \pxtt{1} = \halpha \dPhiop{\htheta}{\cpxtt{0}} + (1-\halpha)\txt, \\
	\label{eq:logit_2_step_pred_cost}
	& \pxtt{2} = \halpha{q_2^0} \dPhiop{\htheta}{\cpxtt{0}} + \halpha{q_2^1} \dPhiop{\htheta}{\cpxtt{1}} + (1-\halpha)\txt.
\end{align}

A significantly large $\hat{\theta}$ can be interpreted as higher-step travelers anticipating that lower-step travelers will strongly favor yesterday's shortest path. This situation bears similarities to the case of a large $\hat{\gamma}$ in the CH-NTP model. In both cases, higher-step travelers are making an extreme prediction about the responsiveness of lower-step travelers to cost differences.

\subsection{Mixed prediction-based stochastic equilibria}
Similar to the CH-NTP dynamic, we term fixed points of the CH-Logit dynamic as mixed prediction-based stochastic equilibria (MPSE) and describe them in the following proposition.
\begin{proposition}\label{prop:MPSE}
	When the route cost function $\ccop{\x}$ is continuous, the dynamical system in Eqs.~\eqref{eq:logit_model}-\eqref{eq:logit_2_step_pred_cost} admits at least one MPSE (i.e., one fixed point). Moreover, a vector $\x^{\circ}\equiv(\xc{k},k=0,1,2)^T$ is an MPSE if $\forall r\in R_w, w\in W, k\in K$,
	\begin{equation}\label{eq:MPSE}
		x_{rw}^{k,\circ} = p^k d_w \frac{\exp\bigbracket{-\theta\pi_{rw}^k(\txc) }}{\sum_{s\in R_w}{\exp\bigbracket{-\theta\pi_{sw}^k(\txc)} }},
	\end{equation}
	where $\pxc{k}\equiv(\pi_{sw}^k,s\in R_w, w\in W)^T$ is a function of the aggregate flow pattern $\txc=\sum_{k}{\xc{k}}$, dictated by Eqs.~\eqref{eq:logit_0_step_pred_cost}-\eqref{eq:logit_2_step_pred_cost}.
\end{proposition}
\proof{
The existence of a fixed point is guaranteed by the continuity of both the closed-form Logit operator and the route cost function. To find the fixed points of the dynamical system, we set the RHS of Eq.~\eqref{eq:logit_model} to zero and rearrange terms, yielding Eq.~\eqref{eq:MPSE}.
}\endproof
We now give the definition of a $|K|$-class SUE in Definition \ref{def:class_sue} and show that it is one of the MPSE in Proposition~\ref{prop:logit_multi_eqs} when $\htheta=\theta$.
\begin{definition}\label{def:class_sue}
	A route flow pattern $\x^{\star} \equiv (\xs{k}, \xs{k}\in \Omega_{p^k}, k\in K)^T$ is said to be a $|K|$-class SUE parameterized by $\theta$ if $\forall r\in R_w, w\in W, k\in K$,
	\begin{equation}
		x_{rw}^{k,\star} = p^k d_w \frac{\exp\bigbracket{-\theta c_{rw}^k(\txs)}}{\sum_{s\in R_w}{\exp\bigbracket{-\theta c_{sw}^k(\txs)}}}.
	\end{equation}
\end{definition}

\begin{proposition}\label{prop:logit_multi_eqs}
	When $\htheta=\theta$, a $|K|(=3)$-class SUE in Definition \ref{def:class_sue} parameterized by $\theta$ is an MPSE of the dynamical system with Eqs.~\eqref{eq:logit_model}-\eqref{eq:logit_2_step_pred_cost}, but not vice versa.
\end{proposition}
\proof{
At the SUE, we first observe from Eq.~\eqref{eq:logit_0_step_pred_cost} that $\pxs{0} = \txs$. Thus, for each $r\in R_w, w\in W$, the 1-step traveler's predicted flow Eq.~\eqref{eq:logit_1_step_pred_cost} becomes:
		\begin{equation}
			\pi^{1,\star}_{rw} = \halpha q_1^0 d_w \frac{\exp\bigbracket{-\theta c_{rw}^k(\txs)}}{\sum_{s\in R_w}{\exp\bigbracket{-\theta c_{sw}^k(\txs)}} } + (1-\halpha)q_1^0\tilde{x}_{rw}^{\star} = \halpha \tilde{x}_{rw}^{\star} + (1-\halpha)\tilde{x}_{rw}^{\star} = \tilde{x}_{rw}^{\star},
		\end{equation}
		where the second equality follows directly from Definition \ref{def:class_sue}. 
		
		Substituting $\pxs{0}=\txs$ and $\pxs{1}=\txs$ into Eq.~\eqref{eq:logit_2_step_pred_cost}, we derive that for each $r\in R_w, w\in W$:
		\begin{equation}
			\pi^{2,\star}_{rw} = \halpha (q_2^0+q_2^1) d_w \frac{\exp(-\theta c_{rw}^k(\txs))}{\sum_{s\in R_w}{\exp(-\theta c_{sw}^k(\txs))} } + (1-\halpha)\tilde{x}^{\star}_{rw} = \tilde{x}^{\star}_{rw}.
		\end{equation}
		
		Hence, $\pxs{0}$, $\pxs{1}$ and $\pxs{2}$ all equal to $\txs$. By substituting these three predicted costs into Eq.~\eqref{eq:logit_model} and applying Definition~\ref{def:class_sue}, Eq.~\eqref{eq:logit_model} equals 0. Therefore, the SUE constitutes an MPSE.\par
		Similar to the CH-NTP dynamic, there exist some MPSE that are not SUE when $|K|\geq 2$. We can easily construct a counter-example demonstrating their existence (similar to Figures~\ref{fig:vis_evolution_k2_hgamma_130}-\subref{fig:vis_evolution_k2_hgamma_250}) to negate the necessity. This example is omitted to conserve space.
}\endproof

\subsection{Local stability}
Due to multiple equilibria, this section examines the CH-Logit dynamic's local stability by first deriving its Jacobian at the MPSE and then analyzing stability near the SUE.

\subsubsection{Jacobian matrix}
\begin{lemma}\label{lemma:Upsilon}
The Jacobian matrix of the Logit operator parameterized by $\theta$ evaluated at a cost vector $\cc$, $\Upsilon^\theta[\cc]$, is a block diagonal matrix:
\begin{equation}
\Upsilon^{\theta}[\cc] = \begin{bmatrix}
   \Upsilon^{\theta}_{w=1}[\cc_{w=1}] &  & \bigzero \\ 
   & \ddots & \\ 
   \bigzero &  & \Upsilon^{\theta}_{w=|W|}[\cc_{w=|W|}]
 \end{bmatrix},
\end{equation}
where $\Upsilon^\theta_w[\cc_w] = -\theta d_w\bigbracket{\text{Diag}\bigbracket{\tPhiop{\theta}{w}{\cc_w}} - \tPhiop{\theta}{w}{\cc_w}\bigbracket{\tPhiop{\theta}{w}{\cc_w}}^T}$. Moreover, given any $\cc\equiv(\cc_w,w\in W)^T$, each $\Upsilon_w^\theta[\cc_w]$ is negative semidefinite (NSD), and so is $\Upsilon^{\theta}[\cc]$.
\end{lemma}
\proof{

The Jacobian matrix formula for an individual OD pair $w$ follows from \citet{gao2017properties}, with a negative sign applied to account for our cost-based formulation rather than their utility-based approach, yielding $\Upsilon^\theta_w[\cc_w]$. The block-diagonal structure results from route independence across OD pairs, and $\Upsilon^{\theta}[\cc]$ is NSD since each diagonal block is NSD.
}\endproof

With Lemma~\ref{lemma:Upsilon}, the Jacobian matrix of Logit dynamic is derived in the following fact. 
\begin{fact}\label{fact:logit_jacob}
The Jacobian matrix for the CH-Logit dynamic with $|K|=3$ is a 3-by-3 block matrix:
	\begin{equation}\label{eq:logit_jacob}
	J\Phi = \begin{bmatrix}
   	J\Phi_{0,0} & J\Phi_{0,1} & J\Phi_{0,2} \\
   	J\Phi_{1,0} & J\Phi_{1,1} & J\Phi_{1,2} \\
   	J\Phi_{2,0} & J\Phi_{2,1} & J\Phi_{2,2} \\
    \end{bmatrix},
\end{equation}
where each block $J\Phi_{i,j}$ is defined as follows.
\begin{align*}
	& J\Phi_{0,0} = \alpha p^{0}\Upsilon^{\theta}_{k=0}D_0 + (1-\alpha)I, \\
	 & J\Phi_{0,j\neq0}=\alpha p^{0}\Upsilon^{\theta}_{k=0}D_0,\\
	 & J\Phi_{1,1} = \alpha p^1 \Upsilon^{\theta}_{k=1} D_1 \bigbracket{\halpha\Upsilon^{\htheta}_{k=0}D_0 + (1-\halpha)I}  + (1-\alpha)I,\\
	 & J\Phi_{1,j\neq1} = \alpha p^1 \Upsilon^{\theta}_{k=1} D_1 \bigbracket{\halpha\Upsilon^{\htheta}_{k=0}D_0 + (1-\halpha)I}, \\
	 & J\Phi_{2,2} = \alpha p^2 \Upsilon^{\theta}_{k=2} D_2\bigsbracket{\halpha q_2^0 \Upsilon^{\htheta}_{k=0}D_0 + \halpha q_2^1 \Phi^{\htheta}_{k=1} D_1 \bigbracket{\halpha\Upsilon^{\htheta}_{k=0} D_0 + (1-\halpha)I} + (1-\halpha)I} + (1-\alpha)I, \\
	 & J\Phi_{2,j\neq2} = \alpha p^2 \Upsilon^{\theta}_{k=2} D_2\bigsbracket{\halpha q_2^0 \Upsilon^{\htheta}_{k=0}D_0 + \halpha q_2^1 \Phi^{\htheta}_{k=1} D_1 \bigbracket{\halpha\Upsilon^{\htheta}_{k=0} D_0 + (1-\halpha)I} + (1-\halpha)I},
\end{align*}
where $D_k$ represents the Jacobian of the route cost functions evaluated at predicted flow $\pxtt{k}$ and $\Upsilon^{\theta}_{k}$ the Jacobian of Logit operator with $\theta$ evaluated at $\cpxtt{k}$.
\end{fact}
\proof{
The derivation can be found in Appendix \ref{apdx:logit_jacob}.
}\endproof

\subsubsection{Stability near the SUE}

When $\theta\neq \htheta$, the block submatrices in Eq.~\eqref{eq:logit_jacob} do not commute with each other. This requires direct calculation of the inverse matrix of one of these submatrices to evaluate $|\lambda{I}- J\Phi|$. However, none of these inverse submatrices has a closed-form. Hence, we analytically examine the case of $\theta=\htheta$ only.

\begin{proposition}\label{prop:logit_stable_general}
Denote $\rho_i$ as the $i$-th eigenvalue of $\dUpop{\theta}{\ccop{\txs}} \sDop{\txs}$ at the SUE under Assumption \ref{assumption:cost_psd} and $\rho_i \leq 0, \forall i$. Define function $\psi(\rho_i;\alpha, \allowbreak \halpha,p^0,p^1,p^2) \equiv \alpha\halpha^2 p^2\frac{p^1}{p^0+p^1}\rho_i^3 + (\alpha\halpha-\alpha\halpha p^0-\alpha\halpha^2p^2\frac{p^1}{p^0+p^1})\rho_i^2 + (\alpha\halpha p^0-\alpha\halpha+\alpha)\rho_i +(1-\alpha)$ where $p^0+p^1+p^2=1$. When $\htheta=\theta$ and $|K|\leq 3$, the SUE is locally asymptotically stable when $-1 < \psi(\rho_i;\alpha,\halpha,p^0,p^1,p^2) < 1, \forall i$. For $|K|=2$, we have $p^2=0$. For $|K|=1$, we have $p^1=p^2=0$, $p^0=1$ and $\halpha=\alpha$.

\end{proposition}
\proof{
Please see Appendix \ref{apdx:logit_stable_general}.
}\endproof
\begin{remark}
	Recall that the stability of the CH-NTP dynamic does not depend on the distribution of strategic thinking levels (i.e., $p^0,p^1$, and $p^2$) when the projection operator always yields positive flows (see Proposition~\ref{prop:tp_ue_stability_eq_gamma}). In contrast, the proportions would affect the stability of the CH-Logit dynamic even though the Logit operator always generates positive flows.
\end{remark}

\begin{figure}[h!]
\centering
\includegraphics[scale=0.62]{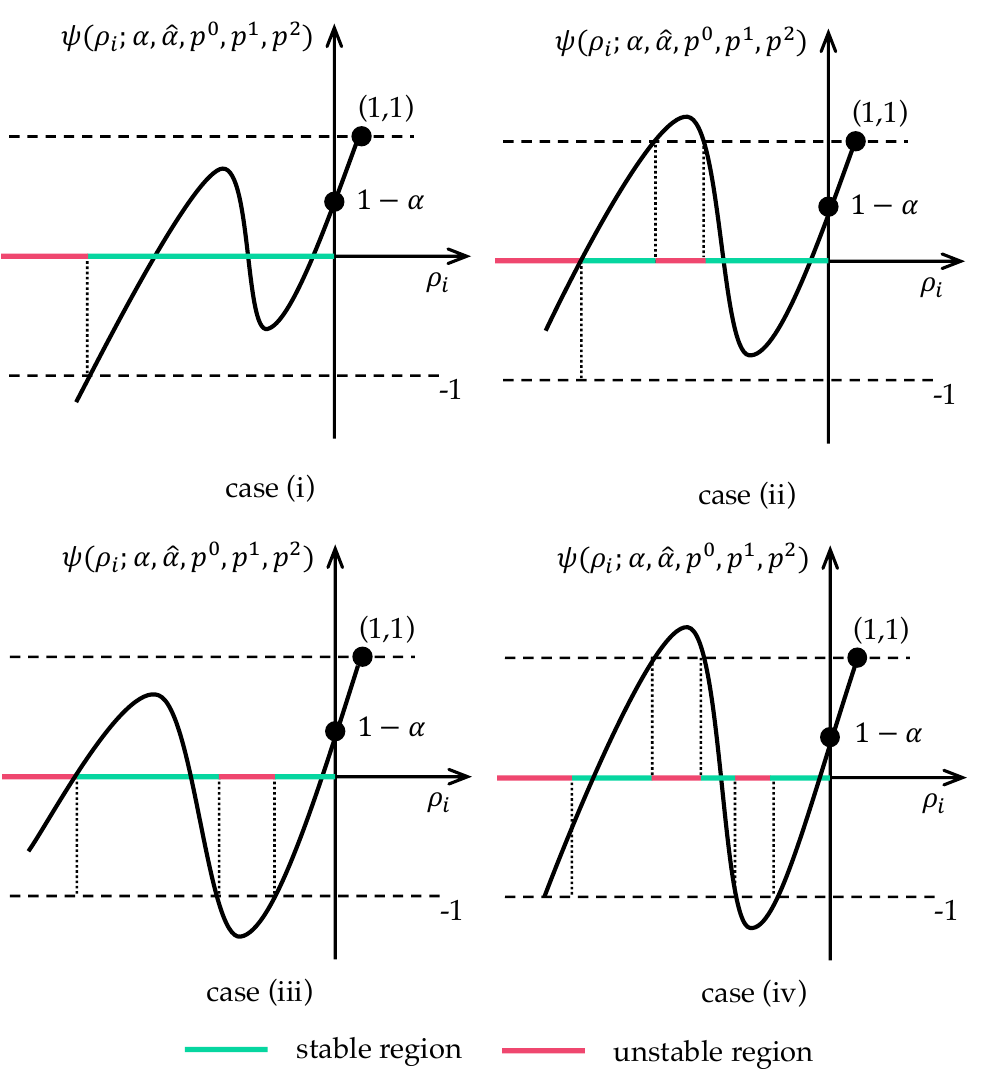}
\caption{\centering Sketches of stable and unstable regions of the CH-Logit dynamic under four possible scenarios of $\psi(\rho_i;\alpha,\halpha,p^0,p^1,p^2)$.}\label{fig:logit_analytical_poly_k3}
\end{figure}\par
Scrutinization of the cubic function $\psi(\rho_i;\alpha, \allowbreak \halpha,p^0,p^1,p^2)$ shows that it crosses points $(1,1)$ and $(0,1-\alpha)$ and exhibits different shapes depending on $\alpha,\halpha,p^0,p^1$ and $p^2$. Figure~\ref{fig:logit_analytical_poly_k3} depicts four possible scenarios when determining the stable region. Here the stable/unstable region refers to the region of the five parameters that make the SUE stable/unstable (i.e., whether or not $\psi(\rho_i;\alpha,\halpha,p^0,p^1,p^2)\in (-1,1),\forall \rho_i$). They are classified based on whether the cubic function has one or three real roots (which may not be distinct) at $1$ or $-1$. The analytical results of the stable region for $|K|=3$ with $\halpha\neq\alpha$ are quite cumbersome and do not reveal any direct insights. (They are thus omitted for simplicity.) Therefore, similar to the case of the CH-NTP dynamic, we turn to analyze two special cases below.

\paragraph{Perfect prediction}
Even under $\halpha=\alpha$, the stability condition of the CH-Logit dynamic given by Proposition~\ref{prop:logit_stable_general} is much more complex than the CH-NTP dynamic. We explore the stability numerically. Given an eigenvalue $\rho_i$, we enumerate all the possible combinations of $p^0$ and $p^1$ in a resolution of 0.001 ($p^2$ is not required as $p^2=1-p^0-p^1$) with $\alpha\in \{0.2,0.5,0.8\}$ and check whether the stability condition is satisfied as per Proposition~\ref{prop:logit_stable_general}. Stable and unstable regions under different values of $\rho_i$ are depicted in Figure~\ref{fig:logit_numerical}. Some observations can be made on these plots:
\begin{figure}[h!]
    \centering
    \begin{subfigure}[t]{0.49\columnwidth}
        \includegraphics[width=\columnwidth]{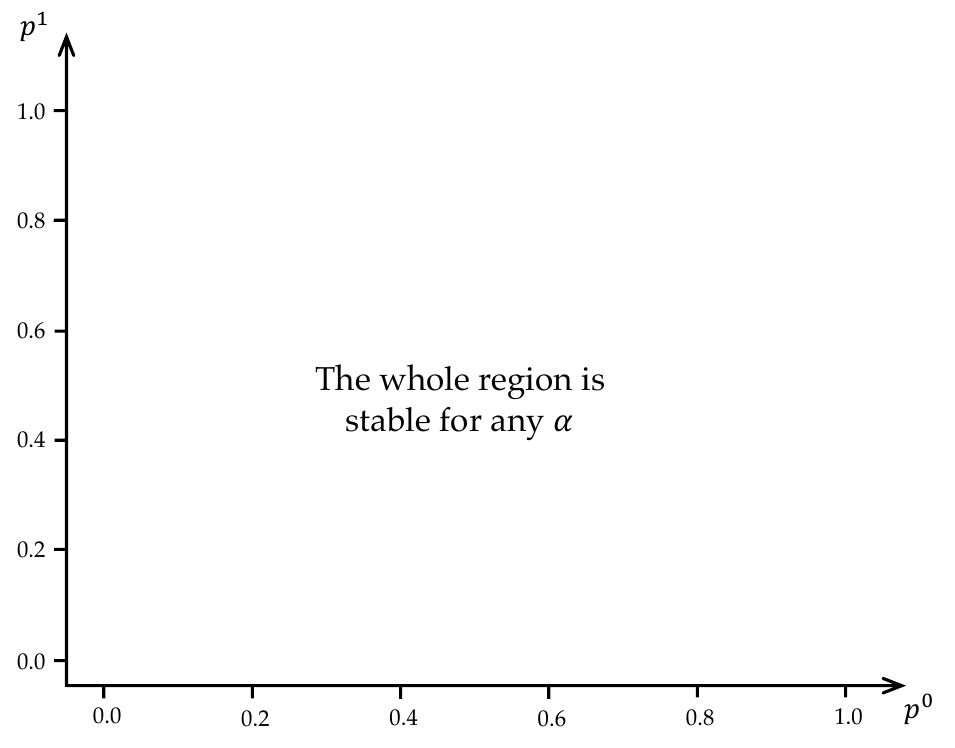}
        \caption{$\rho_i \geq -1$}
        \label{fig:logit_numerical_u1}
    \end{subfigure}
    \begin{subfigure}[t]{0.49\columnwidth}
        \includegraphics[width=\columnwidth]{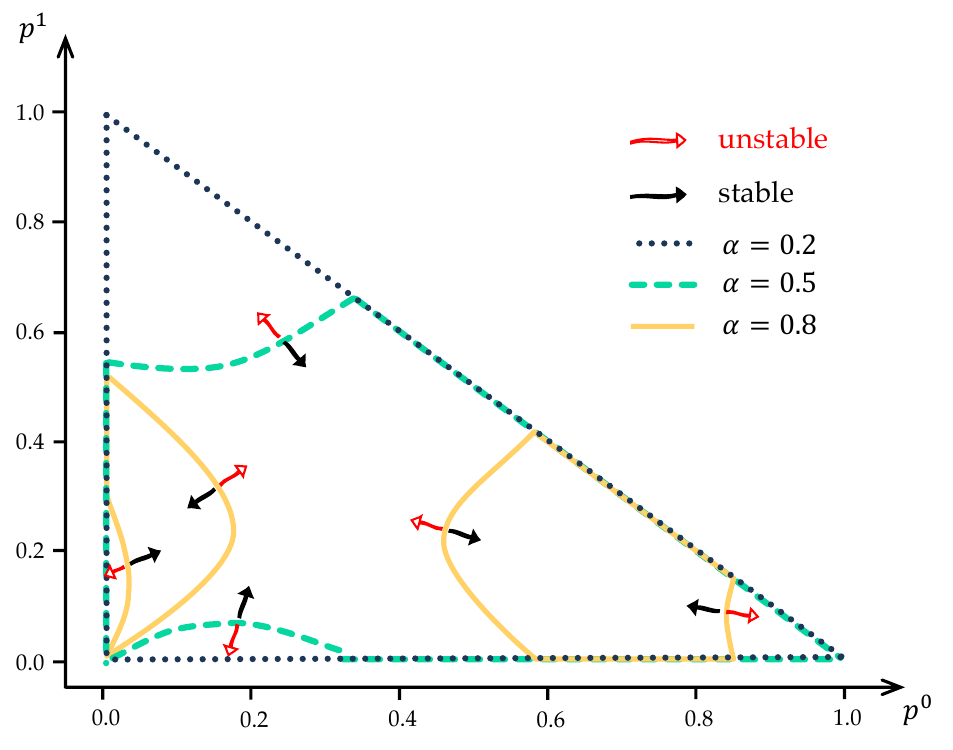}
        \caption{$\rho_i=-3$}
        \label{fig:logit_numerical_u3}
    \end{subfigure}\\
    \begin{subfigure}[t]{0.49\columnwidth}
        \includegraphics[width=\columnwidth]{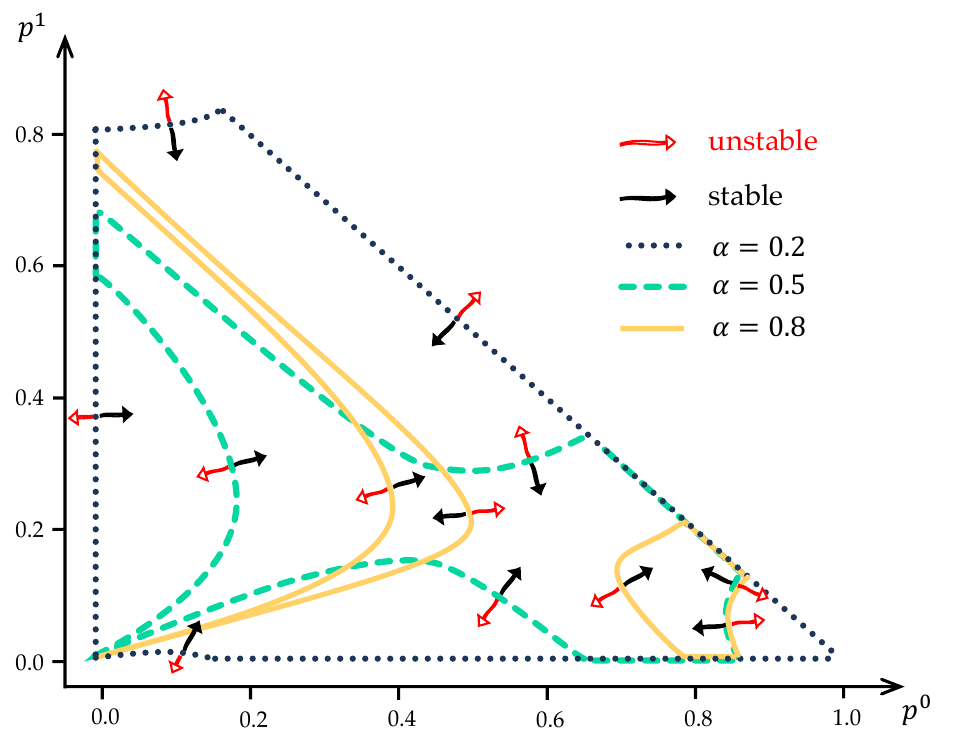}
        \caption{$\rho_i=-6$}
        \label{fig:logit_numerical_u6}
    \end{subfigure}
    \begin{subfigure}[t]{0.49\columnwidth}
        \includegraphics[width=\columnwidth]{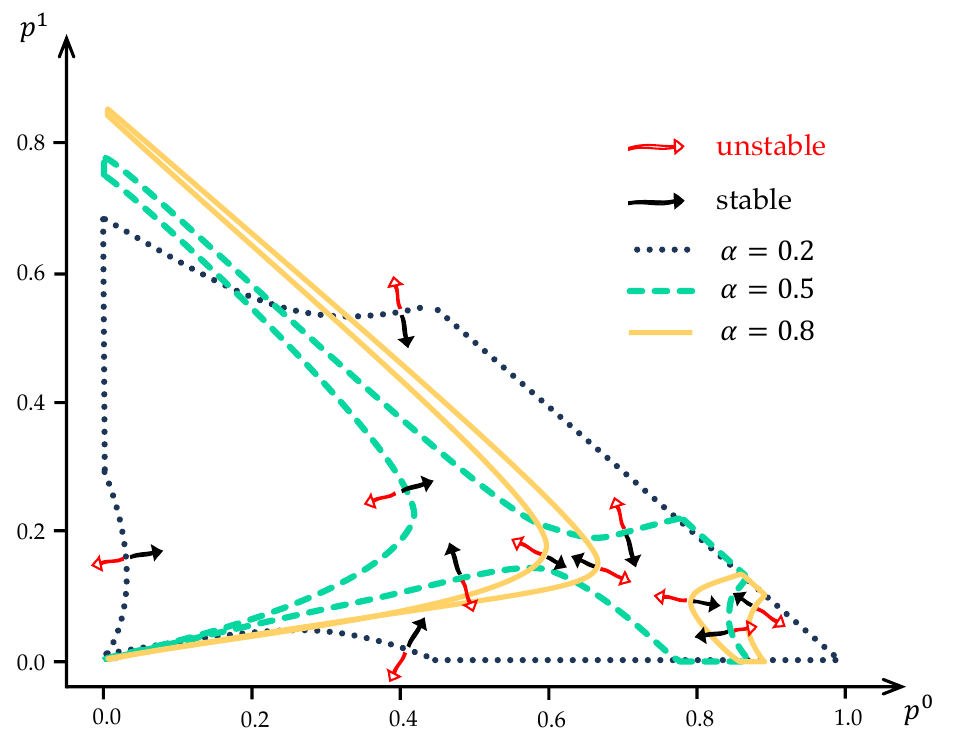}
        \caption{$\rho_i=-9$}
        \label{fig:logit_numerical_u9}
    \end{subfigure} \\
    \caption{\centering Stable and unstable regions of the CH-Logit dynamic with $\halpha=\alpha$ w.r.t. different $\rho_i$.}
    \label{fig:logit_numerical}
\end{figure}
\begin{enumerate}[(i)]
	\item When $\rho_i\geq -1$, the system maintains stability regardless of parameter values. This occurs because $\psi(\rho_i=-1;\alpha,\halpha=\alpha,p^0,p^1,p^2)=2\alpha^2-2(p^0+1)\alpha+1 = (\sqrt{2}\alpha-\frac{\sqrt{2}}{2}(1+p^0))^2 - \frac{1}{2}(1+p^0)^2+1$, which exceeds $-1$ due to constraints $p^0\leq 1$ and $\alpha < 1$. Further note that when $\rho_i\in [-1,0]$, the function monotonically increases in $\rho_i$. Hence, if $\rho_i\geq -1$, the cubic function is always greater than -1 but less than $1-\alpha$;
	\item Given a $\rho_i$ in $(-\infty,-1)$, the stable region shrinks as $\alpha$ increases. Moreover, the stable region splits into two pieces when $\alpha$ exceeds a certain threshold (approximately 0.6);
	\item For a fixed small $\alpha$ that does not separate the stable region, a balanced pattern of $p^0$, $p^1$, and $p^2$ is more likely to be stable. In other words, an extremely small or large $p^0$, $p^1$, or $p^2$ would compromise the stability;
	\item For a fixed large $\alpha$ that separates the stable region, as $\rho_i$ decreases in $(-\infty,-1)$, the two pieces of stable regions tend to move rightward. It indicates that when the stability condition is restricted, a large $p^0$ is more likely to stabilize the system.
\end{enumerate}

\paragraph{Imperfect prediction when $|K|=2$}
The cubic function $\psi(\rho_i;\alpha, \allowbreak \halpha,p^0,p^1,p^2)$ in Proposition~\ref{prop:logit_stable_general} reduces to a quadratic or linear function when $|K|=2$ or 1, respectively. In these two cases, the stable region in terms of $\rho_i$ can be succinctly described by the key parameters. Proposition~\ref{prop:logit_stability_neq_k2} gives the results, followed by the discussions on how over- and under-predictions will affect the stability.

\begin{proposition}\label{prop:logit_stability_neq_k2}
The CH-Logit dynamic with $|K|=1$ is locally asymptotically stable at the SUE under Assumption \ref{assumption:cost_psd} when $\rho_{i} > \frac{\alpha-2}{\alpha}, \forall i$. When $\htheta=\theta$, for $|K|=2$, the SUE under Assumption \ref{assumption:cost_psd} is locally asymptotically stable when one of the following two conditions holds for all $\rho_i$: (i) $1-\alpha + \frac{\alpha(1-\halpha+\halpha p^0)^2}{4\halpha(p^0-1)} > -1$ and $\rho_{i} > \frac{1}{(p^0-1)\halpha}$; or (ii) $1-\alpha + \frac{\alpha(1-\halpha+\halpha p^0)^2}{4\halpha(p^0-1)} < -1$ and $\{\rho_i: \frac{1}{(p-1)\halpha} \leq \rho_i \leq f_{0}(\alpha, \alpha, p^0) \}  \cup \{\rho_i: f_{1}(\alpha, \halpha, p^0) \leq \rho_i < 0 \}$. Here $f_{0}(\alpha,\halpha,p^0)\equiv \frac{1}{2} ( -\sqrt{\frac{\alpha (1+\halpha-\halpha p^0)^2+8 \halpha  (p^0-1)}{\alpha  \halpha ^2 (p^0-1)^2}}+  \frac{1}{\halpha  (p^0-1)}+1 )$, and $f_{1}(\alpha, \halpha, p^0)\equiv \frac{1}{2} ( \sqrt{\frac{\alpha (1+\halpha-\halpha p^0)^2+8 \halpha  (p^0-1)}{\alpha  \halpha ^2 (p^0-1)^2}}+\frac{1}{\halpha  (p^0-1)}+1 )$.
\end{proposition}
\proof{
	Please see Appendix \ref{apdx:logit_stability_neq_k2}.
}\endproof

Proposition~\ref{prop:logit_stability_neq_k2} reveals insights into how the strategic-thinking behavior and associated over- and under-predictions affect the stability by comparing the stable region size between CH-Logit models with $|K|=1$ and $2$. The discussion is as follows.

\paragraph*{Over-prediction $(\halpha \geq \alpha)$}	
	For case (i) of $|K|=2$ in Proposition~\ref{prop:logit_stability_neq_k2}, when we solve $1-\alpha + \frac{\alpha(1-\halpha+\halpha p^0)^2}{4\halpha(p^0-1)} > -1$ with $\halpha \geq \alpha$, we obtain $p^0 \leq 2 \sqrt{2} \sqrt{-\frac{\alpha -2}{\alpha ^2 \halpha ^2}}+\frac{\alpha  \halpha +\alpha -4}{\alpha  \halpha}$. Denote the RHS of this inequality as $h(\alpha,\halpha)$. Since $h(\alpha,\halpha)$ increases monotonically with $\halpha$ and $h(\alpha,\halpha=\alpha) \geq 2(\sqrt{2}-1)$, we have $h(\alpha,\halpha)\geq 2(\sqrt{2}-1)\approx 0.828$. The stability condition for $|K|=1$ requires all the $\rho_i > \frac{\alpha-2}{\alpha}$ while the condition for $|K|=2$ needs all the $\rho_i > \frac{1}{(p^0-1)\halpha}$. Comparison of these two thresholds yields: $\frac{1}{(p^0-1)\halpha}-\frac{\alpha-2}{\alpha} < 0$ when $p^0 > \frac{\alpha \halpha +\alpha -2 \halpha }{\alpha  \halpha -2 \halpha }$ and $\frac{1}{(p^0-1)\halpha}-\frac{\alpha-2}{\alpha} > 0$ when $p^0 < \frac{\alpha \halpha +\alpha -2 \halpha }{\alpha  \halpha -2 \halpha }$. Denote $\frac{\alpha \halpha +\alpha -2 \halpha }{\alpha  \halpha -2 \halpha }$ as $g(\alpha,\halpha)$. Summarizing the above, the stable region expands when $g(\alpha,\halpha) < p^0 < h(\alpha,\halpha)$ and shrinks when $0 < p^0 < g(\alpha,\halpha)$.
	
	For case (ii) of $|K|=2$ in Proposition~\ref{prop:logit_stability_neq_k2}, $p^0 > h(\alpha,\halpha)$. Combined size of the two separated stable regions over $\rho_i$ is $f_0(\alpha, \halpha, p^0) - \frac{1}{(p-1)\halpha} + 0 - f_1(\alpha, \halpha, p^0) = \frac{1}{\halpha(1-p^0)}\cdot \bigbracket{1-\sqrt{\frac{8\halpha(p^0-1) + \alpha(1+\halpha-p^0\halpha)^2}{\alpha}}}$, which is always less than $\frac{2-\alpha}{\alpha}$ when $p^0 > h(\alpha,\halpha)$. Hence, the total size of the two separated stable regions is expanded.\par
\begin{figure}[t!]
    \centering
    \includegraphics[scale=0.65]{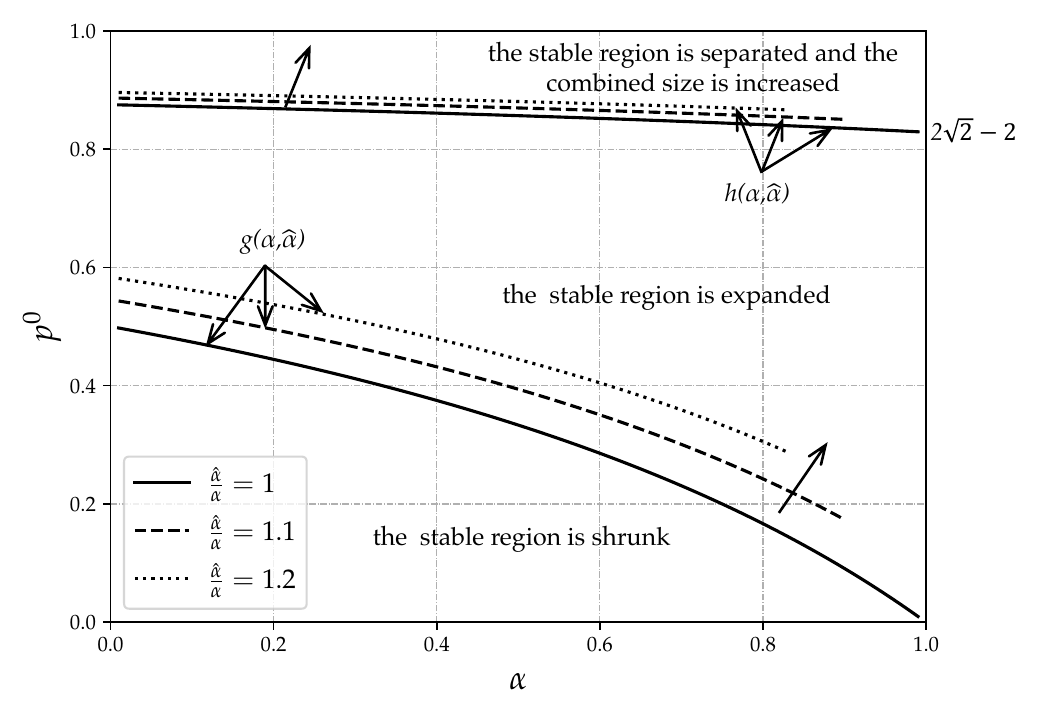}
    \caption{\centering Stable region alterations when $|K|$ changes from 1 to 2 under different $p^0$, $\alpha$, $\halpha$, and $\halpha \geq \alpha$.}\label{fig:logit_k2_region_comp_large_halpha}
    \end{figure}

	Figure~\ref{fig:logit_k2_region_comp_large_halpha} summarizes the above discussion by showing how the stable region changes when $|K|$ shifts from 1 to 2 under different values of $p^0$, $\alpha$, $\halpha$. The parametric space of $p^0$ and $\alpha$ with a shrinking stable region enlarges as over-prediction severity increases (i.e., $\frac{\halpha}{\alpha}$ becomes larger).

\paragraph*{Under-prediction $(\halpha \leq \alpha)$}

\paragraph*{Scenario (1): $\halpha > 3-2\sqrt{2}$}
Consider first case (i) of $|K|=2$ in Proposition~\ref{prop:logit_stability_neq_k2} where $f^{\ast}(\alpha, \halpha, p^0) > -1$ and $p^0 < h(\alpha,\halpha)$. The discussion is divided into two sub-cases:

\indent 1.When $\halpha \geq \frac{\alpha}{2-\alpha}$, solving $\frac{1}{(p^0-1)\halpha} < \frac{\alpha-2}{\alpha}$ yields $p^0 > g(\alpha,\halpha)$ and solving $\frac{1}{(p^0-1)\halpha} > \frac{\alpha-2}{\alpha}$ obtains $p^0 < g(\alpha,\halpha)$. Hence, the region is expanded when $g(\alpha,\halpha) < p^0 < h(\alpha,\halpha)$ and shrunk when $p^0 < g(\alpha,\halpha)$. It can be verified that $g(\alpha,\halpha)$ is monotonically increasing in $\halpha$ and that $h(\alpha,\halpha)-g(\alpha,\halpha)$ is monotonically decreasing in $\halpha$ when $0<\alpha,\halpha<1$. Hence, the parametric space of $p^0$, $\alpha$, and $\halpha$ ($<\alpha$) rendering the stable region shrunk is smaller than the perfect prediction case, while the parametric space leading to an expanded stable region is larger than the perfect prediction case;

\indent 2.When $\halpha < \frac{\alpha}{2-\alpha}$, $\frac{1}{(p^0-1)\halpha} < \frac{\alpha-2}{\alpha}$ always holds and thus the expanded region is simply $p^0 < h(\alpha,\halpha)$. When prediction behaviors are considered, no stable region of $|K|=1$ is shrunk.


For case (ii) where $f^{\ast}(\alpha, \halpha, p^0) < -1$, $p^0 > h(\alpha,\halpha)$, we verify that $f_0(\alpha, \halpha, p^0) - \frac{1}{(p-1)\halpha} + 0 - f_1(\alpha, \halpha, p^0) > \frac{2-\alpha}{\alpha}$. Hence, when $p^0 > h(\alpha,\halpha)$, the stable region separates and increases in combined size. Since $h(\alpha,\halpha)$ is monotonically increasing in $\halpha$, the parametric space rendering the total size of the stable region greater is larger than the perfect prediction case. \par
\begin{figure}[h!]
    \centering
    \includegraphics[scale=0.65]{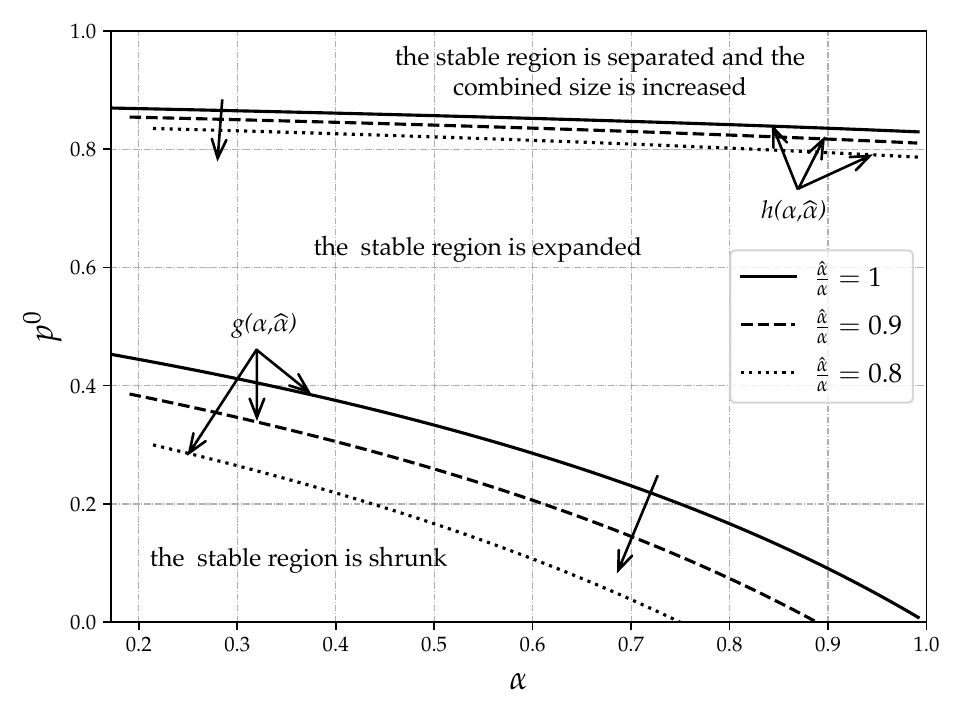}
    \caption{\centering Stable region alterations when $|K|$ changes from 1 to 2 under different $p^0$, $\alpha$, $\halpha$ and $3-2\sqrt{2}< \halpha \leq \alpha$.}\label{fig:logit_k2_region_comp_small_halpha}
    \end{figure}

The above discussion is summarized in Figure~\ref{fig:logit_k2_region_comp_small_halpha}. Note how the parametric space of $p^0$ and $\alpha$ with an expanding stable region enlarges as a ``moderately'' small $\halpha\in (3-2\sqrt{2},\alpha)$ increases. In other words, a mild under-prediction helps stabilize the CH-Logit dynamic, which is very similar to the finding of the CH-NTP dynamic.

\paragraph*{Scenario (2): $\halpha < 3-2\sqrt{2}$}
This scenario depends on the comparison between $\frac{8\halpha}{(1+\halpha)^2}$ and $\alpha$. When $\frac{8\halpha}{(1+\halpha)^2} > \alpha$, the analysis mirrors the case of $\halpha > 3-2\sqrt{2}$; i.e., it helps stabilize the dynamic. When $\frac{8\halpha}{(1+\halpha)^2} < \alpha$, $f^{\ast}(\alpha, \halpha, p^0) < -1$ always holds and thus the stable region is separated, while the combined size is increased. The latter case is similar to the CH-NTP dynamic with $\hgamma < \frac{\gamma}{2}$.




\section{Conclusions and Future Research}\label{sec:conclusion}
This paper developed a new modeling framework of day-to-day network flow dynamics that incorporates cognitive hierarchy theory to better capture travelers' dynamic re-routing behaviors. The classical NTP and Logit dynamics, on behalf of the DUE and SUE cases, respectively, were extended into the framework. By capturing heterogeneity in travelers' strategic-reasoning levels, the proposed CH-NTP model significantly improved the goodness-of-fit for a recent virtual experiment, which traditional day-to-day models could not achieve. 



We find that in addition to converging to the user equilibria, the proposed dynamics may also converge to other equilibria, depending on the initial conditions and parameters. For example, our analysis of the CH-NTP dynamic reveals that non-DUE MPE emerge when travelers extensively anticipate others' behaviors. Interestingly, non-DUE MPE can either improve or worsen system efficiency compared to DUE, depending on the proportion of strategic travelers and the cost functions. Moreover, when higher-step travelers can accurately predict lower-step ones' switching propensities, thinking multiple steps ahead does not affect the local stability of the CH-NTP dynamic around the DUE. In comparison, the stability condition for the CH-Logit dynamic largely depends on the proportion of heterogeneous travelers. Finally, we find that for both dynamics with $|K|=2$: (i) when 1-step travelers over-predict 0-step ones' switching propensity, the stable region is shrunk, and when the over-prediction is severe (e.g., $\hgamma>\bar{\gamma}$ for the CH-NTP dynamic), the UE is always unstable; (ii) when 1-step travelers \textit{moderately} under-predict 0-step ones' re-routing tendency, the stable region is increased; and (iii) when the under-predictions are severe, the stable region is separated into two parts.\par

Despite the fresh findings generated, our models still have limitations, which may direct future research. First, we select two widely-used day-to-day models from the literature as examples of the general modeling framework. Many more day-to-day models can be analyzed in the future \citep[e.g.,][]{smith1979existence,smith1983existence,nagurney1997projected,jin2007dynamical,xiao2019day,li2024day}. We surmise that findings on local stability still generally hold when hierarchical cognitive behaviors are incorporated into these models, provided that the dynamic itself exhibits regular properties (e.g., definiteness of the Jacobian). Second, it would also be interesting to apply the idea of cognitive hierarchy to investigate the departure time choice problem in the day-to-day setting \citep{guo2018we,jin2020stable,jin2021stable,lamotte2021monotonicity,guo2023day}. Exploring how different levels of strategic thinking influence travelers' departure time adjustment may reveal new equilibrium properties in day-to-day bottleneck dynamics. Third, the virtual experiment data suggest that travelers' strategic thinking levels may evolve with experience, as participants appeared to adopt more uniformly random strategies after extended gameplay. Future research could incorporate time-varying parameters, particularly a time-decreasing prediction parameter, to better capture this behavioral adaptation. Fourth, future research can investigate how different measures, such as congestion tolling \citep{tan2015dynamic,guo2016day,han2017discrete,wu2024multiday}, traffic control \citep{smith1979traffic,kohler2019traffic}, and information provision schemes \citep{liu2020learning}, can improve system performance. Due to the multi-equilibria phenomenon, it is be interesting to drive the traffic system to desired states via these measures.


\section*{Acknowledgement}
Correspondence concerning this article should be addressed to Feng Xiao. The research was supported by the National Natural Science Foundation of China (Project No.'s~72201214 and 72025104), the Sichuan Science and Technology Program (Project No.~2023NSFSC1035) and the Central Guidance for Local Science and Technology Development Fund Projects (Project No.~2024ZYD0108).

\newpage
\begin{appendices}

\section{Definitions and theorems on local stability and block matrices}\label{apdx:stabilities}
\setcounter{theorem}{0} \renewcommand{\thetheorem}{\ref{apdx:stabilities}.\arabic{theorem}}
\setcounter{equation}{0} \renewcommand{\theequation}{\ref{apdx:stabilities}.\arabic{equation}}
\setcounter{definition}{0} \renewcommand{\thedefinition}{\ref{apdx:stabilities}.\arabic{definition}}
\newcommand{\subSDet}{\begin{vsmallmatrix}S_{22} & S_{23} \\ \bzero & S_{33}\end{vsmallmatrix}}

\begin{definition}
	Suppose $\x^{\dagger}$ is an equilibrium point (fixed point) of the discrete autonomous system:
	\begin{equation}\label{eq:discrete_system}
		\x^{(t+1)} = \phi(\x^{(t)}).
	\end{equation}
	$\x^\dagger$ is said to be:
	\begin{enumerate}[(i)]
	\item locally stable if given any $\epsilon > 0$, there exists a $\delta(\epsilon)>0$ such that
	\begin{equation}
		||\x^{(0)}-\x^\dagger|| < \delta(\epsilon) \to ||\x^{(t)} - \x^\dagger|| < \epsilon, \forall t=1,2,3,\ldots ;
	\end{equation}
	\item locally asymptotically stable if there exists an $\eta>0$ such that:
	\begin{equation}
		||\x^{(0)}-\x^\dagger|| < \eta \to \lim_{t\to \infty} (\x^{(t)})=\x^\dagger;
	\end{equation}
	and
	\item unstable if (i) is not true.
	\end{enumerate}
\end{definition}

\begin{theorem}\label{thm:eigen_stability}
	\citep{khalil2002nonlinear,antsaklis2006linear} Let $\x^\dagger$ be an equilibrium point of the system in Eq.~\eqref{eq:discrete_system} and $\phi$ be continuously differentiable in a neighborhood of $\x^\dagger$ with radius $\epsilon$, $\mathcal{B}(\x^\dagger, \epsilon)$. Let $J$ be the Jacobian evaluated at $\x^\dagger$, $J = \pder{\phi(\x)}{\x}|_{\x=\x^\dagger}$. Then $\x^\dagger$ is:
	\begin{enumerate}[(i)]
		\item locally asymptotically stable in $\mathcal{B}(\x^\dagger, \epsilon)$ if and only if all eigenvalues of $J$ are within the unit circle of the complex plane (i.e., if $\lambda_1,\ldots,\lambda_n$ denote the eigenvalues of $J$, then $|\lambda_j| < 1, j=1,\ldots,n$);
		\item locally stable if and only if $|\lambda_j| \leq 1, j=1,\ldots,n$, and each eigenvalue with $|\lambda_j|=1$ is associated with a Jordan block of order 1. If the $n \times n$ matrix $J$ has a repeated eigenvalue $|\lambda_j|=1$ of algebraic multiplicity $q_j$, then the Jordan block associated with $\lambda_j$ has order one if and only if rank $(J-\lambda_j I)=n-q_j$; i.e., the dimension of the null space of $(J-\lambda_j I)$ equals $q_j$;\par
		and
		\item unstable if and only if (ii) is not true.
	\end{enumerate}
\end{theorem}

\begin{theorem}\label{thm:block_2_by_2_det}
	(adapted from Equation 0.8.5.13 in \citealp{horn2012matrix}) For block matrices $S_{11}$, $S_{12}$, $S_{21}$ and $S_{22}$, if $S_{11}$ commutes with either $S_{12}$ or $S_{21}$, i.e., $S_{11}S_{12} = S_{12}S_{11}$ or $S_{11}S_{21} = S_{21}S_{11}$, then $\begin{vsmallmatrix}S_{11} & S_{12}\\ S_{21} & S_{22} \end{vsmallmatrix} = |S_{11}S_{22}-S_{21}S_{12}|$.
\end{theorem}

\begin{theorem}\label{thm:upper_block_triangle}
	For upper triangular block matrices like
	\begin{equation}
		S = \begin{vmatrix}
   		S_{11} & S_{12} & S_{13} \\
   		\bzero & S_{22} & S_{23} \\
   		\bzero & \bzero & S_{33}
    	\end{vmatrix},
	\end{equation}
	where each block matrix has the size of $m\times m$, its determinant $|S|=|S_{11}|\cdot|S_{22}|\cdot|S_{33}|$.
\end{theorem}
\proof{Proof.}{
(We give a proof for completeness because most proofs we have found deal with the case of $2\times 2$ matrices only.) If $S_{11}$ is replaced by a $m$-size identity matrix, expanding the first column yields: 
	\begin{equation}\label{eq:recur_3_by_3}
		\begin{vmatrix}
   			I_m & S_{12} & S_{13} \\
   			\bzero & S_{22} & S_{23} \\
   			\bzero & \bzero & S_{33}
    	\end{vmatrix} = 
    	\begin{vmatrix}
    		I_{m-1} & S_{12}^{1-} & S_{13}^{1-} \\
   			\bzero & S_{22} & S_{23} \\
   			\bzero & \bzero & S_{33}
    	\end{vmatrix},
	\end{equation}
	where $S_{12}^{1-}$ and $S_{13}^{1-}$ represent $S_{12}$ and $S_{13}$ with their first rows removed. When $m=1$, Eq.~\eqref{eq:recur_3_by_3} reduces to $\subSDet$. By induction on $m$, the determinant on the left-hand side of Eq.~\eqref{eq:recur_3_by_3} equals $\subSDet$. According to Theorem~\ref{thm:block_2_by_2_det}, this equals $|S_{22}|\cdot |S_{33}|$. Denote $G$ as matrix $\begin{vsmallmatrix} S_{22} & S_{23} \\ \bzero & S_{33}\end{vsmallmatrix}$.\par
	When $G$ is invertible, the following always holds:
	\begin{equation}\label{eq:block_construct}
		\bigsbracket{\begin{array}{ccc}
   			I & 0 & 0 \\
   			0 & & \\
   			0 & \multicolumn{2}{c}{\smash{\raisebox{.5\normalbaselineskip}{$G^{-1}$}}}\\
		\end{array}}
		\cdot 
		\bigsbracket{\begin{array}{ccc}
   			S_{11} & S_{12} & S_{13} \\
   			0 & & \\
   			0 & \multicolumn{2}{c}{\smash{\raisebox{.5\normalbaselineskip}{$G$}}}\\
		\end{array}}
		=
		\bigsbracket{\begin{array}{ccc}
   			S_{11} & S_{12} & S_{13} \\
   			0 & & \\
   			0 & \multicolumn{2}{c}{\smash{\raisebox{.5\normalbaselineskip}{$I_{2m\times 2m}$}}}\\
		\end{array}}.
	\end{equation}\par
	Using the similar mathematical induction technique on RHS of Eq.~\eqref{eq:block_construct}, we know its determinant is simply $|S_{11}|$. Hence, Eq.~\eqref{eq:block_construct} indicates that $|G^{-1}|\cdot |S| = |S_{11}|$. Since $|G^{-1}|=\frac{1}{|G|}$, $|S|=|S_{11}|\cdot\subSDet = |S_{11}|\cdot|S_{22}|\cdot|S_{33}|$.\par
The case of $G$ being not invertible is trivial since $G$ would have at least one non-pivot row, and so does $S$. In this case, $|S|=|S_{11}||S_{22}||S_{33}|=0$.
}\endproof

\section{Proof of Lemma~\ref{lemma:Q}}\label{apdx:jp_jacob}
\setcounter{equation}{0} \renewcommand{\theequation}{\ref{apdx:jp_jacob}.\arabic{equation}}

Projecting a vector $\bz$ onto $\Omega_\eta \equiv \{ \x| \Gamma\x=\eta\boldsymbol{d},\x\geq0\}$ requires solving the optimization problem $\argmin_{y\in \Omega_{\eta}} \distance{\y-\bz}{2}$. The Lagrangian of this optimization problem can be expressed as:
\begin{equation}
	L(\bz, \bmu, \btau) = \frac{1}{2}\distance{\x-\bz}{2} - \bmu^T\x + \btau^T(\Gamma\x-\eta\bd),
\end{equation}
where $\bmu\equiv(\mu_{rw},r\in R_{w},w\in W)^T$ and $\btau\equiv(\tau_{w},w\in W)^T$ represent the Lagrangian multipliers. These multipliers correspond to the non-negative flow constraints and flow conservation constraints, respectively.

The optimal $\x^{\ast}, \bmu^{\ast}, \btau^{\ast}$ must satisfy the following Karush-Kuhn-Tucker conditions:
\begin{equation}\label{eq:tgp_lg_stationary}
	\x^{\ast} - \bz -\bmu^{\ast} + \Gamma^T\btau^{\ast} = \bzero,
\end{equation}
\begin{equation}\label{eq:tgp_lg_cons}
	\Gamma\x = \eta\bd, \x^{\ast}\geq 0, \bmu^{\ast}\geq \bzero,
\end{equation}
\begin{equation}\label{eq:tgp_lg_complement}
	\bmu\odot \x = \bzero,
\end{equation}
where $\odot$ is the Hadamard product operator. Let $E(\bz_w) = \{ r\in R_w|x_{rw}^{\ast}>0 \}$. From Eq.~\eqref{eq:tgp_lg_complement} we obtain that $\mu_{rw}^{\ast}=0, \forall r\in E(\bz_w), \forall w$. Hence, from Eq.~\eqref{eq:tgp_lg_stationary} we further know $x_{rw}^{\ast} = z_{rw}-\tau_{w}^{\ast}, \forall r\in E(\bz_w), \forall w$. Substituting $x_{rw}^{\ast}=z_{rw}-\tau_{w}^{\ast}$ into the equality in Eq.~\eqref{eq:tgp_lg_cons} yields $\sum_{r\in E(\bz_w)}{(z_{rw}-\tau_{w}^{\ast})} = \eta d_w, \forall w$. Hence $\tau_{w}^{\ast} = \frac{(\sum_{r\in E(\bz_w)}{z_{rw}})-\eta d_w}{|E(\bz_w)|}$.\par

Let $\bar{E}(\bz_w) =\{r\in R_w|x_{rw}^{\ast}=0\}$. From Eq.~\eqref{eq:tgp_lg_stationary} we know that $z_{rw}+\mu_{rw}^{\ast}-\tau_{w}^{\ast}=0, \forall r\in \bar{E}(\bz_w)$. Since $\mu_{rw}^{\ast}\geq 0,\forall r$, we can conclude that $z_{rw}\leq \tau_{w}^{\ast}, \forall r\in \bar{E}(\bz_w)$.\par

The above two cases show that the projection, for each $w$, has the form of:
\begin{equation}\label{eq:projection_equiv}
	P_{\Omega_w}[\bz_w] = \text{max}\{\bz_w - \tau_{w}^{\ast}, 0\},
\end{equation}
In a piecewise form, first note that this operator is undifferentiable only when the set $E(\bz_w)$ suddenly changes. The Jacobian of Eq.~\eqref{eq:projection_equiv} can be expressed as:
\begin{equation}
	Q_w[\bz_w] = \text{Diag}(\indone_w) - \frac{\indone_w\indone_w^T}{|E(\bz_w)|},
\end{equation}
where $\indone_w$ is an indicator vector of size $|R_w|$ whose $r$-th entry is 1 if $r\in E(\bz_w)$ and 0 otherwise. When $|E(\bz_w)|$ = $|R_w|$ (i.e., $x_{rw}^{\ast}>0, \forall r$), $Q_w[\bz_w]$ simplifies to $\bQ_w = \text{Diag}(\bone_w) - \frac{\bone_w\bone_w^T}{|R_w|}$ where $\bone_w$ denotes the all-one vector of size $|R_w|$.\par

To prove that $Q_w[\bz_w]$ is PSD for any $\bz_w$, we first observe that the rows or columns corresponding to $\bar{E}(\bz_w)$ (where $\indone_w = 0$) contribute only zeros to $Q_w[\bz_w]$. Thus, for any vector $\bolda$, the quadratic form $\bolda^T Q_w[\bz_w] \bolda$ simplifies to $\bolda_{E(\bz_w)}^T \tilde{Q}_w[\bz_w] \bolda_{E(\bz_w)}$, where $\tilde{Q}_w[\bz_w]$ is the submatrix of $Q_w[\bz_w]$ formed by retaining only the rows and columns indexed by $E(\bz_w)$, and $\bolda_{E(\bz_w)}$ denotes the subvector of $\bolda$ indexed by $E(\bz_w)$. Therefore, it suffices to analyze $\tilde{Q}_w[\bz_w]$, which forms an $|E(\bz_w)|$-by-$|E(\bz_w)|$ matrix, $\tilde{Q}_w[\bz_w] \equiv \text{Diag}(\bone_{E_w}) - \frac{\bone_{E_w}\bone_{E_w}^T}{|E(\bz_w)|}$, where $\bone_{E_w}$ is the all-one vector of size $|E(\bz_w)|$. Note that matrix $|E(\bz_w)|\cdot \tilde{Q}_w[\bz_w]$ can be viewed as the Laplacian of a fully connected graph with the number of nodes being $|E(\bz_w)|+1$. It is well-known that the Laplacian of any graph is PSD. Therefore, $\tilde{Q}_w[\bz_w]$ is also PSD, since scaling by $\frac{1}{|E(\bz_w)|}$ preserves the PSD.

To show $(Q_w[\bz_w])^2=Q_w[\bz_w]$ (i.e., idempotent), we again only need to focus on the non-zero submatrix $\tilde{Q}_w[\bz_w]$. It suffices to show $\tilde{Q}_w[\bz_w](\tilde{Q}_w[\bz_w]-\text{Diag}(\bone_{E_w}))=0$. This holds true because all the entries of $\tilde{Q}_w[\bz_w]-\text{Diag}(\bone_{E_w}$) are the same and entries of each column of $\tilde{Q}_w[\bz_w]$ sum up to 0. To show $ Q_w[\bz_w] $ is symmetric, we note that $ \text{Diag}(\indone_w) $ and $ \indone_w\indone_w^T $ are symmetric, so their difference $ Q_w[\bz_w] $ is symmetric.

The idempotence and symmetry of $ Q_w[\bz_w] $ tell that the matrix is an orthogonal projection matrix. So $\text{rank}(Q_w[\bz_w]) = tr(Q_w[\bz_w]) = |E(\bz_w)|-1$. Since rank is the number of nonzero eigenvalues, the algebraic multiplicity for $\lambda=1$ is also $|E(\bz_w)|-1$. Thanks to $(Q_w[\bz_w])^2=Q_w[\bz_w]$, every vector in the range of $Q_w[\bz_w]$ is an eigenvector with eigenvalue 1 and thus the associated eigenvalue's geometric multiplicity is $|E(\bz_w)|-1$. By the rank-nullity theorem, rank $(Q_w[\bz_w]-I)$ is $|R_w|-|E(\bz_w)|+1$.

\section{Derivation of Fact~\ref{fact:tp_jacob}}\label{apdx:tp_jacob}
\setcounter{equation}{0} \renewcommand{\theequation}{\ref{apdx:tp_jacob}.\arabic{equation}}
Denote $\zt{k}$ as $\xt{k}-\gamma\cpxtt{k},k=0,1,2$. For the 0-step travelers, take the Jacobian, $J[\cdot]$, for RHS of Eq.~\eqref{eq:tp_model} w.r.t. $\xt{0}$:
\begin{align*}
	\sJop{\frac{\alpha \sPop{\zt{0}}+(1-\alpha)\xt{0}}{\xt{0}}} & = \sJop{\frac{\sPop{\zt{0}}}{\zt{0}}} \cdot \sJop{\frac{\zt{0}}{\xt{0}}} + (1-\alpha)I \\
	&= \alpha \sQop{\xt{0}-\gamma\ccop{\txt}}\cdot \bigbracket{I - \gamma \sJop{\frac{\ccop{\txt}}{\txt}} \cdot \sJop{\frac{\txt}{\xt{0}}}} + (1-\alpha)I\\
	&= \alpha \sQop{\xt{0}-\gamma\ccop{\txt}} \cdot \bigbracket{I-\gamma \sDop{\txt}} + (1-\alpha)I, \numberthis
\end{align*}
and
\begin{align*}
	\sJop{\frac{\alpha \sPop{\zt{0}}+(1-\alpha)\xt{0}}{\xt{k\neq 0}}} &= \sJop{\frac{\alpha \sPop{\zt{0}}}{\zt{0}}}\cdot \sJop{\frac{\zt{0}}{\xt{k\neq 0}}} \\
	&= \alpha \sQop{\xt{0}-\gamma\ccop{\txt}} \cdot \bigbracket{-\gamma \sJop{\frac{\ccop{\txt}}{\txt}}\cdot \sJop{\frac{\txt}{\xt{k\neq 0}}}} \\
	&=  \alpha \sQop{\xt{0}-\gamma\ccop{\txt}} \cdot \bigbracket{-\gamma \sDop{\txt}}. \numberthis
\end{align*}\par
For 1-step travelers, first note that for $k=0,1$ and $2$,
\begin{align*}\label{eq:tp_jacobian_k1_temp}
	\sJop{\frac{\cpxtt{1}}{\xt{k}}} & = \sJop{\frac{\cpxtt{1}}{\pxtt{1}}} \bigbracket{\halpha \sJop{\frac{\sPop{\txt-\hgamma\ccop{\txt}}}{\txt-\hgamma\ccop{\txt}}}\sJop{\frac{\txt-\hgamma\ccop{\txt}}{\xt{k}}} + (1-\halpha)I } \\
	& = \sDop{\pxtt{1}}\cdot \bigbracket{ \halpha \sQop{\txt-\hgamma\ccop{\txt}} \cdot \bigbracket{I-\hgamma\sDop{\txt}} + (1-\halpha)I }. \numberthis
\end{align*}\par
Then applying the chain rule, we get:
\begin{align*}
	\sJop{\frac{\alpha \sPop{\zt{1}}+(1-\alpha)\xt{1}}{\xt{1}}} &= \alpha \sJop{\frac{\sPop{\zt{1}}}{\zt{1}}} \cdot \sJop{\frac{\zt{1}}{\xt{1}}} + (1-\alpha)I \\
	&= \alpha \sQop{\xt{1}-\gamma\cpxtt{1}}\cdot \bigbracket{I-\gamma \sJop{\frac{\cpxtt{1}}{\xt{1}}}}+(1-\alpha)I, \numberthis
\end{align*}
and
\begin{align*}
	\sJop{\frac{\alpha \sPop{\zt{1}}+(1-\alpha)\xt{1}}{\xt{k\neq 1}}} &= \sJop{\frac{\sPop{\zt{1}}}{\zt{1}}}\cdot \sJop{\frac{\zt{1}}{\xt{k\neq 1}}} \\
	& = \alpha \sQop{\xt{1}-\gamma\cpxtt{1}} \cdot \bigbracket{-\gamma \sJop{\frac{\cpxtt{1}}{\xt{k\neq 1}}}}, \numberthis
\end{align*}
where $\sJop{\frac{\cpxtt{1}}{\xt{k}}}, k=0,1,2,$ is given by Eq.~\eqref{eq:tp_jacobian_k1_temp}.\par
For 2-step travelers, note that for $k=0,1$ and $2$,
\begin{align*}\label{eq:tp_jacobian_k2_temp}
	&\sJop{\frac{\pxtt{2}(\txt)}{\xt{k}}}  = \sJop{\frac{\halpha \dPop{\Omega_{q_2^0}}{q_2^0\txt-\hgamma\ccop{\txt}}}{q_2^0\txt-\hgamma\ccop{\txt}}} \cdot \sJop{\frac{q_2^0\txt-\hgamma\ccop{\txt}}{\xt{k}}} +(1-\halpha)q_2^0 I \\
	&\quad + \sJop{\frac{\halpha \dPop{\Omega_{q_2^1}}{q_2^1\txt-\hgamma \cpxtt{1}}}{q_2^1\txt-\hgamma\cpxtt{1}}} \cdot \sJop{\frac{q_2^1\txt-\hgamma\cpxtt{1}}{\xt{k}}} +(1-\halpha)q_2^1 I \\
	&\quad = \halpha \sQop{q_2^0\txt-\hgamma\ccop{\txt}}\cdot \bigbracket{q_2^0I-\hgamma \sDop{\txt}} +(1-\halpha) + \halpha \sQop{q_2^1\txt-\hgamma\cpxtt{1}}\cdot \\
	&\quad\quad \bigbracket{q_2^1 I-\hgamma \halpha \bigsbracket{\sDop{\pxtt{1}}\cdot \bigbracket{ \halpha \sQop{\txt-\hgamma\ccop{\txt}} \cdot \bigbracket{I-\hgamma \sDop{\txt}} + (1-\halpha)I}}}.\numberthis
\end{align*}\par
Hence, 
\begin{align*}
	\sJop{\frac{\alpha \sPop{\zt{2}}+(1-\alpha)\xt{2}}{\xt{2}}} &= \alpha \sJop{\frac{\sPop{\zt{2}}}{\zt{2}}} \cdot \sJop{\frac{\zt{2}}{\xt{2}}} + (1-\alpha)I \\
	&= \alpha \sQop{\xt{2}-\gamma\cpxtt{2}}\cdot \bigbracket{I-\gamma \sDop{\pxtt{2}}\sJop{\frac{\pxtt{2}}{\xt{2}}} } +(1-\alpha)I, \numberthis
\end{align*}
and	
\begin{align*}
	\sJop{\frac{\alpha \sPop{\zt{2}}+(1-\alpha)\xt{2}}{\xt{k\neq 2}}} &= \sJop{\frac{\sPop{\zt{2}}}{\zt{2}}}\cdot \sJop{\frac{\zt{2}}{\xt{k\neq 2}}} \\
	& = \alpha \sQop{\xt{2}-\gamma\cpxtt{2}} \cdot \bigbracket{-\gamma \sDop{\pxtt{2}}\sJop{\frac{\pxtt{2}}{\xt{k\neq 2}}}}, \numberthis
\end{align*}
where $\sJop{\frac{\cpxtt{2}}{\xt{k}}}, k=0,1,2,$ is given by Eq.~\eqref{eq:tp_jacobian_k2_temp}.\par
Assembling the above blocks into $J\Phi$ yields Fact~\ref{fact:tp_jacob}.

\section{Proof of Proposition~\ref{prop:tp_ue_stability_eq_gamma}}\label{apdx:tp_ue_stability_eq_gamma}
\setcounter{equation}{0} \renewcommand{\theequation}{\ref{apdx:tp_ue_stability_eq_gamma}.\arabic{equation}}
According to Proposition~\ref{prop:tp_multi_eqs}, at the DUE, $\pxtt{0}=\pxtt{1}=\pxtt{2}$ and thus the Jacobians of route cost function $D_2=D_1=D_0=D_{\ast}$, where $D_{\ast}$ is the Jacobian of the route cost function evaluated at the DUE $\txa$. As per the assumption that the projection operator always yields positive route flow, all the $Q$'s and $\hat{Q}$'s in the Jacobian matrix in Eq.~\eqref{eq:tp_jacob} reduce to $\bQ$ (defined in Lemma~\ref{lemma:Q}). Moreover, since $\alpha=\halpha=1$ and $\hgamma=\gamma$, each sub-block matrix in Eq.~\eqref{eq:tp_jacob} reduces to:
\begin{align*}
	& JP_{0,0} = \bQ(I-\gamma D_{\ast}), \\
	 & JP_{0,j\neq0}=\bQ(-\gamma D_{\ast}),\\
	 & JP_{1,1} = \bQ(-\gamma (D_{\ast}\bQ(I-\gamma D_{\ast}))),\\
	 & JP_{1,j\neq1}=\bQ(I-\gamma (D_{\ast}\bQ(I-\gamma D_{\ast}))), \\
	 & JP_{2,2} = \bQ(I-\gamma D_{\ast}[\bQ(q_2^0I-\gamma D_{\ast}) + D_{\ast}\bQ(q_2^1I-\gamma \bQ(I-\gamma D_{\ast}))] ),\\
	 & JP_{2,j\neq2} = \bQ(-\gamma D_{\ast}[\bQ(q_2^0I-\gamma D_{\ast}) + D_{\ast}\bQ(q_2^1I-\gamma \bQ(I-\gamma D_{\ast}))] ).\numberthis
\end{align*}\par
Denote $A$ as $\bQ(I-\gamma D_{\ast})$. Thanks to the idempotence of $\bQ$ (Lemma~\ref{lemma:Q}), we have $\bQ A=\bQ \bQ(I-\gamma D_{\ast}) = A$. With this identity and after performing matrix algebraic calculations, the Jacobian in Eq.~\eqref{eq:tp_jacob} can be further represented as:

\begin{equation}
	JP = \begin{bmatrix}
   	A & A-\bQ & A-\bQ \\
   	(A-\bQ)A & (A-\bQ)A+\bQ & (A-\bQ)A \\
   	(A-\bQ)A^2 & (A-\bQ)A^2 & (A\b-Q)A^2+\bQ \\
    \end{bmatrix}.
\end{equation}\par
For the case of $|K|=1$, $JP$ reduces to $A$ and the DUE is stable when each eigenvalue of $A$ is within $(-1, 1)$.\par
For the case of $|K|=2$, $JP$ takes the four sub-block matrices in the upper left corner, and its eigenvalues are the roots of the following characteristic polynomial:
\begin{equation}
	\begin{vmatrix}
   	\lambda I-A & -A+\bQ \\
   	-(A-\bQ)A & \lambda I-(A-\bQ)A-\bQ
    \end{vmatrix} = 0.
\end{equation}\par

Using the identity $\bQ A=A$, one can verify that $(\lambda I-A)\cdot (-(A-\bQ)A) = (-(A-\bQ)A)\cdot (\lambda I-A)$. Hence, $\lambda I-A$ commutes with $-(A-\bQ)A$. Referring to Theorem~\ref{thm:block_2_by_2_det}, we have:
\begin{align*}\label{eq:tp_k2_final_det}
	& \bigvbracket{(\lambda I-A)(\lambda I-A^2+A-\bQ) + (A^2-A)(-A+\bQ)} = \bigvbracket{(\lambda I-A^2)(\lambda I-\bQ)} \\
	& \quad = \bigvbracket{\lambda I-A^2} \bigvbracket{\lambda I-\bQ} = 0. \numberthis
\end{align*}\par
Eq.~\eqref{eq:tp_k2_final_det} shows that for $|K|=2$, $JP$ has $2\cdot\sum_w{|R_w|}$ eigenvalues, half of which are the eigenvalues of $\bQ$ and the other half are the ones of $A^2$. For each OD pair $w$, we know from Lemma~\ref{lemma:Q} that $\bQ_w$ has $|R_w|$ eigenvalues with $|R_w|-1$ of them being 1 and the other one being 0. So $JP$ will have $\sum_w{(|R_w|-1)}$ eigenvalues being 1 and $|W|$ ones being 0. Due to the geometric and algebraic multiplicities with eigenvalue one both being $|R_w|-1$, every eigenvalue on the unit circle has an associated Jordan block of order 1, which satisfies the condition (ii) in Theorem~\ref{thm:eigen_stability}. As a result, the stability only depends on the eigenvalues of $A^2$.\par
The eigenvalues of $A^2$ are simply the squares of $A$'s. As long as $A$'s eigenvalues are within $(-1,1)$, the eigenvalues of $A^2$ are within $[0,1)$, which guarantee the local stability. (Note that asymptotic stability cannot be achieved due to eigenvalue 1 of $\bQ$.)\par
For the case of $|K|=3$, $J\Phi$'s eigenvalues are the roots of the following characteristic polynomial for $J\Phi$:
\begin{equation}\label{eq:tp_k3_det_origin}
	|\lambda I-J\Phi| = \begin{vmatrix}
   	\lambda I-A & -A+\bQ & -A+\bQ \\
   	-(A-\bQ)A & \lambda I-(A-\bQ)A-\bQ & -(A-\bQ)A \\
   	-(A-\bQ)A^2 & -(A-\bQ)A^2 & \lambda I-(A-\bQ)A^2-\bQ \\
    \end{vmatrix} = 0.
\end{equation}\par
Row and column operations do not change the determinant. Subtracting the third ``block column'' of the determinant from the first and second block columns and then adding the first and second ``block rows'' to the third block row, the determinant in Eq.~\eqref{eq:tp_k3_det_origin} becomes an upper triangular matrix:
\begin{equation}
	|\lambda I-J\Phi| = \begin{vmatrix}
   	\lambda I-\bQ & 0 & -(A-\bQ) \\
   	0 & \lambda I-\bQ & -(A-\bQ)A \\
   	0  & 0 & \lambda I-A^3 \\
    \end{vmatrix} = 0.
\end{equation}\par
Invoking Theorem~\ref{thm:upper_block_triangle}, 
\begin{equation}
	\bigvbracket{\lambda I-J\Phi} = \bigvbracket{\lambda I-\bQ}^2 \cdot
	\bigvbracket{\lambda I-A^3} = 0. \numberthis
\end{equation}\par
Similar to the reasoning of case $|K|=2$, $JP$ with $|K|=3$ has $3\cdot\sum_w{|R_w|}$ eigenvalues, two-thirds of which are equal to 1 or 0 (i.e., the eigenvalues of $\bQ$) and one-third of which are the ones of $A^3$. Similar to the case of $|K|=2$, every eigenvalue on the unit circle has an associated Jordan block of order 1. When $A$'s eigenvalues are in $(-1,1)$, the eigenvalues of $A^3$ are also in $(-1,1)$, which satisfies condition (ii) in Theorem~\ref{thm:eigen_stability}.

\section{Proof of Lemma~\ref{lemma:A_property}}\label{apdx:A_property}
\setcounter{equation}{0} \renewcommand{\theequation}{\ref{apdx:A_property}.\arabic{equation}}
Lemma~\ref{lemma:Q} says that $\sQop{\bz}$ is PSD and by Assumption \ref{assumption:cost_psd}, $\sDop{\x}$ is PSD. The product of two PSD matrices is always diagonalizable, and its eigenvalues are nonnegative. Hence, $\sQop{\bz}\sDop{\x}$ can be decomposed as $V \Lambda V^{-1}$ where $V$ is the eigenvector matrix and $\Lambda$ is the eigenvalue matrix. By the idempotent nature of $\sQop{\bz}$ shown in Lemma~\ref{lemma:Q},
		\begin{align*}\label{eq:QD_eigen_decomp}
			\sQop{\bz}\bigbracket{I-\gamma \sDop{\x}} \cdot (-\gamma \sQop{\bz}\sDop{\x}) &= -\gamma \sQop{\bz}\sDop{\x} + \gamma^2\sQop{\bz}\sDop{\x}\sQop{\bz}\sDop{\x} \\
			&= V\cdot(-\gamma\Lambda+\gamma^2\Lambda^2)\cdot V^{-1}.\numberthis
		\end{align*}\par
			
	$\sQop{\bz}(I-\gamma \sDop{\x})$ is not diagonalizable because $\sQop{\bz}(I-\gamma \sDop{\x})$ does not commute with $-\gamma \sQop{\bz}\sDop{\x}$. Still, $\sQop{\bz}(I-\gamma \sDop{\x})$ share the same eigenvector with $-\gamma \sQop{\bz}\sDop{\x}$ for nonzero eigenvalues. Suppose $\mu_i$ is a nonzero eigenvalue of $\sQop{\bz}\sDop{\x}$ with corresponding eigenvector $\bv$, i.e., $\sQop{\bz}\sDop{\x}\bv = \mu_i\bv,\mu_i\neq 0$. Then we have from Eq.~\eqref{eq:QD_eigen_decomp} that $(-\gamma\xi_i\mu_i)\bv = (-\gamma\mu_i + \gamma^2\mu_i^2)\bv$ where $\xi_i$ is the corresponding eigenvalue for $\sQop{\bz}(I-\gamma \sDop{\x})$. Solving it yields $\xi_i = 1 - \gamma\mu_i, \mu_i\neq 0$.\par
	As shown in Lemma~\ref{lemma:Q}, $\sQop{\bz}(I-\gamma \sDop{\x})$ has a rank of $|E_{z_w}|-1$ for each $w$ and thus the number of zero eigen values are $\sum_w(|R_w|-|E_{z_w}|+1)$. They don't contribute to the maximum modulus of eigenvalues of $\sQop{\bz}(I-\gamma \sDop{\x})$.

\section{Proof of Proposition~\ref{prop:tp_ue_stability_neq_gamma}}\label{apdx:tp_ue_stability_neq_gamma}
\setcounter{equation}{0} \renewcommand{\theequation}{\ref{apdx:tp_ue_stability_neq_gamma}.\arabic{equation}}
For $|K|=2$ with $\hgamma\neq \gamma$,
\begin{equation}
	\bigvbracket{\lambda I - JP} = \begin{vmatrix}
   	\lambda I-A & -A+\bQ \\
   	-(A-\bQ)\hat{A} & \lambda I-(A-\bQ)\hat{A}-\bQ
    \end{vmatrix} = 0,
\end{equation}
where $A = \bQ(I-\gamma D_{\ast})$ and $\hat{A} = \bQ(I-\hgamma D_{\ast})$. Using again the idempotence of $\bQ$, one can verify that $(\lambda I - A)$ and $-(A-\bQ)\hat{A}$ commute. Referring to Theorem~\ref{thm:block_2_by_2_det}, we have $|(\lambda I-A\hat{A}+\hat{A}-A)| |(\lambda I-\bQ)| = 0$. $A$ and $\hat{A}$ share the same eigenvector $\bv$ with $\bQ D_{\ast}$ for the eigenvalue $\lambda$ of $\bQ D_{\ast}$ being nonzero. Assume $\bQ D_{\ast}\bv = \lambda\bv, \lambda\neq 0$. The matrix $A\hat{A}-\hat{A}+A$ determining the stability multiplied by $\bv$ becomes:
\begin{align*}
	&(A\hat{A}-\hat{A}+A)\bv = [(\bQ-\gamma \bQ D_{\ast})(-\hgamma \bQ D_{\ast}) + \bQ - \gamma \bQ D_{\ast}\bQ + (\hgamma-\gamma)\bQ D_{\ast}]\bv \\
	& = \bigsbracket{(1-\gamma\lambda)(-\hgamma\lambda) + (1-\gamma\lambda) + (\hgamma-\gamma)\lambda} \bv \\
	& = \bigsbracket{\gamma\hgamma\lambda^2 - 2\gamma\lambda + 1} \bv, \numberthis
\end{align*}
where the first equality is also due to the property of $\bQ$ in Lemma~\ref{lemma:Q}. The above equation represents the nonzero eigenvalue of $A\hat{A}-\hat{A}+A$ by the eigenvalue $\lambda$ of $\bQ D_{\ast}$ as $\gamma\hgamma\lambda^2 - 2\gamma\lambda + 1, \lambda > 0$. Again, zero eigenvalue(s) of $A\hat{A}-\hat{A}+A$ don't contribute to the maximum modulus of eigenvalues.

\section{Proof of Proposition~\ref{prop:non_due_mpe}}\label{apdx:non_due_mpe}
\setcounter{equation}{0} \renewcommand{\theequation}{\ref{apdx:non_due_mpe}.\arabic{equation}}

We prove it by contradiction for each class.\par
\textbf{Class-0.} Assume $x_i^{0,\dagger} > 0$ for all routes $i$. The equilibrium condition for class $k=0$ is:
		\begin{equation}
		\xd{0} = \dPop{\Omega_{p^0}}{\xd{0}-\gamma\ccop{\txd}}.
	\end{equation}
	 Since $x_i^{0,\dagger} > 0, \forall i$, the first-order optimality condition implies that for each OD pair $w$, travel costs are equalized across all routes:
	 \begin{equation}
	 	c_i(\txd) = c_j(\txd), \forall i, j\in R_w.
	 \end{equation}
	 Two subcases arise: (i) If the DUE requires zero flow on some routes, no $\txd\in \Omega_1$ can satisfy these equalities, which results in a contradiction; (ii) If the DUE solution requires strictly positive flows on all routes, the equalities reduces to the DUE condition, contradicting the assumption that $\txd$ is not DUE. Thus, class $k=0$ must abandon at least one route.
	
\textbf{Class-1.} Assume $x_i^{1,\dagger} > 0$ for all routes $i$. The equilibrium condition for class $k=1$ is:
	\begin{equation}
		\xd{1} = \dPop{\Omega_{p^1}}{\xd{1}-\gamma\ccop{\pxd{1}\bigbracket{\txd}}},
	\end{equation}
	where $\pxd{1} = \halpha \dPop{\Omega_{1}}{\txd-\hgamma\ccop{\txd}} + (1-\halpha)\txd$. 
	
	If all routes in class $k=1$ have positive flow, then the first-order optimality condition implies that for each OD pair $w$:
	\begin{equation}
		c_i\bigbracket{\pxd{1}\bigbracket{\txd}} = c_j\bigbracket{\pxd{1}\bigbracket{\txd}}, \forall i,j \in R_w.
	\end{equation}

	Similar to the analysis of class $k=0$, we consider two subcases. 
	
	\begin{enumerate}[(i)]
		\item If the DUE admits zero flow on some routes, then no $\pxd{1}\in \Omega_1$ can satisfy the equalities, leading to a contradiction.
		\item If the DUE admits strictly positive flow on all routes, then $\pxd{1}$ is a DUE flow $\txa$. This case requires more detailed analysis, as there could be many non-DUE total flow patterns ($\txd$) that, after the prediction step Eq.~\eqref{eq:tp_k1_pred_cost}, become a DUE ($\pxd{1}$). In other words, $\pxd{1}$ could be the DUE while $\txd$ is not. This occurs because the projection operation in Eq.~\eqref{eq:tp_k1_pred_cost} is surjective but not injective.)
		
		To derive a contradiction, we substitute $\pxd{0}=\txd$ and $\pxd{1}=\txa$ into the joint VI characterization of the MPE in Eq.~\eqref{eq:MPE} with $|K|=2$:
		\begin{equation}\label{eq:joint_vi_k2}
			\ccop{\txd}^T\bigbracket{\x^0-\xd{0}} + \ccop{\txa}^T \bigbracket{\x^1-\xd{1}} \geq 0, \forall \x^0\in \Omega_{p^0}, \x^1\in \Omega_{p^1}.
		\end{equation}
		Since $\txd$ is not a DUE, there exists at least one OD pair $w$ with routes $r,s\in R_w$ such that $c_r\bigbracket{\txd} > c_s\bigbracket{\txd}$. Note that $\ccop{\txa}$ is constant across routes in each OD pair, making the term $\ccop{\txa}^T \bigbracket{\x^1-\xd{1}} = 0$. Therefore, the cost difference must come from class $k=0$ travelers: they must be using at least two routes in OD pair $w$ with unequal costs, i.e., $x_{r}^{0,\dagger} > 0, x_{s}^{0,\dagger} > 0$ with $c_r(\txd) > c_s(\txd)$.
		
		We construct a perturbed flow $\x^0$ by reallocating a small amount $\epsilon > 0$ from route $r$ to route $s$, while maintaining the flow on all the other routes:
		\begin{equation}
			x_r^0 = x_{r}^{0,\dagger}-\epsilon, x_s^0 = x_{s}^{0,\dagger}+\epsilon.
		\end{equation}
		For this perturbed flow pattern (which remains in $\Omega_{p^0}$), we obtain:
		\begin{equation}
			\ccop{\txd}^T \bigbracket{\x^0 - \xd{0}} = \epsilon\bigbracket{c_s\bigbracket{\txd}-c_r\bigbracket{\txd}} < 0.
		\end{equation}
		This inequality shows there exists a flow pattern $\x^0$ that violates the joint VI of Eq.~\eqref{eq:joint_vi_k2}, i.e., 
		\begin{equation}
			\ccop{\txd}^T\bigbracket{\x^0-\xd{0}} + \ccop{\txa}^T \bigbracket{\x^1-\xd{1}} < 0, \exists \x^0\in \Omega_{p^0}, \x^1\in \Omega_{p^1}.
		\end{equation}
		This contradiction completes the proof for the second subcase.
	\end{enumerate}
	In summary, we have proven that for each traveler class $k \in \{0,1\}$, there exists at least one route $j(k)$ such that $x_{j(k)}^{k,\dagger}=0$.

\section{Proof of Example~\ref{example:two_route_cost_comparison}\label{apdx:two_route_cost_comparison}}
\setcounter{equation}{0} \renewcommand{\theequation}{\ref{apdx:two_route_cost_comparison}.\arabic{equation}}

For a DUE with positive flows on both routes, the equilibrium condition requires:
		\begin{equation}\label{eq:two_route_ue_condition}
			\tilde{x}_1^\ast = \frac{a_2 d + b_2 - b_1}{a_1+a_2}, \tilde{x}_2^\ast = \frac{a_1 d + b_1 - b_2}{a_1+a_2}, -a_1 d < b_1 - b_2 < a_2 d.
		\end{equation}
			
		The total cost at DUE equals:
		\begin{equation}
			TC^\ast = c_1(\tilde{x}_1^\ast)\cdot \tilde{x}_1^\ast + c_2(\tilde{x}_2^\ast)\cdot \tilde{x}_2^\ast = \frac{a_1a_2d^2+a_1b_2d+a_2b_1d}{a_1+a_2}.
		\end{equation}
			
	At non-DUE MPE, Proposition \ref{prop:non_ue_mpe_zero_flow} implies each traveler class abandons one route. We first analyze the case where 0-step travelers exclusively use Route 1 (i.e., $x_1^{0,\dagger} = p^0 d$ and $x_2^{0,\dagger}=0$). This indicates that $c_1(\tilde{x}_{1}^{\dagger}) \leq c_2(\tilde{x}_{2}^{\dagger})$. For 1-step travelers, $\xd{1}$ can only take two possible values: $(0,p^1 d)^T$ or $(p^1 d,0)^T$. In the latter case, $c_1(\tilde{x}_{1}^{\dagger})=a_1d + b_1$ and $c_2(\tilde{x}_{2}^{\dagger})=b_2$. Due to the third inequality of Eq.~\eqref{eq:two_route_ue_condition}, we have $c_1(\tilde{x}_{1}^{\dagger}) > c_2(\tilde{x}_{2}^{\dagger})$, which contradicts $c_1(\tilde{x}_{1}^{\dagger}) \leq c_2(\tilde{x}_{2}^{\dagger})$. Therefore, $\xd{1}$ must equal $(0,p^1 d)^T$. As a result, we can deduce that $\xd{0}=(p^0d, 0)$ and $\xd{1}=(0, p^1d)$ constitute a non-DUE MPE.

	At the non-DUE MPE of $\xd{0}=(p^0d, 0)$ and $\xd{1}=(0, p^1d)$, $c_1(\tilde{x}_{1}^{\dagger}) \leq c_2(\tilde{x}_{2}^{\dagger})$ indicates that $a_1p^0d+b_1\leq a_2(1-p^0)d+b_2$, which simplifies to $p^0\leq \frac{b_2-b_1+a_2d}{(a_1+a_2)d}$. The total cost at this non-DUE MPE becomes:
	\begin{equation}
		TC^\dagger = (a_1p^0d+b_1)p^0d + (a_2(1-p^0)d+b_2)(1-p^0)d.
	\end{equation}
			
	The difference between $TC^\dagger$ and $TC^\ast$ can be expressed as a quadratic function of $p^0$:
	\begin{equation}
		\Delta TC = TC^\dagger - TC^\ast = (a_1 + a_2)d^2 \cdot (p^0)^2 - [2a_2d^2 - (b_1 - b_2)d] \cdot p^0 + \frac{a_2d(a_2d + b_2 - b_1)}{a_1 + a_2}.
	\end{equation}
	$\Delta TC$ has two roots: $\frac{a_2}{a_1+a_2}$ and $\frac{b_2-b_1+a_2d}{(a_1+a_2)d}$. Since the coefficient $(a_1+a_2)d^2$ is positive ($a_1,a_2>0$), the parabola opens upward. If $b_2 < b_1$, the left root is $\frac{b_2-b_1+a_2d}{(a_1+a_2)d}$. Any $p^0\leq \frac{b_2-b_1+a_2d}{(a_1+a_2)d}$ would lie to the left of this root, resulting in $TC^\dagger > TC^\ast$. If $b_2 > b_1$, the right root is $\frac{b_2-b_1+a_2d}{(a_1+a_2)d}$. Any value $p^0\in \bigbracket{\frac{a_2}{a_1+a_2},\frac{b_2-b_1+a_2d}{(a_1+a_2)d}}$ would yield a lower $TC^\dagger$ than $TC^\ast$.
			
	The derivation for the case where 0-step travelers use only route 2 follows analogous logic. In this case, $\xd{0}=(0, p^0d)$ and $\xd{1}=(p^1d, 0)$ constitute the other non-DUE MPE. Through similar deduction logic, we can prove that at this second non-DUE MPE, $TC^\dagger < TC^\ast$ when $b_2 < b_1$ and $p^1\in \bigbracket{\frac{b_2-b_1+a_2d}{(a_1+a_2)d}, \frac{a_2}{a_1+a_2}}$.

\section{Derivation of Fact~\ref{fact:logit_jacob}}\label{apdx:logit_jacob}
\setcounter{equation}{0} \renewcommand{\theequation}{\ref{apdx:logit_jacob}.\arabic{equation}}
For the 0-step travelers, take the Jacobian, $J[\cdot]$, of the dynamical system in Eq.~\eqref{eq:logit_model} w.r.t. $\xt{k}$:
\begin{align*}
	\sJop{\frac{\alpha{p^0} \dPhiop{\theta}{\cpxtt{0}} +(1-\alpha) \xt{0}}{\xt{0}}} &= \alpha \sJop{\frac{p^0\dPhiop{\theta}{\cpxtt{0}}}{\xt{0}}} + (1-\alpha)I \\
	&= \alpha p^{0}\dUpop{\theta}{\ccop{\txt}}\cdot \sDop{\txt} + (1-\alpha)I, \numberthis
\end{align*}
and
\begin{align*}
	\sJop{\frac{\alpha{p^0} \dPhiop{\theta}{\cpxtt{0}} +(1-\alpha)\xt{0}}{\xt{k\neq 0}}} &= \alpha \sJop{\frac{p^0\Phi^\theta[\cpxtt{0}]}{\xt{k\neq 0}}} \\
	&= \alpha p^{0}\dUpop{\theta}{\ccop{\txt}}\cdot D[\txt]. \numberthis
\end{align*}\par
For the 1-step travelers, first note that for $k=0,1,2$,
\begin{align*}
	\sJop{\frac{\cpxtt{1}}{\xt{k}}} &= \sJop{\frac{\cpxtt{1}}{\pxtt{1}}} \cdot \bigbracket{ \sJop{\frac{\halpha\dPhiop{\htheta}{\cpxtt{0}}}{\txt}}\cdot \sJop{\frac{\txt}{\xt{k}}} + (1-\halpha)I} \\
	& = \sDop{\pxtt{1}}  \cdot \bigbracket{\halpha\dUpop{\htheta}{\ccop{\txt}}\cdot \sDop{\txt} + (1-\halpha)I}. \numberthis
\end{align*}\par
Hence,
\begin{align*}
	& \sJop{\frac{\alpha{p^1} \dPhiop{\theta}{\cpxtt{1}} + (1-\alpha)\xt{1}}{\xt{1}}} = \alpha{p^1}\sJop{\frac{\dPhiop{\theta}{\cpxtt{1}}}{\cpxtt{1}}}\cdot \sJop{\frac{\cpxtt{1}}{\xt{1}}} + (1-\alpha)I \\
	&\quad = \alpha p^1 \dUpop{\theta}{\cpxtt{1}}\cdot \sDop{\pxtt{1}}  \cdot \bigbracket{\halpha\dUpop{\htheta}{\ccop{\txt}}\cdot \sDop{\txt} + (1-\halpha)I} + (1-\alpha)I, \numberthis
\end{align*}
and
\begin{align*}
	& \sJop{\frac{\alpha{p^1} \dPhiop{\theta}{\cpxtt{1}} + (1-\alpha)\xt{1}}{\xt{k\neq 1}}} \\
	&\quad = \alpha{p^1} \dUpop{\theta}{\cpxtt{1}}\cdot \sDop{\pxtt{1}}  \cdot \bigbracket{\halpha\dUpop{\htheta}{\ccop{\txt}}\cdot \sDop{\txt} + (1-\halpha)I}. \numberthis
\end{align*}\par
For the 2-step travelers, note that for $k=0,1,2$,
\begin{align*}\label{eq:logit_jacobian_k2_temp}
	\sJop{\frac{\pxtt{2}}{\xt{k}}} &= \sJop{\frac{\halpha q_2^0\dPhiop{\htheta}{\cpxtt{0}}}{\txt}}\cdot \sJop{\frac{\txt}{\xt{k}}} + \sJop{\frac{\halpha{q_2^1}\dPhiop{\htheta}{\cpxtt{1}}}{\txt}}\cdot \sJop{\frac{\txt}{\xt{k}}} + (1-\halpha)I \\
	& = \halpha{q_2^0}\dUpop{\htheta}{\ccop{\txt}}\cdot \sDop{\txt} + \halpha{q_2^1} \dPhiop{\htheta}{\cpxtt{1}} \\
	&\cdot \bigbracket{ \sDop{\pxtt{1}}\cdot \bigbracket{\halpha\dUpop{\htheta}{\ccop{\txt}} \sDop{\txt} + (1-\halpha)I} } + (1-\halpha)I. \numberthis
\end{align*}\par
Finally,
\begin{align*}
	&\sJop{\frac{\alpha{p^2} \dPhiop{\theta}{\cpxtt{2}} + (1-\alpha)\xt{2}}{\xt{k\neq 2}}} = \alpha{p^2} \dUpop{\theta}{\cpxtt{2}}\cdot \sDop{\pxtt{2}}\cdot \sJop{\frac{\pxtt{2}}{\xt{k}}},\numberthis
\end{align*}
and
\begin{align*}
	& \sJop{\frac{\alpha{p^2} \dPhiop{\theta}{\cpxtt{2}} + (1-\alpha)\xt{2}}{\xt{k=2}}} = \alpha{p^2} \dUpop{\theta}{\cpxtt{2}}\cdot \sDop{\pxtt{2}}\cdot \sJop{\frac{\pxtt{2}}{\xt{k}}} + (1-\alpha)I, \numberthis
\end{align*}
where $\sJop{\frac{\pxtt{2}}{\xt{k}}}$ is given by Eq.~\eqref{eq:logit_jacobian_k2_temp}.\par
Integrating the above blocks into $J\Phi$ yields Fact~\ref{fact:logit_jacob}.

\section{Proof of Proposition~\ref{prop:logit_stable_general}}\label{apdx:logit_stable_general}
\setcounter{equation}{0} \renewcommand{\theequation}{\ref{apdx:logit_stable_general}.\arabic{equation}}
It is well-known that the eigenvalues of the product of two positive semidefinite matrices are real and nonnegative. Thanks to the negative semidefiniteness of the Jacobian of Logit operator (Lemma~\ref{lemma:projection_property}) and positive semidefiniteness of the Jacobian of the route cost function (Assumption \ref{assumption:cost_psd}), $\rho_i \leq 0, \forall i$.\par
When $\theta=\htheta$, according to Proposition~\ref{prop:logit_multi_eqs}, at the SUE, the predicted flow $\pxs{0} = \pxs{1} = \pxs{2} = \txs$. Hence, the Jacobians of route cost function $D_2=D_1=D_0=D_{\star}$, where $D_{\star}$ is the Jacobian of the route cost function evaluated at the SUE $\txs$. Using Fact~\ref{fact:logit_jacob}, and denoting $A$ as $\dUpop{\theta}{\ccop{\txt}}\sDop{\txt}$, the eigenvalues of $J\Phi$, denoted as $\lambda$, are the roots of the following characteristic polynomial:
\begin{equation}\label{eq:logit_k3_det_origin}
	\bigvbracket{\lambda I -J\Phi} = \begin{vmatrix}
   	& \lambda I - X - (1-\alpha)I, & -X & -X \\
   	& -Y & \lambda I -Y -(1-\alpha)I & -Y\\
   	& -Z & -Z & \lambda I -Z -(1-\alpha)I\\
    \end{vmatrix} = 0,
\end{equation}
where $X=\alpha p^{0}A$, $Y=\alpha\halpha p^1A^2+\alpha(1-\halpha)p^1A$ and $Z=\alpha\halpha^2p^2q_2^1A^3 + (\alpha\halpha p^2-\alpha\halpha^2p^2q_2^1)A^2 + \alpha p^2(1-\halpha)A$ with $q^1_2=\frac{p^1}{p^0+p^1}$. Assume that $\rho_i$ is the $i$-th eigenvalue of $A$.\par
Consider first the case of $|K|=1$ where $p^0=1$ and $p^1=p^2=0$. The above determinant degenerates into $|\lambda I-\alpha A-(1-\alpha)I|=0$. Thus $\rho_i$ of the matrix $A$ is related with $\lambda_i$ via:
\begin{equation}\label{eq:logit_k0_stable_relation}
	\lambda_i - f(\rho_i; \alpha) = 0,
\end{equation}
where $f(\rho_i; \alpha) \equiv \alpha\rho_i + (1-\alpha)$.\par
For the case of $|K|=2$ where $0<p^0<1$, $0<p^1<1$ and $p^2=0$. $J\Phi$ takes the four block submatrices in the upper left corner. Since $\lambda I-X-(1-\alpha)I$ commutes with $-X$, we can use Theorem~\ref{thm:block_2_by_2_det} to obtain $\bigvbracket{\lambda I - J\Phi}=\bigvbracket{(\lambda-1+\alpha)I(\lambda I- \halpha\alpha p^1A^2 - \alpha(1-\halpha)p^1A - \alpha p^0A - (1-\alpha)I)}$. Then $\rho_i$ corresponds to the two eigenvalues of $J\Phi$, $\lambda_{i,1}$ and $\lambda_{i,2}$, by the following equation:
\begin{equation}\label{eq:logit_k1_stable_relation}
	\bigbracket{\lambda_{i,1}-1+\alpha} \bigbracket{\lambda_{i,2} - g(\alpha,\halpha,p^0,p^1)} = 0,
\end{equation}
where $g(\rho_i; \alpha,\halpha,p^0,p^1) \equiv \halpha\alpha p^1\rho_i^2 + (\alpha-\alpha\halpha p^1)\rho_i + (1-\alpha)$.\par
For $|K|=3$, subtracting the third ``block column'' of the determinant from the first and second block columns, and then adding the first and second ``block rows'' to the third block row, the determinant in Eq.~\eqref{eq:logit_k3_det_origin} becomes an upper triangular matrix:
\begin{equation}
	|\lambda I -J\Phi| = \begin{vmatrix}
   	& \lambda I - (1-\alpha)I, & 0 & -X \\
   	& 0 & \lambda I -(1-\alpha)I & -Y\\
   	& 0 & 0 & \lambda I -X-Y-Z -(1-\alpha)I\\
    \end{vmatrix} = 0,
\end{equation}
which, according to Theorem~\ref{thm:upper_block_triangle}, yields:
\begin{equation}
	\bigvbracket{\lambda I-J\Phi} = \bigvbracket{
		\lambda I - (1-\alpha)I
	}^2 \cdot
	\bigvbracket{\lambda I -X-Y-Z-(1-\alpha)I} = 0. \numberthis
\end{equation}\par
Each $\rho_i$ is then related to the corresponding three eigenvalues of $J\Phi$, $\lambda_{i,1}$, $\lambda_{i,2}$ and $\lambda_{i,3}$, by the following equation:
\begin{equation}\label{eq:logit_k2_stable_relation}
	\bigbracket{\lambda_{i,1}-1+\alpha} \bigbracket{\lambda_{i,2}-1+\alpha} \bigbracket{\lambda_{i,3} - \psi(\rho_i;\alpha,\halpha,p^0,p^1,p^2)} = 0,
\end{equation}
where $\psi(\rho_i;\alpha,\halpha,p^0,p^1,p^2) \equiv \alpha\halpha^2 p^2q_2^1\rho_i^3 + (\alpha\halpha-\alpha\halpha p^0-\alpha\halpha^2p^2q_2^1)\rho_i^2 + (\alpha\halpha p^0-\alpha\halpha+\alpha)\rho_i +(1-\alpha)$.\par
Referring to Theorem~\ref{thm:eigen_stability}, the system is stable if and only if all the $\lambda_{i}'\text{s} \in (-1,1)$. Observing Eqs.~\eqref{eq:logit_k0_stable_relation}, \eqref{eq:logit_k1_stable_relation} and \eqref{eq:logit_k2_stable_relation}, since $1 - \alpha \in (0, 1)$ when $0<\alpha< 1$, the stability depends on $f(\rho_i; \alpha)$, $g(\rho_i; \alpha,\halpha,p^0,p^1)$ and $\psi(\rho_i;\alpha,\halpha,p^0,p^1,p^2)$, respectively, for $|K|=1$, $|K|=2$ and $|K|=3$. Note that $\psi(\rho_i;\alpha,\halpha,p^0,p^1,p^2)$ contains $g(\rho_i; \alpha,\halpha,p^0,p^1)$ as a special case when $p^2=0$ (and thus $q_2^1=\frac{p^1}{p^0+p^1}=p^0$). Moreover, $g(\rho_i; \alpha,\halpha,p^0,p^1)$ contains $f(\rho_i; \alpha)$ as a special case when $p^0=1$, $p^1=0$ and $\halpha=\alpha$. Therefore, we can simply consult whether $\psi(\rho_i;\alpha,\halpha,p^0,p^1,p^2) \in (-1,1)$ or not to check the local stability.

\section{Proof of Proposition~\ref{prop:logit_stability_neq_k2}}\label{apdx:logit_stability_neq_k2}
\setcounter{equation}{0} \renewcommand{\theequation}{\ref{apdx:logit_stability_neq_k2}.\arabic{equation}}

When $|K|=1$, $p^1=p^2=0$ and thus $-1 < \psi(\rho_i;\alpha,\halpha,p^0,p^1,p^2) < 1$ becomes $-1 < \alpha\rho_i+(1-\alpha)$. The stablility condition associated with $\rho_i$'s is $\frac{\alpha-2}{\alpha} < \rho_i < 0, \forall i$.\par
When $|K|=2$, $p^2=0$ and thus $\psi(\rho_i;\alpha,\halpha,p^0,p^1,p^2)$ becomes $\halpha\alpha (1-p^0)\rho_i^2 + (\alpha\halpha p^0-\alpha\halpha+\alpha)\rho_i +1-\alpha$, which is denoted as $f(\rho_i; \alpha,\halpha, p^0)$. For this quadratic function of $\rho_i$, we know that $f(0; \alpha, p^0) = 1-\alpha \in [0,1)$ and axis of symmetry is $\rho_i=\frac{1}{2} \bigbracket{\frac{1}{\halpha (p^0-1)}+1} < 0$ for $0<\halpha<1$ and $0<p^0<1$. Two roots of $f(\rho_i; \alpha,\halpha,p^0)=1$ are 1 and $\frac{1}{(p^0-1)\halpha}$, respectively. Determining the stable region (i.e., $\rho_i$'s range satisfying $-1<f(\rho_i; \alpha, \halpha, p^0)<1,\forall i$) thus encounters two cases, depending on the minimum, $f^{\ast}(\rho_i; \alpha, \halpha, p^0) = 1-\alpha + \frac{\alpha(1-\halpha+\halpha p^0)^2}{4\halpha(p^0-1)}$.
\begin{itemize}
	\item When $f^{\ast}(\rho_i; \alpha, \halpha, p^0) > -1$, as long as $\rho_i > \frac{1}{(p^0-1)\halpha}$, $f(\rho_i; \alpha, p^0)$ lies within $(-1,1)$.
	\item When $f^{\ast}(\rho_i; \alpha, p^0) < -1$. The quadratic function always has two roots at $f(\rho_i; \alpha, \halpha,p^0)=-1$, denoted $\rho_{i-}$ and $\rho_{i+}$, respectively. Some algebraic calculations yield that $\rho_{i-}=\frac{1}{2} \bigbracket{-\sqrt{\frac{\alpha (1+\halpha-\halpha p^0)^2+8 \halpha  (p^0-1)}{\alpha  \halpha ^2 (p^0-1)^2}}+\frac{1}{\halpha  (p^0-1)}+1}$ and $\rho_{i+}=\frac{1}{2} \bigbracket{\sqrt{\frac{\alpha (1+\halpha-\halpha p^0)^2+8 \halpha  (p^0-1)}{\alpha  \halpha ^2 (p^0-1)^2}}+\frac{1}{\halpha (p^0-1)}+1}$. In this case, the stable region is separated. One region ranges $\frac{1}{(p-1)\halpha}$ to $\rho_{i-}$ and the other ranges from $\rho_{i+}$ to 0.
\end{itemize}

 \end{appendices}

\onehalfspacing
\begin{footnotesize}
\bibliographystyle{informs2014trsc.bst}
\bibliography{literature.bib}

\begin{thebibliography}{65}
\providecommand{\natexlab}[1]{#1}
\providecommand{\url}[1]{\texttt{#1}}
\providecommand{\urlprefix}{URL }

\bibitem[{Antsaklis \protect\BIBand{} Michel(2006)}]{antsaklis2006linear}
Antsaklis PJ, Michel AN, 2006 \emph{Linear systems} (Springer Science \& Business Media).

\bibitem[{Bie \protect\BIBand{} Lo(2010)}]{bie2010stability}
Bie J, Lo HK, 2010 \emph{Stability and attraction domains of traffic equilibria in a day-to-day dynamical system formulation}. \emph{Transportation Research Part B: Methodological} 44(1):90--107.

\bibitem[{Brown, Camerer, \protect\BIBand{} Lovallo(2013)}]{brown2013estimating}
Brown AL, Camerer CF, Lovallo D, 2013 \emph{Estimating structural models of equilibrium and cognitive hierarchy thinking in the field: The case of withheld movie critic reviews}. \emph{Management Science} 59(3):733--747.

\bibitem[{Camerer, Ho, \protect\BIBand{} Chong(2004)}]{camerer2004cognitive}
Camerer CF, Ho TH, Chong JK, 2004 \emph{A cognitive hierarchy model of games}. \emph{The Quarterly Journal of Economics} 119(3):861--898.

\bibitem[{Cantarella \protect\BIBand{} Cascetta(1995)}]{cantarella1995dynamic}
Cantarella GE, Cascetta E, 1995 \emph{Dynamic processes and equilibrium in transportation networks: towards a unifying theory}. \emph{Transportation Science} 29(4):305--329.

\bibitem[{Chong, Ho, \protect\BIBand{} Camerer(2016)}]{chong2016generalized}
Chong JK, Ho TH, Camerer C, 2016 \emph{A generalized cognitive hierarchy model of games}. \emph{Games and Economic Behavior} 99:257--274.

\bibitem[{Costa-Gomes \protect\BIBand{} Crawford(2006)}]{costa2006cognition}
Costa-Gomes MA, Crawford VP, 2006 \emph{Cognition and behavior in two-person guessing games: An experimental study}. \emph{American Economic Review} 96(5):1737--1768.

\bibitem[{Cui \protect\BIBand{} Zhang(2017)}]{cui2017cognitive}
Cui TH, Zhang Y, 2017 \emph{Cognitive hierarchy in capacity allocation games}. \emph{Management Science} 64(3):1250--1270.

\bibitem[{Facchinei \protect\BIBand{} Pang(2007)}]{facchinei2007finite}
Facchinei F, Pang JS, 2007 \emph{Finite-dimensional variational inequalities and complementarity problems} (Springer Science \& Business Media).

\bibitem[{Friesz et~al.(1994)Friesz, Bernstein, Mehta, Tobin, \protect\BIBand{} Ganjalizadeh}]{friesz1994day}
Friesz TL, Bernstein D, Mehta NJ, Tobin RL, Ganjalizadeh S, 1994 \emph{Day-to-day dynamic network disequilibria and idealized traveler information systems}. \emph{Operations Research} 42(6):1120--1136.

\bibitem[{Fudenberg \protect\BIBand{} Levine(1998)}]{fudenberg1998learning}
Fudenberg D, Levine D, 1998 \emph{Learning in games}. \emph{European Economic Review} 42(3-5):631--639.

\bibitem[{Gao \protect\BIBand{} Pavel(2017)}]{gao2017properties}
Gao B, Pavel L, 2017 \emph{On the properties of the softmax function with application in game theory and reinforcement learning}. \emph{arXiv preprint arXiv:1704.00805} .

\bibitem[{Gaode-Map(2020)}]{pengpai2020gaode}
Gaode-Map, 2020 \emph{Gaode map: Promoting proactive and refined new traffic management to empower smart new travel (translated from chinese)}. \urlprefix\url{https://www.thepaper.cn/newsDetail_forward_10285810}, [Online; accessed 2025-03-19].

\bibitem[{Goldfarb \protect\BIBand{} Xiao(2011)}]{goldfarb2011thinks}
Goldfarb A, Xiao M, 2011 \emph{Who thinks about the competition? managerial ability and strategic entry in us local telephone markets}. \emph{American Economic Review} 101(7):3130--61.

\bibitem[{Guo, Yang, \protect\BIBand{} Huang(2018)}]{guo2018we}
Guo RY, Yang H, Huang HJ, 2018 \emph{Are we really solving the dynamic traffic equilibrium problem with a departure time choice?} \emph{Transportation Science} 52(3):603--620.

\bibitem[{Guo, Yang, \protect\BIBand{} Huang(2023)}]{guo2023day}
Guo RY, Yang H, Huang HJ, 2023 \emph{The day-to-day departure time choice of heterogeneous commuters under an anonymous toll charge for system optimum}. \emph{Transportation Science} 57(3):661--684.

\bibitem[{Guo et~al.(2015)Guo, Yang, Huang, \protect\BIBand{} Tan}]{guo2015link}
Guo RY, Yang H, Huang HJ, Tan Z, 2015 \emph{Link-based day-to-day network traffic dynamics and equilibria}. \emph{Transportation Research Part B: Methodological} 71:248--260.

\bibitem[{Guo et~al.(2016)Guo, Yang, Huang, \protect\BIBand{} Tan}]{guo2016day}
Guo RY, Yang H, Huang HJ, Tan Z, 2016 \emph{Day-to-day flow dynamics and congestion control}. \emph{Transportation Science} 50(3):982--997.

\bibitem[{Han \protect\BIBand{} Du(2012)}]{han2012link}
Han L, Du L, 2012 \emph{On a link-based day-to-day traffic assignment model}. \emph{Transportation Research Part B: Methodological} 46(1):72--84.

\bibitem[{Han et~al.(2017)Han, Wang, Lo, Zhu, \protect\BIBand{} Cai}]{han2017discrete}
Han L, Wang DZ, Lo HK, Zhu C, Cai X, 2017 \emph{Discrete-time day-to-day dynamic congestion pricing scheme considering multiple equilibria}. \emph{Transportation Research Part B: Methodological} 104:1--16.

\bibitem[{He, Guo, \protect\BIBand{} Liu(2010)}]{he2010link}
He X, Guo X, Liu HX, 2010 \emph{A link-based day-to-day traffic assignment model}. \emph{Transportation Research Part B: Methodological} 44(4):597--608.

\bibitem[{He \protect\BIBand{} Liu(2012)}]{he2012modeling}
He X, Liu HX, 2012 \emph{Modeling the day-to-day traffic evolution process after an unexpected network disruption}. \emph{Transportation Research Part B: Methodological} 46(1):50--71.

\bibitem[{He \protect\BIBand{} Peeta(2016)}]{he2016marginal}
He X, Peeta S, 2016 \emph{A marginal utility day-to-day traffic evolution model based on one-step strategic thinking}. \emph{Transportation Research Part B: Methodological} 84:237--255.

\bibitem[{Ho \protect\BIBand{} Su(2013)}]{ho2013dynamic}
Ho TH, Su X, 2013 \emph{A dynamic level-k model in sequential games}. \emph{Management Science} 59(2):452--469.

\bibitem[{Horn \protect\BIBand{} Johnson(2012)}]{horn2012matrix}
Horn RA, Johnson CR, 2012 \emph{Matrix analysis} (Cambridge university press).

\bibitem[{Horowitz(1984)}]{horowitz1984stability}
Horowitz JL, 1984 \emph{The stability of stochastic equilibrium in a two-link transportation network}. \emph{Transportation Research Part B: Methodological} 18(1):13--28.

\bibitem[{Hossain \protect\BIBand{} Morgan(2013)}]{hossain2013markets}
Hossain T, Morgan J, 2013 \emph{When do markets tip? a cognitive hierarchy approach}. \emph{Marketing Science} 32(3):431--453.

\bibitem[{Jin(2007)}]{jin2007dynamical}
Jin WL, 2007 \emph{A dynamical system model of the traffic assignment problem}. \emph{Transportation Research Part B: Methodological} 41(1):32--48.

\bibitem[{Jin(2020)}]{jin2020stable}
Jin WL, 2020 \emph{Stable day-to-day dynamics for departure time choice}. \emph{Transportation Science} 54(1):42--61.

\bibitem[{Jin(2021)}]{jin2021stable}
Jin WL, 2021 \emph{Stable local dynamics for day-to-day departure time choice}. \emph{Transportation Research Part B: Methodological} 149:463--479.

\bibitem[{Khalil(2002)}]{khalil2002nonlinear}
Khalil HK, 2002 \emph{Nonlinear Systems, 3rd Edition} (Prentice Hall, Upper Saddle River).

\bibitem[{Kinderlehrer \protect\BIBand{} Stampacchia(2000)}]{kinderlehrer2000introduction}
Kinderlehrer D, Stampacchia G, 2000 \emph{An introduction to variational inequalities and their applications} (SIAM).

\bibitem[{K{\"o}hler \protect\BIBand{} Strehler(2019)}]{kohler2019traffic}
K{\"o}hler E, Strehler M, 2019 \emph{Traffic signal optimization: combining static and dynamic models}. \emph{Transportation Science} 53(1):21--41.

\bibitem[{Krichene, Drigh{\`e}s, \protect\BIBand{} Bayen(2015)}]{krichene2015online}
Krichene W, Drigh{\`e}s B, Bayen AM, 2015 \emph{Online learning of nash equilibria in congestion games}. \emph{SIAM Journal on Control and Optimization} 53(2):1056--1081.

\bibitem[{Kumar \protect\BIBand{} Peeta(2015)}]{kumar2015day}
Kumar A, Peeta S, 2015 \emph{A day-to-day dynamical model for the evolution of path flows under disequilibrium of traffic networks with fixed demand}. \emph{Transportation Research Part B: Methodological} 80:235--256.

\bibitem[{Lamotte \protect\BIBand{} Geroliminis(2021)}]{lamotte2021monotonicity}
Lamotte R, Geroliminis N, 2021 \emph{Monotonicity in the trip scheduling problem}. \emph{Transportation Research Part B: Methodological} 146:14--25.

\bibitem[{Lau(2020)}]{google_maps_ai_2020}
Lau J, 2020 \emph{Google maps 101: How ai helps predict traffic and determine routes}. \url{https://blog.google/products/maps/google-maps-101-how-ai-helps-predict-traffic-and-determine-routes/}, accessed: [Sep 18, 2024].

\bibitem[{Li et~al.(2024)Li, Wang, Feng, Xie, \protect\BIBand{} Nie}]{li2024day}
Li J, Wang Q, Feng L, Xie J, Nie Y, 2024 \emph{A day-to-day dynamical approach to the most likely user equilibrium problem}. \emph{Transportation Science} 58(6):1193--1213.

\bibitem[{Li, Liu, \protect\BIBand{} Nie(2018)}]{li2018managing}
Li R, Liu X, Nie YM, 2018 \emph{Managing partially automated network traffic flow: Efficiency vs. stability}. \emph{Transportation Research Part B: Methodological} 114:300--324.

\bibitem[{Liu et~al.(2020)Liu, Guo, Easa, Yan, Wei, \protect\BIBand{} Tang}]{liu2020experimental}
Liu S, Guo L, Easa SM, Yan H, Wei H, Tang Y, 2020 \emph{Experimental study of day-to-day route-choice behavior: evaluating the effect of atis market penetration}. \emph{Journal of Advanced Transportation} 2020(1):8393724.

\bibitem[{Liu \protect\BIBand{} Szeto(2020)}]{liu2020learning}
Liu W, Szeto WY, 2020 \emph{Learning and managing stochastic network traffic dynamics with an aggregate traffic representation}. \emph{Transportation Research Part B: Methodological} 137:19--46.

\bibitem[{Lo(2017)}]{lo2017adaptive}
Lo AW, 2017 \emph{Adaptive markets: Financial evolution at the speed of thought} (Princeton University Press).

\bibitem[{Nagurney(2000)}]{nagurney2000multiclass}
Nagurney A, 2000 \emph{A multiclass, multicriteria traffic network equilibrium model}. \emph{Mathematical and Computer Modelling} 32(3-4):393--411.

\bibitem[{Nagurney \protect\BIBand{} Zhang(1997)}]{nagurney1997projected}
Nagurney A, Zhang D, 1997 \emph{Projected dynamical systems in the formulation, stability analysis, and computation of fixed-demand traffic network equilibria}. \emph{Transportation Science} 31(2):147--158.

\bibitem[{Penot(2005)}]{penot2005continuity}
Penot JP, 2005 \emph{Continuity properties of projection operators}. \emph{Journal of Inequalities and Applications} 2005(5):1--13.

\bibitem[{Qi et~al.(2023)Qi, Jia, Qu, \protect\BIBand{} He}]{qi2023investigating}
Qi H, Jia N, Qu X, He Z, 2023 \emph{Investigating day-to-day route choices based on multi-scenario laboratory experiments, part i: Route-dependent attraction and its modeling}. \emph{Transportation Research Part A: Policy and Practice} 167:103553.

\bibitem[{Sandholm(2010)}]{sandholm2010population}
Sandholm WH, 2010 \emph{Population games and evolutionary dynamics} (MIT press).

\bibitem[{Selten et~al.(2007)Selten, Chmura, Pitz, Kube, \protect\BIBand{} Schreckenberg}]{selten2007commuters}
Selten R, Chmura T, Pitz T, Kube S, Schreckenberg M, 2007 \emph{Commuters route choice behaviour}. \emph{Games and Economic Behavior} 58(2):394--406.

\bibitem[{Smith(1979{\natexlab{a}})}]{smith1979traffic}
Smith M, 1979{\natexlab{a}} \emph{Traffic control and route-choice; a simple example}. \emph{Transportation Research Part B: Methodological} 13(4):289--294.

\bibitem[{Smith(1979{\natexlab{b}})}]{smith1979existence}
Smith MJ, 1979{\natexlab{b}} \emph{The existence, uniqueness and stability of traffic equilibria}. \emph{Transportation Research Part B: Methodological} 13(4):295--304.

\bibitem[{Smith(1983)}]{smith1983existence}
Smith MJ, 1983 \emph{The existence and calculation of traffic equilibria}. \emph{Transportation Research Part B: Methodological} 17(4):291--303.

\bibitem[{Smith(1984)}]{smith1984stability}
Smith MJ, 1984 \emph{The stability of a dynamic model of traffic assignment: an application of a method of {L}yapunov}. \emph{Transportation Science} 18(3):245--252.

\bibitem[{Sun et~al.(2020)Sun, Shi, Han, Wang, \protect\BIBand{} Shu}]{sun2020dynamic}
Sun N, Shi H, Han G, Wang B, Shu L, 2020 \emph{Dynamic path planning algorithms with load balancing based on data prediction for smart transportation systems}. \emph{Ieee Access} 8:15907--15922.

\bibitem[{Tan, Yang, \protect\BIBand{} Guo(2015)}]{tan2015dynamic}
Tan Z, Yang H, Guo RY, 2015 \emph{Dynamic congestion pricing with day-to-day flow evolution and user heterogeneity}. \emph{Transportation Research Part C: Emerging Technologies} 61:87--105.

\bibitem[{Wang, Guo, \protect\BIBand{} Huang(2021)}]{wang2021day}
Wang SY, Guo RY, Huang HJ, 2021 \emph{Day-to-day route choice in networks with different sets for choice: experimental results}. \emph{Transportmetrica B: Transport Dynamics} 9(1):712--745.

\bibitem[{Waze-user(2024)}]{reddit2024waze}
Waze-user, 2024 \emph{Waze not always giving me the fastest route.} \urlprefix\url{https://www.reddit.com/r/waze/comments/1ego5z4/waze_not_always_giving_me_the_fastest_route_is_it/}, [Online; accessed 2025-03-19].

\bibitem[{Waze-user(2025)}]{reddit2025waze}
Waze-user, 2025 \emph{Why doesn't waze give me the quickest route?} \urlprefix\url{https://www.reddit.com/r/waze/comments/1hdnkcz/why_doesnt_waze_give_me_the_quickest_route/}, [Online; accessed 2025-03-19].

\bibitem[{Wu, Yin, \protect\BIBand{} Lynch(2024)}]{wu2024multiday}
Wu M, Yin Y, Lynch JP, 2024 \emph{Multiday user equilibrium with strategic commuters}. \emph{Transportation Science} .

\bibitem[{Xiao et~al.(2019)Xiao, Shen, Xu, Li, Yang, \protect\BIBand{} Yin}]{xiao2019day}
Xiao F, Shen M, Xu Z, Li R, Yang H, Yin Y, 2019 \emph{Day-to-day flow dynamics for stochastic user equilibrium and a general {L}yapunov function}. \emph{Transportation Science} 53(3):683--694.

\bibitem[{Xiao, Yang, \protect\BIBand{} Ye(2016)}]{xiao2016physics}
Xiao F, Yang H, Ye H, 2016 \emph{Physics of day-to-day network flow dynamics}. \emph{Transportation Research Part B: Methodological} 86:86--103.

\bibitem[{Yang \protect\BIBand{} Zhang(2009)}]{yang2009day}
Yang F, Zhang D, 2009 \emph{Day-to-day stationary link flow pattern}. \emph{Transportation Research Part B: Methodological} 43(1):119--126.

\bibitem[{Ye, Xiao, \protect\BIBand{} Yang(2018)}]{ye2018exploration}
Ye H, Xiao F, Yang H, 2018 \emph{Exploration of day-to-day route choice models by a virtual experiment}. \emph{Transportation Research Part C: Emerging Technologies} 94:220--235.

\bibitem[{Ye, Xiao, \protect\BIBand{} Yang(2021)}]{ye2021day}
Ye H, Xiao F, Yang H, 2021 \emph{Day-to-day dynamics with advanced traveler information}. \emph{Transportation Research Part B: Methodological} 144:23--44.

\bibitem[{Zhang et~al.(2015)Zhang, Guan, Ma, \protect\BIBand{} Tian}]{zhang2015nonlinear}
Zhang W, Guan W, Ma J, Tian J, 2015 \emph{A nonlinear pairwise swapping dynamics to model the selfish rerouting evolutionary game}. \emph{Networks and Spatial Economics} 15(4):1075--1092.

\bibitem[{Zhou, Xu, \protect\BIBand{} Meng(2020)}]{zhou2020discrete}
Zhou B, Xu M, Meng Q, 2020 \emph{A discrete day-to-day link flow dynamic model considering travelers’ heterogeneous inertia patterns}. \emph{Transportmetrica A: Transport Science} 16(3):1400--1428.

\end{thebibliography}
\end{footnotesize}
\end{document}